\documentclass[a4paper,11pt,nofootinbib]{article}
\pdfoutput=1

\usepackage{cite}
\usepackage{footmisc}
\usepackage{jheppub,fancyhdr,graphicx,epstopdf,color,amsmath,cases,slashed,hyperref}
\usepackage{makecell}
\usepackage{caption, subcaption}  
\usepackage{dcolumn}   
\usepackage{amsmath, amssymb}        
\usepackage{color,latexsym, array,multirow,  verbatim, enumerate,cancel}
\usepackage{slashed}
\usepackage{bigcenter}
\usepackage[utf8]{inputenc}
\usepackage{footnote}
\makesavenoteenv{tabular}
\makesavenoteenv{table}

\def\issue(#1,#2,#3){{\bf #1}, #2 (#3)}

\catcode`\@=11
\def\lsim{\mathrel{\mathpalette\@versim<}}
\def\gsim{\mathrel{\mathpalette\@versim>}}
\def\@versim#1#2{\vcenter{\offinterlineskip
\ialign{$\m@th#1\hfil##\hfil$\crcr#2\crcr\sim\crcr } }}
\catcode`\@=12

\catcode`@=12

\newcommand{\met}{$\cancel E_T$}

\newcommand{\newc}{\newcommand}
\newc{\wt}{\widetilde}
\newc{\ra}{\rightarrow}

\def\beq {\begin{equation}}
\def\eeq {\end{equation}}
\def\bi {\begin{itemize}}
\def\ei {\end{itemize}}
\def\bea {\begin{eqnarray}}
\def\eea {\end{eqnarray}}

\def \met{\slashed{E}_T}

\newcommand{\br}{\begin{eqnarray}}
\newcommand{\er}{\end{eqnarray}}
\newcommand{\be}{\begin{equation}}
\newcommand{\ee}{\end{equation}}

\newcommand{\ch}{\widetilde \chi^\pm}

\def \ch2p {{\wt\chi_2^+}}

\def \ch2m {{\wt\chi_2^-}}

\newc{\dmchi}{\Delta m_{\wt\chi}}

\def\issue(#1,#2,#3){{\bf #1}, #2 (#3)}

\hypersetup{
   colorlinks=true,       
   linkcolor=blue,        
   citecolor=blue,         
   filecolor=magenta,      
   }

\title{Prospects of non-resonant di-Higgs searches and Higgs boson self-coupling measurement at the HE-LHC using machine learning techniques}

\author[a]{Amit Adhikary}
\author[b]{Rahool Kumar Barman}
\author[a]{Biplob Bhattacherjee}
\affiliation[a]{Centre for High Energy Physics, Indian Institute of Science, Bangalore 560012, India}
\affiliation[b]{School of Physical Sciences, Indian Association for the Cultivation of Sciences, Kolkata, 700040, India}
\emailAdd{amitadhikary@iisc.ac.in}
\emailAdd{psrkb2284@iacs.res.in}
\emailAdd{biplob@iisc.ac.in}
\date{\today}

\abstract
{The prospects of observing the non-resonant di-Higgs production in the Standard Model at the proposed high energy upgrade of the LHC, $viz.$ the HE-LHC~($\sqrt{s}=27~{\rm TeV}$ and $\mathcal{L} = 15~{\rm ab^{-1}}$) is studied. Various di-Higgs final states are considered based on their cleanliness and production rates. The search for the non-resonant double Higgs production at the HE-LHC is performed in the $b\bar{b}\gamma\gamma$, $b\bar{b}\tau^{+}\tau^{-}$, $b\bar{b}WW^{*}$, $WW^{*}\gamma\gamma$, $b\bar{b}ZZ^{*}$ and $b\bar{b}\mu^{+}\mu^{-}$ channels. The signal-background discrimination is performed through multivariate analyses using the Boosted Decision Tree Decorrelated~(BDTD) algorithm in the~\texttt{TMVA} framework, the \texttt{XGBoost} toolkit and Deep Neural Network~(DNN). The variation in the kinematics of Higgs pair production as a function of the self-coupling of the Higgs boson, $\lambda_{h}$, is also studied. The ramifications of varying $\lambda_{h}$ on the $b\bar{b}\gamma\gamma$, $b\bar{b}\tau^{+}\tau^{-}$ and $b\bar{b}WW^{*}$ search analyses optimized for the SM hypothesis is also explored. }

\begin{document}
\maketitle

\section{Introduction}
\label{intro}

The discovery of the Higgs boson in 2012, by the ATLAS~\cite{Aad:2012tfa} and CMS~\cite{Chatrchyan:2012xdj} collaborations, laid the foundation stone to a new era in the field of particle physics. Since its discovery, numerous measurements have been performed to unravel the properties of the observed 125 GeV resonance, and these results indicate towards its compatibility with the properties of the Higgs boson as predicted by the Standard Model~(SM) of particle physics. The couplings of the observed Higgs boson with the third generation quarks and leptons, and the gauge bosons, fall within the allowed SM uncertainties and have been measured with considerable precision~\cite{Sirunyan:2018koj,Aaboud:2018gay,Aaboud:2018zhk,Aaboud:2017jvq,Aaboud:2018jqu,Aaboud:2018xdt,Sirunyan:2018kst,Sirunyan:2019twz}. However, we still have a long way to go along the precision frontier in the measurement of the first and second generation Yukawa couplings. Another important aspect of the the Standard Model which has remained elusive till now is the Higgs potential. The location of the minimum of the Higgs potential has been successfully measured, however, we are yet to observe any clear signature of the self-coupling of the Higgs boson ($\lambda_{h}$) which is another key ingredient in the understanding of the stability of the Higgs potential.  

In the SM, the only direct probe to measure $\lambda_{h}$ is the non-resonant pair production of the Higgs boson. The difficulty in its measurement emerges from the smallness of the di-Higgs production cross-section. At the leading order (LO), the Feynman diagrams which dominantly contribute to the non-resonant Higgs pair production process proceeds through top quark loops in the gluon fusion channel with a destructive interference between the triangle and the box diagrams. The fine cancellation results in a smaller production cross-section. At the centre of mass energy~($\sqrt{s}$) of $13~{\rm TeV}$, the SM di-Higgs production cross-section in the gluon fusion channel stands at $ {31.05}^{+2.2\%}_{-5.0\%}$~fb at the NNLO level~\cite{Dawson:1998py,Borowka:2016ehy,Baglio:2018lrj,deFlorian:2013jea,Shao:2013bz,deFlorian:2015moa,Grazzini:2018bsd,Baglio:2020ini}, while the production rate at $\sqrt{s} = 14~{\rm TeV}$ increases to only ${36.69}^{+2.1\%}_{-4.9\%}$~fb~\cite{Dawson:1998py,Borowka:2016ehy,Baglio:2018lrj,deFlorian:2013jea,Shao:2013bz,deFlorian:2015moa,Grazzini:2018bsd,Baglio:2020ini} at the NNLO level\footnote{The ${\rm N^{3}LO}$ QCD corrections to the di-Higgs production cross-section in the gluon fusion channel have been computed in Refs.~\cite{Chen:2019lzz,Chen:2019fhs}.}. The cross-section of di-Higgs production through other modes namely vector boson fusion, associated production with vector bosons and associated production with top and bottom pairs are much smaller than the production rate in the gluon fusion channel, and, are generally ignored. The small signal production rate is however compensated by the presence of two Higgs bosons which can give rise to phenomenologically rich final states. 

Various beyond the SM~(BSM) scenarios can also enhance the di-Higgs production cross-section. New physics~(NP) cases, such as the presence of an extended Higgs sector with a heavier resonance which can decay into a pair of SM-like Higgs bosons, composite Higgs models, strongly interacting theories, effective field theories (EFT) with modified top Yukawa coupling, Supersymmetric and extra-dimension theories with heavy colored states, can potentially increase the di-Higgs production cross-section. A wide array of studies pertaining to the case of di-Higgs production in various NP models can be found in Refs.~\cite{Baglio:2014nea,Hespel:2014sla,Lu:2015qqa,Bian:2016awe,Kribs:2012kz,Dawson:2012mk,Pierce:2006dh,Kanemura:2008ub,
Nishiwaki:2013cma,Ellwanger:2013ova,Chen:2014xra,Liu:2014rba,Slawinska:2014vpa,Goertz:2014qta,Azatov:2015oxa,Lu:2015jza,
Carvalho:2015ttv,Gorbahn:2016uoy,Carvalho:2016rys,Banerjee:2016nzb,Cao:2016zob,Dolan:2012ac,Contino:2010mh,Grober:2010yv,Contino:2012xk,
Grober:2016wmf,Liu:2004pv,Dib:2005re,Wang:2007zx,Barger:2014taa,Nakamura:2017irk,Crivellin:2016ihg,Cao:2013si,Sun:2012zzm,
Costa:2015llh,No:2013wsa,Kotwal:2016tex,Gao:2019uco,Cheung:2020xij,Huang:2019bcs,Barducci:2019xkq,Basler:2019nas,Alves:2019igs,Englert:2019eyl,Babu:2018uik,Basler:2018dac,Bauer:2017cov,Flores:2019hcf,Englert:2019xhz,DiMicco:2019ngk,Alasfar:2019pmn,Capozi:2019xsi,Li:2019uyy,Alves:2018jsw,Adhikary:2018ise,Borowka:2018pxx,Chen:2018uim,Alves:2018oct,Buchalla:2018yce,Heng:2018kyd,Kim:2018uty,Cao:2015oaa}.

The ATLAS and CMS collaborations have also performed numerous searches in various non-resonant di-Higgs final states using both Run-I and the Run-II dataset. While none of these searches have reported any excess over the SM background, upper limits have been derived on the di-Higgs production cross-section. The most stringent upper limit has been derived by ATLAS~\cite{Aad:2019uzh} through a combination of searches in the $4b$, $b\bar{b}WW^{*}$, $b\bar{b}\tau^{+}\tau^{-}$, $4W$, $b\bar{b}\gamma\gamma$ and the $WW^{*}\gamma\gamma$ channels using the Run-II data collected at an integrated luminosity~($\mathcal{L}$) of $\sim 36~{\rm fb^{-1}}$. The upper limit stands at $6.9$ times the SM production cross-section and the Higgs self-coupling is constrained within $-5.0 < k_{\lambda} < 12.0$ (where, $k_{\lambda}$ is the ratio of $\lambda_{h}$ over the SM value of Higgs self-coupling), at $95\%$ confidence level (CL). The same dataset has also been used by ATLAS and CMS to perform non-resonant di-Higgs searches in the $4W$~\cite{Aaboud:2018ksn}, $b\bar{b}WW^{*}$~\cite{Aaboud:2018zhh}, $b\bar{b}\tau^{+}\tau^{-}$~\cite{Aaboud:2018sfw,Sirunyan:2017djm}, $WW^{*}\gamma\gamma$~\cite{Aaboud:2018ewm}, $b\bar{b}\gamma\gamma$~\cite{Aaboud:2018ftw,CMS:2017ihs} and the $4b$~\cite{Aaboud:2018knk} final states. Among these results, searches in the $b\bar{b}\tau^{+}\tau^{-}$~\cite{Aaboud:2018sfw}, $4b$~\cite{Aaboud:2018knk} and the $b\bar{b}\gamma\gamma$~\cite{Aaboud:2018ftw} channels yield an upper limit of $12.7$, 22 and $13$ times the SM cross-section value at $\sqrt{s}=13~{\rm TeV}$, respectively, at $95\%$ CL,  while the other channels result in weaker upper limits. Similarly, the analogous searches by CMS in the $b\bar{b}\gamma\gamma$~\cite{CMS:2017ihs} and the $b\bar{b}\tau^{+}\tau^{-}$~\cite{Sirunyan:2017djm} channels have reported an upper limit of 19.2 and 30 times the SM di-Higgs production cross-section, respectively, at $95\%$ CL. Searches have also been performed in the $b\bar{b}l\nu l\nu$ final state by both ATLAS and CMS using the LHC Run-II data collected at $\sim 139~{\rm fb^{-1}}$ and $\sim 36~{\rm fb^{-1}}$, respectively, and the corresponding upper limits at $95\%$ CL stand at 40~\cite{Aad:2019yxi} and 79~\cite{Sirunyan:2017guj} times the SM production cross-section, respectively. CMS has also combined the search results from $b\bar{b}\tau^{+}\tau^{-}$, $b\bar{b}\gamma\gamma$, $4b$ and the $b\bar{b}VV$ ($V= W^{\pm},Z$) channels and derived an upper limit at 22.2 times the SM non-resonant di-Higgs production cross-section at $95\%$ CL~\cite{Sirunyan:2018two}. The non-resonant di-Higgs searches performed by ATLAS and CMS using Run-I dataset and the Run-II dataset collected at $\mathcal{L} \sim 2~{\rm fb^{-1}}$ can be found in Refs.~\cite{CMS:2016foy,Aad:2014yja,Aad:2015uka,Aad:2015xja,Aaboud:2016xco}. Indirect measurement of $\lambda_{h}$ has also been performed by ATLAS using the LHC Run-II dataset ($\mathcal{L}\sim 80~{\rm fb^{-1}}$)~\cite{ATL-PHYS-PUB-2019-009} and has excluded $k_{\lambda}$ outside $-3.2 < k_{\lambda} < 11.9$ at $95\%$ CL. The indirect probe makes use of single Higgs production at next-to-leading order (NLO) where the electroweak loop corrections are sensitive to the self-coupling of the Higgs boson. In Ref.~\cite{ATLAS-CONF-2019-049}, a combination of indirect searches through single Higgs production using Run-II data collected at $\mathcal{L}\sim 80~{\rm fb^{-1}}$ and non-resonant di-Higgs searches in the $4b$, $b\bar{b}\tau^{+}\tau^{-}$ and $b\bar{b}\gamma\gamma$ channels using Run-II data collected at $\mathcal{L}\sim 36~{\rm fb^{-1}}$, results in the allowed range of $k_{\lambda}$ to be within $[-2.3:10.3]$ at $95\%$ CL. The potential reach of the high luminosity upgrade of the LHC~(HL-LHC: $\sqrt{s}=14~{\rm TeV}$, $\mathcal{L}\sim 3000~{\rm fb^{-1}}$) in probing $k_{\lambda}$ has also been studied by ATLAS in the $b\bar{b}\gamma\gamma$ search channel~\cite{ATL-PHYS-PUB-2017-001}. This search reports a signal significance of $1.05\sigma$ and a projected exclusion region of $-0.8 \leq k_{\lambda} \leq 7.7$ at $95\%$ CL. Through a combination of searches in the $4b$, $b\bar{b}\gamma\gamma$ and the $b\bar{b}\tau^{+}\tau^{-}$ channels, ATLAS has reported a projected signal significance of $3.5\sigma$~\cite{ATL-PHYS-PUB-2018-053}.  

Numerous phenomenological studies have also focused on exploring the future prospects of non-resonant di-Higgs searches and $\lambda_{h}$ measurement at the HL-LHC, and, have exhibited encouraging results through the use of novel kinematic variables, variables constructed from boosted objects, jet substructure techniques, precision calculations and multivariate analysis techniques~\cite{Kim:2018cxf, Kim:2019wns,Dolan:2012rv,Barr:2013tda,Barger:2013jfa,Kling:2016lay,Alves:2017ued,Adhikary:2017jtu,Amacker:2020bmn,Abdughani:2020xfo,Heinrich:2019bkc,Arganda:2018ftn,Chang:2018uwu,Cao:2015oxx}. In addition, the prospects of measuring $\lambda_{h}$ in the future lepton colliders has also been studied (see Refs.~\cite{Asner:2013psa,Barklow:2017awn,Maltoni:2018ttu}). Indirect measurements of $\lambda_{h}$ at the HL-LHC and the future lepton colliders have also been performed through the study of single Higgs boson production~(see Refs.~\cite{Li:2019jba,McCullough:2013rea,Maltoni:2017ims,DiVita:2017eyz,Gorbahn:2016uoy,Bizon:2016wgr,Degrassi:2016wml,Goertz:2013kp}). The electroweak oblique parameters have also been used to probe $\lambda_{h}$ in Ref.~\cite{Kribs:2017znd}. The complementarity between the direct and indirect measurements of $\lambda_{h}$ offer an interesting avenue for advancement of our understanding about the Higgs potential, however, we must note that the indirect probes would be more vulnerable to the NP couplings present in the loop corrections and offers a higher degree of complicacy in extracting model-independent limits. Therefore, the direct probe of $\lambda_{h}$ offers more viable options for model-independent studies. However, we must also note that the numerous direct searches of $\lambda_{h}$ performed in the context of the HL-LHC point towards a common conclusion that the future prospect of observing the non-resonant di-Higgs signal is rather bleak unless new physics effects are present or the signal-background discrimination efficiency of the analysis techniques improves considerably. We would also like to point out that an extensive list of studies exist in the literature where the potential capability of the ambitiously proposed $\sqrt{s}=100~{\rm TeV}$ hadron collider in probing $\lambda_{h}$ has been explored~\cite{Mangano:2020sao,Banerjee:2019jys,Banerjee:2018yxy,Kling:2016lay,Bizon:2018syu,Goncalves:2018qas,Barr:2014sga,Chang:2018uwu,Contino:2016spe,Park:2020yps}. The future prospects of di-Higgs searches at the 100~TeV hadron collider have also been studied in the context of Higgs portal models and in the Higgs EFT framework in Refs.~\cite{He:2015spf,Azatov:2015oxa,Cao:2016zob,Corbett:2017ieo,Kuday:2017vsh,Gao:2019uco}. On the contrary, mush less impetus has been given towards studying the potential reach of the more realizable proposed energy upgrade to the HL-LHC, $viz.$ the HE-LHC which is projected to reach $\sqrt{s} = 27~{\rm TeV}$ at an integrated luminosity of $\mathcal{L} \sim 15~{\rm ab^{-1}}$, in constraining $\lambda_{h}$, and the corresponding exhaustive studies are by and large missing from literature. This stimulates the necessity for a comprehensive study of the future potential of the HE-LHC.

In this work, we perform a systematic evaluation of various final states emerging from non-resonant di-Higgs production at the HE-LHC, with inclusion of an exhaustive list of backgrounds, assessment of detector effects and implementation of advanced multivariate analysis techniques. Before giving a detailed outline of this paper, we briefly revisit the existing works where the potential capability of the HE-LHC in probing the Higgs boson self-coupling has been studied. The SM non-resonant di-Higgs production cross-section at $\sqrt{s}=27$~TeV in the gluon fusion channel is $139.9^{+1.3\%}_{-3.9\%}~{\rm fb}$ at NNLO~\cite{Dawson:1998py,Borowka:2016ehy,Baglio:2018lrj,deFlorian:2013jea,Shao:2013bz,deFlorian:2015moa,Grazzini:2018bsd,Baglio:2020ini} and is roughly $\sim 3.5$ times larger than its 14~TeV counterpart. In the context of HE-LHC, the $b\bar{b}\gamma\gamma$ channel has been studied in detail in Refs.~\cite{Goncalves:2018qas,Homiller:2018dgu,Cepeda:2019klc}. Ref.~\cite{Goncalves:2018qas} reports a $5\sigma$ observation at the HE-LHC for $\mathcal{L} \sim 2.5~{ab^{-1}}$ and a potential measurement accuracy of $15\%$ and $30\%$ at $68\%$ and $95\%$ CL, respectively. Ref.~\cite{Cepeda:2019klc} also reports a $5\sigma$ observation and a projected exclusion reach of $0.2 < k_{\lambda} < 1.8$ at $95\%$ CL. A summary of the indirect probes of $\lambda_{h}$ through single Higgs production and differential distribution measurements can also be found in Ref.~\cite{Cepeda:2019klc}. Lastly, Ref.~\cite{Homiller:2018dgu} outlines a $\geq 4.5\sigma$ observation at $\sim 40\%$ measurement accuracy for $\mathcal{L} \sim 15~{\rm ab^{-1}}$. The potential reach of other di-Higgs search channels have, however, remained largely unexplored. Furthermore, the combination of search results from multiple final states was observed to yield the most stringent constraints on $\lambda_{h}$ for the case of LHC Run-I and Run-II measurements, as well as the HL-LHC projections, and we aim to explore this facet for the HE-LHC in the present study.

In the first part of our analysis, we study the $b\bar{b}\gamma\gamma$, $b\bar{b}\tau^{+}\tau^{-}$, $b\bar{b}WW^{*}$, $WW^{*}\gamma\gamma$, $b\bar{b}ZZ^{*}$ and the $b\bar{b}\mu^{+}\mu^{-}$ final states emerging from the decay of the Higgs pair. The choice of these final states is motivated either by the cleanliness with respect to the backgrounds and/or larger branching ratio of the decay modes of the Higgs boson along with non-availability of existing studies. Previous studies have indicated the effectiveness of multivariate analysis in performing signal-background discrimination, and at times have exhibited an enhanced efficiency in improving the statistical signal significance compared to the more traditional cut-based analysis techniques~\cite{Baglio:2012np,Alves:2017ued,Kling:2016lay,Barger:2013jfa,Adhikary:2017jtu}. Consequently, we adhere to multivariate analysis in this work, and choose a variety of kinematic variables in each of the aforesaid analysis channels chosen through a wide literature survey. Our main goal is to improve the signal significance. In our study, we have considered three different multivariate analysis techniques. The first one is based on the Boosted Decision Tree Decorrelated (BDTD) algorithm in the \texttt{TMVA} framework~\cite{2007physics3039H}, in the second case, we use the \texttt{XGBoost}~\cite{Chen_2016} toolkit which utilizes a gradient tree boosting algorithm, and thirdly, we use Deep Neural Network~(DNN)~\cite{Goodfellow-et-al-2016,Tensorflow,Keras}. 

The objective of this work can be broadly classified into two different sections: to estimate the discovery reach of non-resonant di-Higgs searches at the HE-LHC through multivariate analysis using the BDTD algorithm, the XGBoost toolkit, and the DNN framework in various well-motivated di-Higgs final states~(Sec.~\ref{Sec:non_res_dihiggs}) and  assess HE-LHC’s future potential in constraining $\lambda_{h}$~(Sec.~\ref{sec:Cons-selfC}). We also study the ramifications of varying $\lambda_{h}$ on the di-Higgs signal sensitivity in Sec.~\ref{sec:Cons-selfC}. We conclude in Sec.~\ref{sec:summary}.

\section{Non-resonant di-Higgs production at the HE-LHC}
\label{Sec:non_res_dihiggs}

In this analysis, we consider such di-Higgs final states which have photons and/or leptons in the final state, $viz.$ $b\bar{b}\gamma\gamma$, $b\bar{b}\tau^{+}\tau^{-}$, $b\bar{b}WW^{*}$, $WW^{*}\gamma\gamma$, $b\bar{b}ZZ^{*}$ and $b\bar{b}\mu^{+}\mu^{-}$. Final states like $\tau^{+}\tau^{-}WW^{*}$, $\tau^{+}\tau^{-}ZZ^{*}$, $4\tau$, $ZZ^{*}\gamma\gamma$, $4\gamma$, $4Z$ and $4W$ have not been considered due to negligible production rates at the HE-LHC. However, it must be noted that some of these channels might have crucial implications on di-Higgs searches at the 100~TeV collider.

The di-Higgs signal and the background events have been generated at the leading order (LO) parton level with {\tt MG5\_aMC@NLO}~\cite{Alwall:2014hca}. The {\tt NNPDF2.3LO} PDF set~\cite{Ball:2014uwa} has been used with the {\tt A14} tune~\cite{ATL-PHYS-PUB-2014-021}. The showering and hadronization has been done with {\tt Pythia8}~\cite{Sjostrand:2014zea}. Fast detector response has been simulated using {\tt Delphes-3.4.1}~\cite{deFavereau:2013fsa}. Since the detector configuration at the HE-LHC is not known, we simulate the detector effects by assuming the resolution and tagging efficiencies of the present ATLAS detector.

The $jet$ reconstruction is done using the  anti-$kT$~\cite{Cacciari:2008gp} algorithm with the jet radius parameter, $R = 0.4$, and the minimum jet transverse momentum, $p_{T}^{j} > 20~{\rm GeV}$, in the {\tt FastJet}~\cite{Cacciari:2011ma} framework. For the leptons ($l$, $l = e,\mu$) and photons to be isolated, it is required that the sum of transverse momenta of the surrounding  objects within a cone of radius $\Delta R = 0.2$ must be less that $20\%$ of the transverse momentum~($p_{T}$) of the lepton or photon under consideration. Furthermore, the leptons and photons are also required to lie within a pseudorapidity~($\eta$) range of $|\eta| \leq 4.0$ and must carry $p_{T} > 5~{\rm GeV}$ and $> 10~{\rm GeV}$, respectively. A flat $b$-tagging efficiency of $70\%$ has been considered while the $c \to b$ and $j \to b$ fake rates have been fixed at $3\%$~($4\%$) and $0.15\%$~($0.12\%$) provided the $p_{T}$ of the $c$ and $j$ are in between $30-90~{\rm GeV}$ ($> 90~{\rm GeV}$), respectively~\cite{Sirunyan:2017ezt}. We must mention that the fake rates are a function of the transverse momenta and pseudorapidity, and must be dealt with more sophistication in a more precise analysis.

We perform a multivariate analysis in the \texttt{TMVA} framework to efficiently discriminate the signal and the background events. In this respect, multifarious kinematic variables are chosen depending upon the di-Higgs final state. We use the BDTD algorithm in all these analyses. Overtraining of the signal and background samples has been avoided by requiring the results of the Kolmogorov-Smirnov test to be always $>$0.1. We also perform a detailed collider analysis using the XGBoost package and DNN. The XGBoost training is performed by optimizing the learning rate, max-depth, and the regularization parameters, $\eta$ and $\lambda$. We also use early stopping for regularization. The goal is to minimize the loss function and avoid overtraining by maintaining a comparable validation loss and training loss. The XGBoost classifier is trained with the signal and background event samples, and a lower cut is applied on the signal-like probability of the XGBoost output. We refer to this lower limit as the probability cut on the XGBoost output. We also use a DNN classifier to train the signal and the background events. The DNN training is performed on a three-layered network structure with Adam optimizer, and regularization is performed using batch normalization. The categorical cross-entropy loss function is considered, and similar to the case of XGBoost, the aim is to minimize the loss function while maintaining a similar training loss and validation loss to avoid overtraining. 

In this work, the following formula has been used to compute the signal significance~\cite{Cowan:2010js, Cowan:2012}:
\begin{equation}
\label{S}
{\cal S} = \sqrt{2 \left((S+B) \log \left(1+\frac{S}{B}\right)-S \right)} \,,
\end{equation}
where $S$ is the signal yield and $B$ is the total background yield, from the multivariate analyses. Upon assuming a systematic uncertainty, $\sigma_{sys\_un}$, the signal significance formula is modified as the following form:
\begin{equation}
\label{S_sys}
{\cal S}_\text{sys} = \sqrt{2 \left((S+B) \log \left(\frac{(S+B)(B+\sigma_{B}^{2})}{B^{2}+(S+B)\sigma_{B}^{2}}\right)-\frac{B^{2}}{\sigma_{B}^{2}}\log \left(1+\frac{\sigma_{B}^{2}S}{B(B+\sigma_{B}^{2})} \right) \right)} \,,
\end{equation}
where $\sigma_{B}=\sigma_{sys\_un}\times B$. The performance of the multivariate analyses was optimized to maximize the signal significance while also maintaining a reasonably good value of $S/B$.

In the following subsections, we discuss the kinematic features of the numerous di-Higgs final states, provide a detailed outline of the analysis strategies and present our results from the detailed collider search. We begin our discussion by studying the features and prospects of the $b\bar{b}\gamma\gamma$ channel (which is one of the most promising channels for non-resonant di-Higgs searches at the HL-LHC) in the context of searches at the HE-LHC.  

\subsection{The $b\bar{b}\gamma\gamma$ channel}
\label{sec:bbaa}

The small branching ratio of $h \to \gamma\gamma$ puts the signal ($pp \to hh \to b\bar{b}\gamma\gamma$) production rate at a downside, however, the relatively low production rate is compensated by the cleanliness of this channel. Numerous previous works have studied the prospects of this search channel in probing non-resonant di-Higgs production at the HL-LHC. In our previous work~\cite{Adhikary:2017jtu}, where the HL-LHC prospects of the $b\bar{b}\gamma\gamma$ search channel was analyzed through both, cut-based and multivariate techniques, we had obtained a signal significance of $\sim 1.46$ and $\sim 1.76$, respectively, thus, exhibiting a promising prospect for the high energy upgrade of the HL-LHC, $viz.$ the HE-LHC.

The QCD-QED $b\bar{b}\gamma\gamma$ process is the most dominant source of background and has been generated upon merging with an additional $jet$ through the MLM merging scheme~\cite{Mangano:2006rw}. Substantial contribution to the background also arises from the fakes, $viz.$ the $c\bar{c}\gamma\gamma$ and $jj\gamma\gamma$ processes, when the $c$ $jets$ and the light $jets$ ($j$), respectively, gets faked as a $b$ $jet$. These three background processes ($b\bar{b}\gamma\gamma + c\bar{c}\gamma\gamma + jj\gamma\gamma$) will be collectively referred to as $b\bar{b}\gamma\gamma^{*}$ in the remainder of this analysis. Other fake backgrounds which can contaminate the signal are: $b\bar{b}j\gamma$ and $c\bar{c}j\gamma$ (collectively referred to as Fake 1), and, $b\bar{b}jj$ (referred to as Fake 2). In the Fake 1 category, one light $jet$ gets faked as a photon while two light $jets$ are required to get misidentified as photons in the Fake 2 category. Another source of fake background is the $hjj^{*}$ process ($hjj + hcc$) when $h$ decays into photon pair and the $j/c$ gets faked as a $b$ $jet$. The other processes which contribute to the background are $t\bar{t}h$, $Zh$, $b\bar{b}h$ and $Z\gamma\gamma$. The $Z\gamma\gamma$ process has also been generated upon matching with an additional $jet$. The leading order (LO) cross-sections are obtained from \texttt{MG5\_aMC@NLO} and the generation cuts for these background processes have been outlined in Appendix~\ref{sec:appendixA}.

We select events with exactly two $b$ $jets$ and two photons in the final state. The leading and the sub-leading ($p_{T}$ ordered) $b$ $jets$, $b_{1}$ and $b_{2}$ respectively, must carry $p_{T} > 30~{\rm GeV}$ and lie within a pseudorapidity range of $|\eta| < 4.0$. The two photons are also required to carry $p_{T} > 30~{\rm GeV}$ and must fall within a pseudorapidity coverage of $|\eta| < 4.0$. A veto is applied if any $\tau$ $jets$ or isolated leptons ($l$, $l= e,\mu$) are present with $p_{T} > 20~{\rm GeV}$ and $|\eta| < 4.0$. This veto helps to reduce the $t\bar{t}h$ background where leptons can come from top decay. To offset the large $b\bar{b}\gamma\gamma^{*}$ background, we restrict the invariant mass of the photon pair ($m_{\gamma\gamma}$) to lie between 122~GeV and 128~GeV. We impose a lower cut on the $\Delta R$\footnote{The distance between two final state particles a and b in the $\eta-\phi$ plane is calculated as $\Delta R_{ab}^2 = \Delta\eta_{ab}^2 + \Delta\phi_{ab}^2$ where $\Delta\eta_{ab}$ and $\Delta\phi_{ab}$ are the differences in pseudorapidity and azimuthal angle of the particles a,b respectively.} between the photons and the $b$ $jets$, $\Delta R_{b\gamma} > 0.2$, and require the invariant mass of the $b\bar{b}$ pair ($m_{bb}$) to be greater than 50~GeV to further tackle the QCD-QED $b\bar{b}\gamma\gamma^{*}$ background. The acceptance cuts have been summarized in Table~\ref{Table:bbgaga_acceptance_cuts}. We would like to note that the generation level cuts have also been applied alongside the acceptance cuts. 

\begin{table}
\begin{center}
\begin{tabular}{|c|}\hline
Acceptance cuts \\ \hline
$N_{b\ jets} = 2, ~N_{\gamma} = 2$ \\
122~GeV $< m_{\gamma\gamma} <$ 128~GeV \\
$\Delta R_{b\gamma} > 0.2$ \\
$m_{bb} >$ 50~GeV \\  \hline 
\end{tabular}
\caption{\it Acceptance cuts in the $b\bar{b}\gamma\gamma$ channel.}
\label{Table:bbgaga_acceptance_cuts}
\end{center}
\end{table}

The signal and background event samples, after being passed through the aforementioned cuts, are subjected to multivariate analysis using the BDTD algorithm, the XGBoost toolkit, and DNN, in order to discriminate between the background and signal samples and maximize the signal significance. The following 19 kinematic variables are used as inputs in the optimization procedure: 

\begin{equation}
\begin{split}
m_{bb},~\Delta R_{\gamma \gamma},~\Delta R_{bb},~p_{T,bb},~p_{T,\gamma \gamma},~\Delta R_{bb \; \gamma \gamma},~p_{T,hh},~\Delta R_{b_i \gamma_i},\\~m_{hh},~p_{T,b_{1,2}},~p_{T,\gamma_{1,2}},~\met,~\cos{\theta^{*}},~\cos{\theta_{\gamma_1 h}}
\label{bbaa:eq1}
\end{split}
\end{equation}

\begin{figure}
\centering
\includegraphics[scale=0.37]{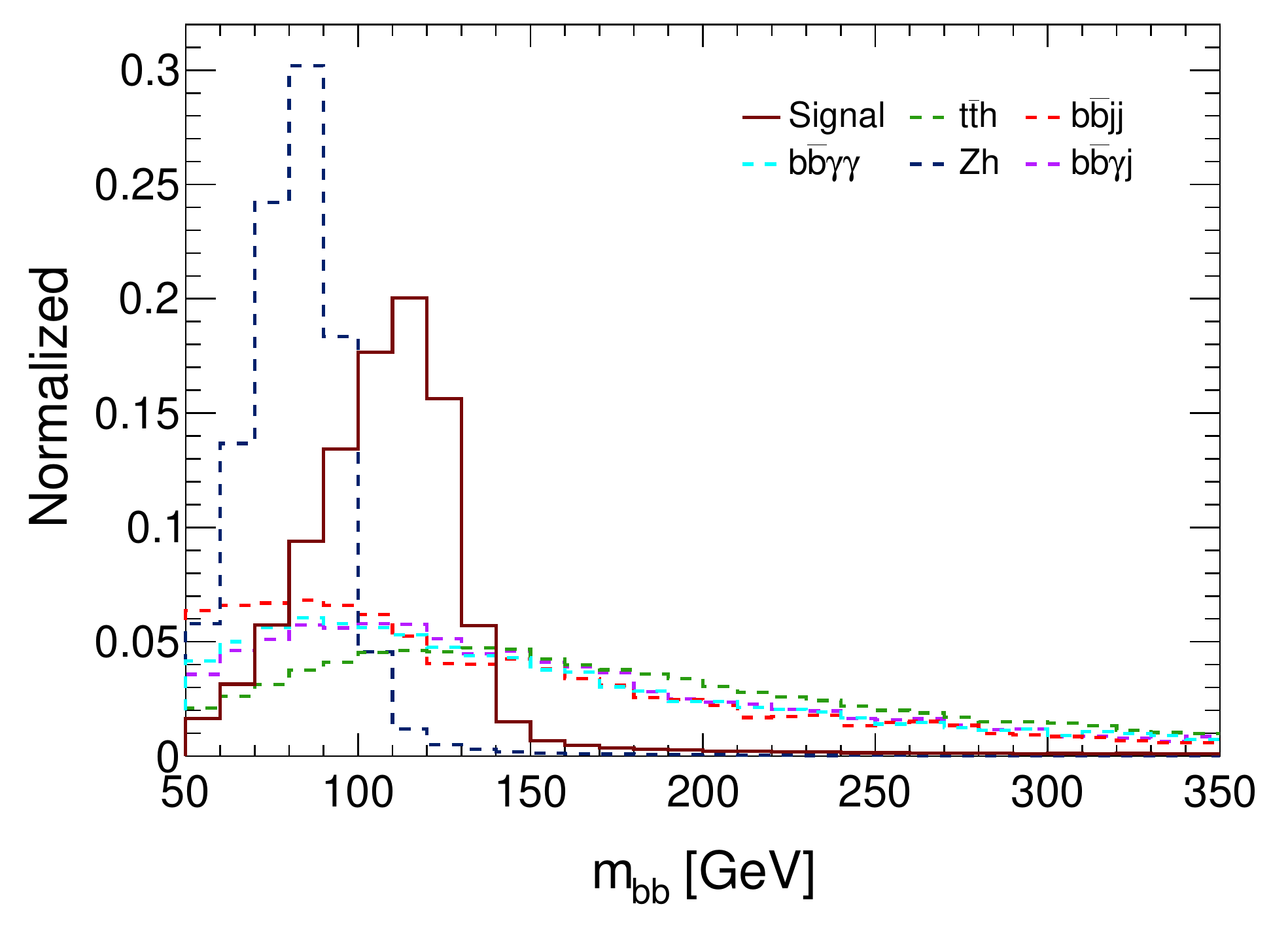}
\includegraphics[scale=0.37]{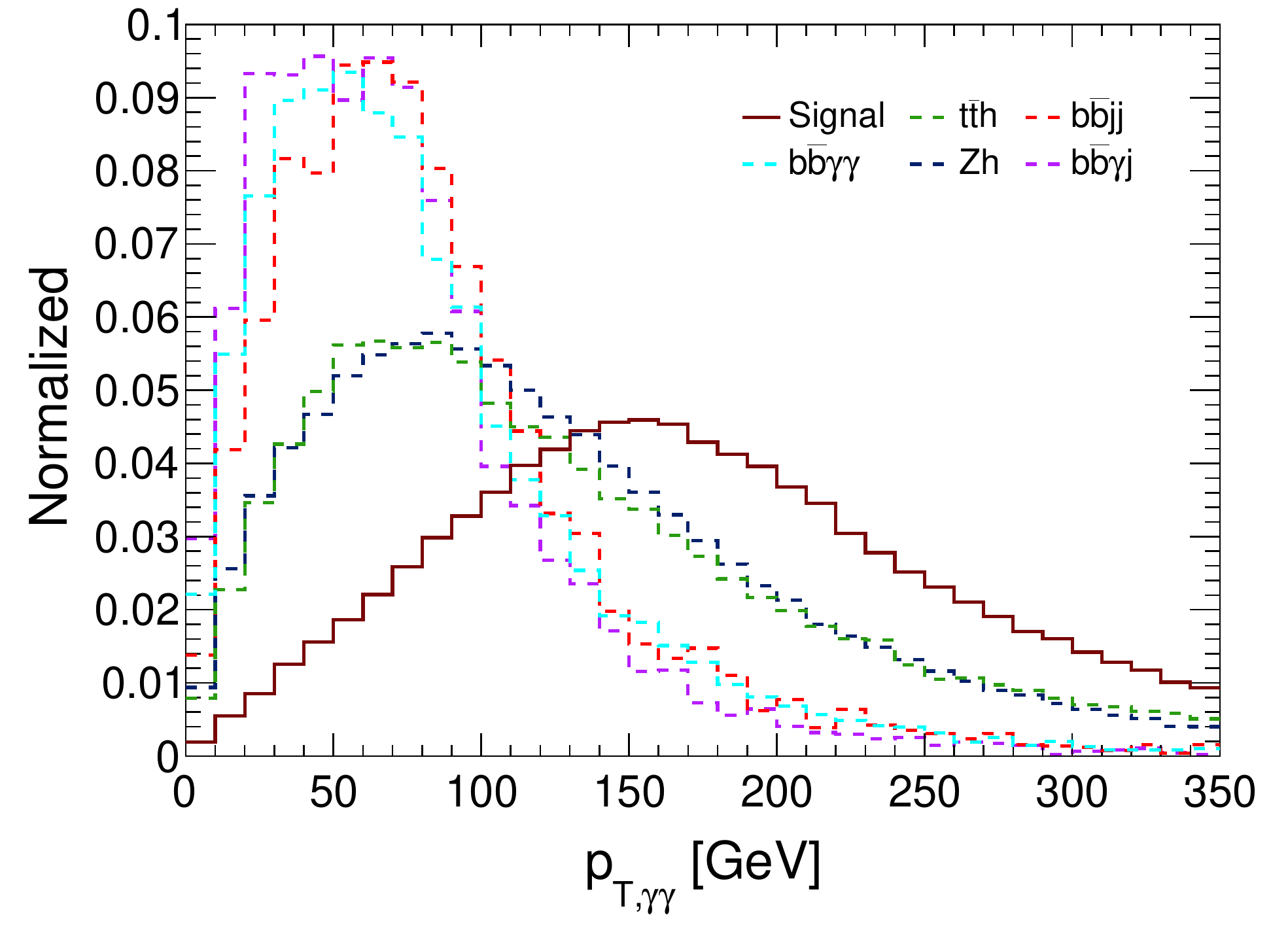}\\
\includegraphics[scale=0.37]{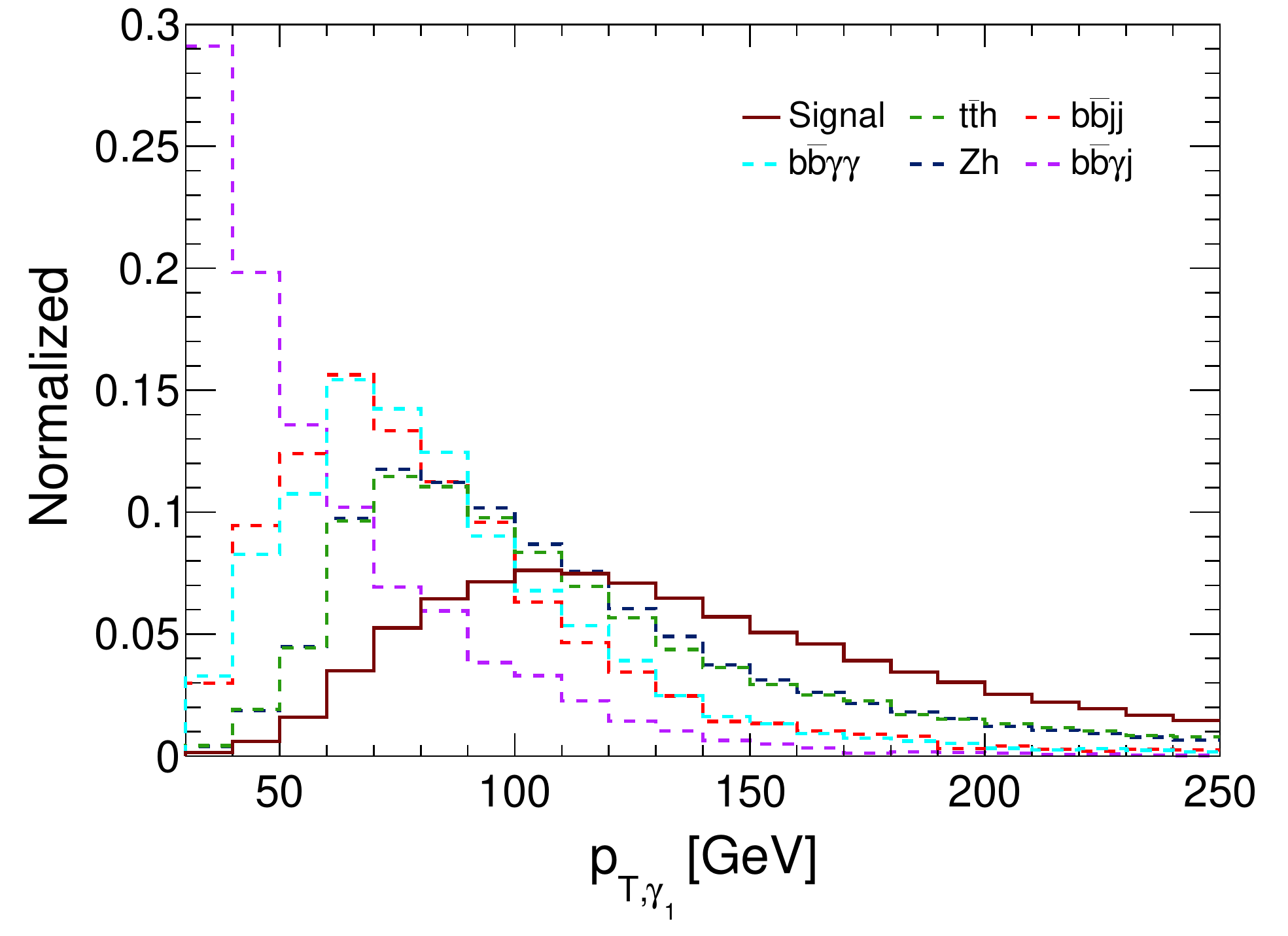}
\includegraphics[scale=0.37]{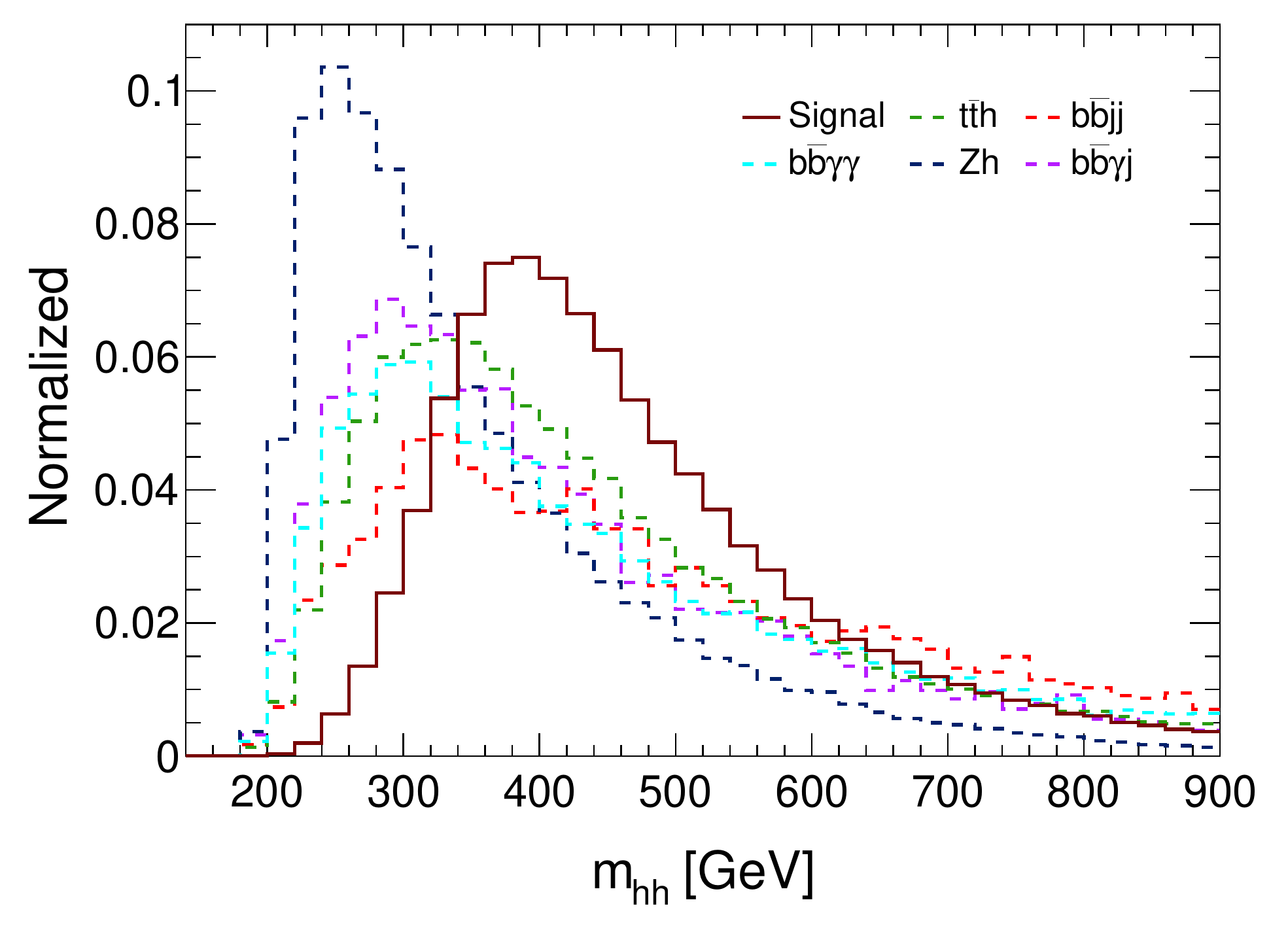}
\caption{\it Normalized distributions of $m_{bb},~p_{T,\gamma\gamma},~p_{T,\gamma_1}$ and $m_{hh}$ for the $pp \to hh \to b\bar{b}\gamma\gamma$ signal and the dominant backgrounds after the acceptance cuts and the generation level cuts.}
\label{bbaa:fig1}
\end{figure}

Here, $p_{T,bb}$ and $p_{T,\gamma\gamma}$ represents the transverse momentum of the system of $b\bar{b}$ pair and the photon pair, respectively, while $\Delta R_{bb\; \gamma\gamma}$ represents the distance in the $\eta-\phi$ plane between the $b\bar{b}$ and $\gamma\gamma$ system. $\theta^{*}$ represents the angle between the outgoing photon and the $Z$-axis in the Collins-Soper~(CS) frame~\cite{PhysRevD.16.2219,Richter-Was:2016mal} and $\cos{\theta^{*}}$ is defined as:
\begin{equation}
\cos\theta^{*}=\frac{Sinh(\eta_{\gamma 1} - \eta_{\gamma 2})}{\sqrt{1+(p_{T,\gamma \gamma}/m_{\gamma\gamma})^2}}~\frac{2~p_{T,\gamma 1}~p_{T,\gamma 2}}{m_{\gamma\gamma}^2}
\end{equation}
The advantage of this variable is its less sensitivity to initial state radiations (ISR) in the CS frame. The final variable is $\cos{\theta_{\gamma_{1}h}}$ where $\theta_{\gamma_{1}h}$ represents the angle between the direction of the leading $p_{T}$ photon ($\gamma_{1}$) in the rest frame of the Higgs boson reconstructed from the diphoton system and the direction of the Higgs boson (reconstructed from the $\gamma\gamma$ system) in the lab frame. We must note that the numerical ordering of particles in our notation represents the $p_{T}$ ordering with the subscript $1$ assigned to the leading $p_{T}$ object in its class.  

\begin{figure}
\centering
\includegraphics[scale=0.42]{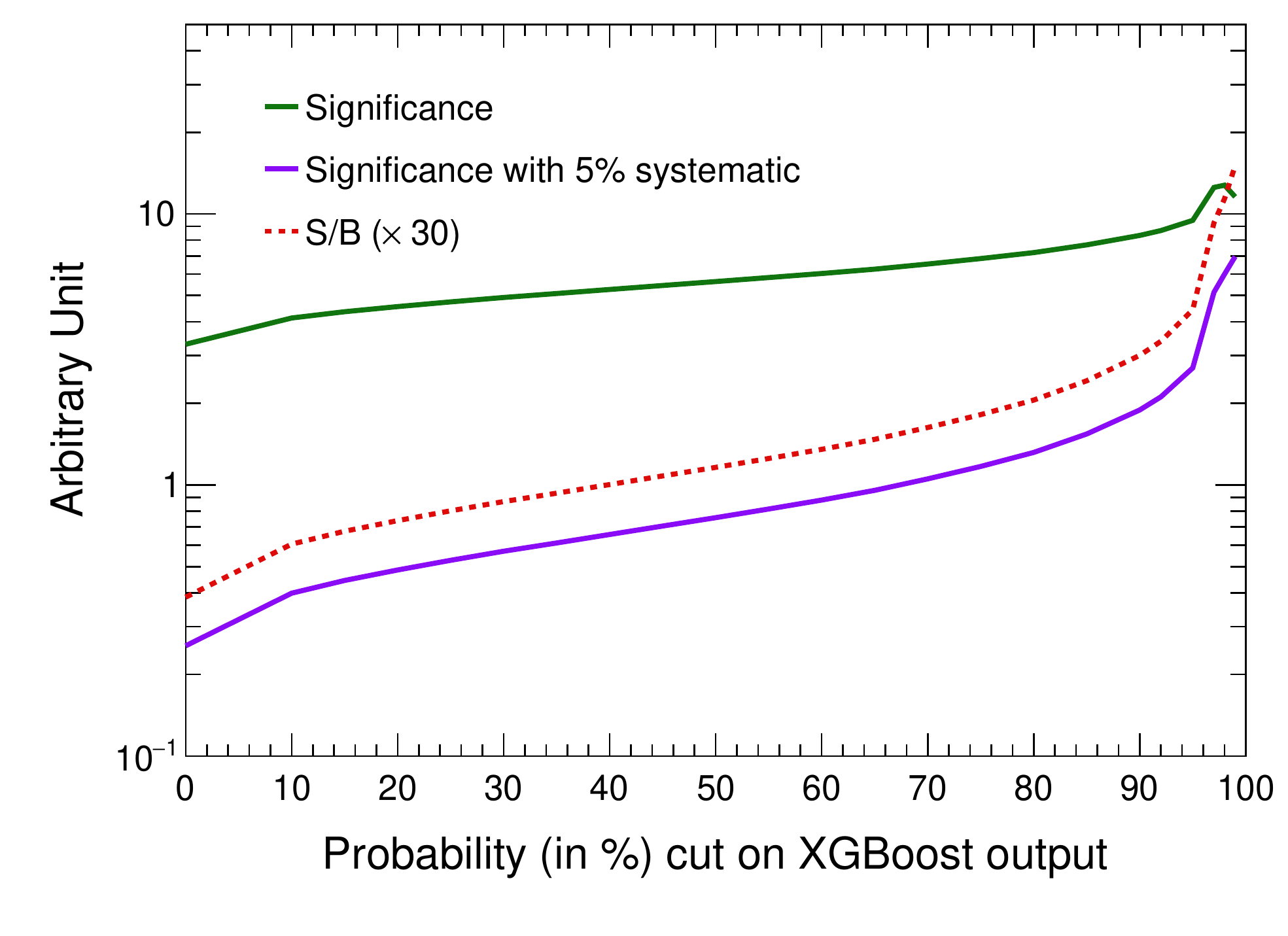}
\caption{\it The variation of significance~(with~($5\%$) and without systematic uncertainty) and $S/B$ is shown as a function of the probability cut on the XGBoost output for the $b\bar{b}\gamma\gamma$  channel.}
\label{bbaa:fig3}
\end{figure}

The kinematic variables which exhibited the maximal capability in discriminating the signal and background event samples during the BDTD optimization procedure are: $m_{bb}$, $p_{T,bb}$, $p_{T,\gamma_{1}}$ and $m_{hh}$. The normalized distribution of these variables for the signal and the dominant background processes, after passing through the acceptance cuts and the generation cuts, are illustrated in Fig.~\ref{bbaa:fig1}. In Table~\ref{bbaa:tab1}, we show the signal and background yields together with the signal significance obtained from the multivariate BDTD analysis. As shown in Table~\ref{bbaa:tab1}, the BDTD optimization yields a signal significance of $9.42$ in the absence of any systematic uncertainty ($\sigma_{sys\_un}$), exhibiting a roughly $\sim 6$ times improvement over its HL-LHC counterpart~\cite{Adhikary:2017jtu}. Upon assuming $\sigma_{sys\_un} = 5\%$, the signal significance reduces to $\sim 2.91$, still within the projected exclusion reach ($> 2\sigma$) at the HE-LHC. We also show the signal and background yields, and the signal significance, obtained from the XGBoost optimization, in Table~\ref{bbaa:tab1}. The kinematic variables listed in Eqn.~\ref{bbaa:eq1} have also been used in the multivariate analysis performed using the XGBoost toolkit. Here, we have applied the probability cut on the XGBoost output at $97\%$ and we obtain a signal significance of $12.46$ ($5.11$ with $\sigma_{sys\_un} = 5\%$). The signal and background yields from the DNN analysis are also listed in Table~\ref{bbaa:tab1}. The DNN training and optimization results in a signal significance of $10.03$  ($3.9$ with $\sigma_{sys\_un} = 5\%$) which is slightly lower than the XGBoost estimation but larger than the BDTD output. We also list the signal and background yields, and the associated signal significance, obtained from the XGBoost analysis upon imposing the probability cut at $95\%$ in Table~\ref{bbaa:tab2}. Reducing the probability cut from $97\%$ to $95\%$ reduces the signal significance to $9.47$. It must also be noted that the yield of the dominant QCD-QED $b\bar{b}\gamma\gamma$ background increases from $\lesssim 1$ to $1633$ when the probability cut on the XGBoost output is reduced from $97\%$ to $95\%$. We would like to acknowledge that the choice of the probability cut is subjective and the results would differ with a different choice of the probability cut. Therefore, for the sake of completeness, we also illustrate the signal significance and the $S/B$ value as a function of probability cut (in $\%$) on the XGBoost output in Fig.~\ref{bbaa:fig3}. The solid green and purple lines in Fig.~\ref{bbaa:fig3} represent the signal significance obtained by assuming zero and $5\%$ systematic uncertainty, respectively. The dashed red line represents the ratio of signal over background, scaled by a factor of $30$\footnote{The scaling has been done for illustrative reasons.}.  

\begin{center}
\begin{table}[htb!]
\centering
\scalebox{0.7}{%
\begin{tabular}{|c|c|c|c|c|c|}\hline
       & Process & Cross section order     & \multicolumn{3}{c|}{Event yield after the analysis with} \\ \cline{4-6}
 &         &           & BDTD & XGBoost & DNN \\ \hline\hline
\multirow{8}{*}{Background}  
 & $t\bar{t}h$                & NLO~\cite{bkg_twiki_cs}                    & $988$   & $8\times 10^{-3}$ & $7\times 10^{-3}$\\  
 & $b\bar{b}\gamma\gamma*$    & LO                                         & $1423$  & $266$ & $1191$\\  
 & $Fake~1$                   & LO                                         & $358$   & $2.4$ & $2.8$\\
 & $Fake~2$                   & LO                                         & $361$   & $36$ & $45$\\  
 & $Zh$                       & NNLO (QCD) + NLO (EW)~\cite{bkg_twiki_cs}  & $297$   & $1149$ & $615$\\ 
 & $b\bar{b}h$                & LO                                         & $92$    & $3\times 10^{-4}$ & $0.21$\\ 
 & $Z\gamma\gamma$            & LO                                         & $99$    & $66$ & $52$\\ 
 & $hjj*$                     & LO                                         & $0.22$     & $286$ & $211$\\ 
  \cline{2-6} 
 & \multicolumn{2}{c|}{Total}                                              & $3618$  & $1806$ & $2117$\\ \hline
\multicolumn{2}{|c|}{Signal ($hh \to b\bar{b}\gamma\gamma$)} & NNLO~\cite{hhtwiki}  & $581$ & $555$ & $478$\\\hline 
\multicolumn{2}{|c|}{\multirow{2}{*}{Significance}} & $0\%$ $\sigma_{sys\_un}$ & $9.42$ & $12.46$ & $10.03$\\ \cline{3-6}
\multicolumn{2}{|c|}{} & $2\%~(5\%)$ $\sigma_{sys\_un}$ & $5.93~(2.91)$ & $9.31~(5.11)$ & $7.26~(3.9)$ \\ \hline
\end{tabular}}
\caption{\it The signal and background yields at the HE-LHC along with the signal significance in the $b\bar{b}\gamma\gamma$ channel from the analysis using BDTD, XGBoost and DNN classifiers.}
\label{bbaa:tab1}
\end{table}
\end{center}

\begin{center}
\begin{table}[htb!]
\centering
\scalebox{0.7}{%
\begin{tabular}{|c|c|c|}\hline
       & Process &  Event yield after the analysis with \\ 
 &               & XGBoost with $95\%$ probability cut\\ \hline\hline
\multirow{8}{*}{Background}  
 & $t\bar{t}h$                & $0.01$   \\  
 & $b\bar{b}\gamma\gamma*$    & $2083$  \\  
 & $Fake~1$                   & $4.6$   \\
 & $Fake~2$                   & $62$   \\  
 & $Zh$                       & $1624$   \\ 
 & $b\bar{b}h$                & $0.34$    \\ 
 & $Z\gamma\gamma$            & $111$    \\ 
 & $hjj*$                     & $419$     \\ 
  \cline{2-3} 
 & Total                      & $4304$  \\ \hline
\multicolumn{2}{|c|}{Signal ($hh \to b\bar{b}\gamma\gamma$)}& $636$ \\\hline 
\multirow{2}{*}{Significance} & $0\%$ $\sigma_{sys\_un}$ & $9.47$ \\ \cline{2-3}
 & $2\%~(5\%)$ $\sigma_{sys\_un}$ & $5.66~(2.7)$  \\ \hline
\end{tabular}}
\caption{\it The signal and background yields at the HE-LHC along with the signal significance for the $b\bar{b}\gamma\gamma$ channel upon applying the probability cut on the XGBoost output at $95\%$.}
\label{bbaa:tab2}
\end{table}
\end{center}

\subsection{The $b\bar{b}\tau\tau$ channel}

Compared to the $b\bar{b}\gamma\gamma$ search channel, the $b\bar{b}\tau\tau$ final state suffers a drawback in terms of signal clarity, however, it gains a favorable stance on account of its considerably large production rate. An added advantage is the possibility to obtain three different final states depending on the decay modes of the $\tau$ leptons: fully leptonic ($b\bar{b}ll+\met$), semi-leptonic ($b\bar{b}l\tau_{h}+\met$) and fully hadronic ($b\bar{b}\tau_{h}\tau_{h}$). In a previous analysis~\cite{Adhikary:2017jtu} where we had studied the future reach of all three decay modes of $b\bar{b}\tau\tau$ channel at the HL-LHC, the fully hadronic decay mode had exhibited the strongest sensitivity. Keeping this observation in mind, in the current subsection we exclusively focus on studying the potential reach of di-Higgs searches at the HE-LHC in the fully hadronic decay mode: $pp \to hh \to b \bar{b}\tau_{h}\tau_{h}$. In similarity with the previous subsection, we perform a detailed multivariate collider analysis with the BDTD algorithm, XGBoost and DNN.

The $t\bar{t}$ process is the leading contributor to the background. In order to generate sufficient statistics of all relevant $t\bar{t}$ decay modes, we generate the event samples for fully leptonic, semi-leptonic, and fully hadronic decay modes of $t\bar{t}$, separately. The QCD-QED $pp \to b\bar{b}Z^\star/\gamma^{\star} \to b\bar{b}\tau\tau$ and the fake $b\bar{b}jj$ (when $jets$ fake as a $\tau_{h}$) processes also provides a sizable contribution to the background. Sub-dominant contributors to the background are: $Zh$, $b\bar{b}h$, $t\bar{t}h$, $t\bar{t}W$ and $t\bar{t}Z$. In the case of $Zh$, contributions can arise from $(Z \to b\bar{b})(h \to \tau^{+}\tau^{-})$ as well as from $(Z \to \tau^{+}\tau^{-})(h \to b\bar{b})$. The large cross-section of the background processes also demands a large event statistics. To efficiently achieve larger statistics, we impose hard cuts at the generation level which have been listed in Appendix~\ref{sec:appendixA}.

The substantial difference in the production cross-section of the signal and the dominant $t\bar{t}$ background necessitates the use of efficient kinematic variables to discriminate the two. The di-tau invariant mass ($m_{\tau\tau}$) is one such variable however the reconstruction of $m_{\tau\tau}$ at the LHC has always been wrapped with complications due to the accompanying $\met$. A number of $\tau\tau$ reconstruction techniques have been studied and applied in previous analyses~\cite{Barr:2003rg,Lester:1999tx,Elagin:2010aw}. In the present work, we use the collinear mass approximation technique~\cite{Elagin:2010aw} to reconstruct $m_{\tau\tau}$. This technique assumes the following: the visible decay products and the neutrinos from a $\tau$ lepton are approximately collinear ($\theta_{vis} = \theta_{\nu}$, $\phi_{vis} = \phi_{\nu}$), and the neutrinos are the only source of $\met$. Based on these assumptions, the x- and y-components of $\met$ can be defined as functions of the neutrino momenta, and individual contribution of the neutrinos to the $\met$ can be ascertained upon solving these relations. We must note that this approximation technique is capable to reconstruct $m_{\tau\tau}$ correctly only under circumstances where the di-tau system is highly boosted and has recoiled against a hard object. This technique is applicable in the context of $pp \to hh \to b\bar{b}\tau\tau$ channel since the $h \to \tau\tau$ system is boosted against the $h \to b\bar{b}$ system. However, it is possible that this technique may overestimate the value of $m_{\tau\tau}$ when the $\met$ resolution is not accurate. Similar to our previous analysis~\cite{Adhikary:2017jtu}, we consider modified lepton selection criteria following~\cite{TheATLAScollaboration:2015rzr}. The total energy deposition of all the stable particles within $\Delta R_{\ell,~\text{particle}} < 0.2$ of the electron or muon is required be at most $15$ GeV. The tagging efficiency of a $\tau$ jet with $p_T>20$ GeV and $|\eta|<4.0$ is fixed at $55\%$ and $50\%$~\cite{CMS-PAS-TAU-16-002} for one pronged and three pronged $\tau$'s, respectively. Also, the rate of a light jet with $p_T>20$ GeV and $|\eta|<4.0$ getting mistagged as a $\tau$ tagged jet is assumed to be $0.35\%$~\cite{CMS-PAS-TAU-16-002}.

We select events containing exactly two $b$-tagged $jets$ and two $\tau$-tagged $jets$ in the final state. The $b$ $jets$ and the $\tau$ $jets$ must carry $p_{T} > 30~{\rm GeV}$ and $p_{T} > 20~{\rm GeV}$, respectively, and both must lie within a pseudorapidity range of $|\eta| < 4.0$. A veto is applied on events containing isolated leptons with $p_{T} > 20~{\rm GeV}$ and $|\eta| < 4.0$. Furthermore, the distance in the $\eta-\phi$ plane between the $b$ $jets$ ($\Delta R_{bb}$), the $\tau$ $jets$ ($\Delta R_{\tau_{1}\tau_{2}}$) and in between all possible pairs of $b$ and $\tau$ $jets$ ($\Delta R_{b_{i}\tau_{j}}$ ($i,j =1,2$)) is required to be greater than $0.2$. Here, the superscripts $1$ and $2$ represents the $p_{T}$ ordered leading and sub-leading objects. Additionally, we also impose the generation level cuts, $viz.$ a lower limit on the invariant mass of the $b$ $jets$, $m_{bb} > 50~{\rm GeV}$ and the invariant mass of the visible decay products of the $\tau$'s, $m_{\tau\tau}^{vis.}> 50~{\rm GeV}$.

The selected signal and background events are subjected to multivariate analysis with the following input kinematic variables:

\begin{equation}
\begin{split}
p_{T,bb},~\Delta R_{bb},~m_{bb},~p_{T,\tau_h\tau_h},~\Delta R_{\tau_h\tau_h},~\Delta\phi_{\tau_{h1}\met},~\Delta\phi_{\tau_{h2}\met},\\~m_{T,\tau_h\tau_h},~m_{\tau_h\tau_h}^{col},~m_{T2},~m_{\textrm{eff}},~\Delta R_{b_1\tau_{h1}},~p_{T,hh}^{\textrm{vis}},~m_{hh}^{\textrm{vis}},~\Delta R_{hh}^{\textrm{vis}} \nonumber
\end{split}
\end{equation}

\begin{center}
\begin{table}[htb!]
\centering
\scalebox{0.7}{%
\begin{tabular}{|c|c|c|c|c|c|}\hline
       & Process & Cross section order     & \multicolumn{3}{c|}{Event yield after the analysis with} \\ \cline{4-6}
 &         &           & BDTD & XGBoost & DNN\\ \hline\hline
\multirow{10}{*}{Background} 
 & $t\bar{t}$ had               	& NNLO~\cite{ttbarNNLO}                       & $377$    & $127$ & $5.1$\\ 
 & $t\bar{t}$ semi-lep             & NNLO~\cite{ttbarNNLO}                       & $28315$  & $9037$ & $11447$\\  
 & $t\bar{t}$ lep                  & NNLO~\cite{ttbarNNLO}                       & $31239$  & $5513$ & $14701$\\   
 & $b\bar{b}\tau\tau$        		& LO                                          & $23443$  & $6055$ & $1708$\\ 
 & $t\bar{t}h$              		& NLO~\cite{bkg_twiki_cs}                     & $3368$   & $1180$ & $1544$\\  
 & $t\bar{t}Z$             	    	& NLO~\cite{Lazopoulos:2008de}                & $2034$   & $443$ & $652$\\  
 & $t\bar{t}W$       		        & LO                                          & $89$     & $15$ & $37$\\ 
 & $b\bar{b}h$                     & LO                                          & $20$     & $9.8$ & $4.7$\\
 & $Zh$                            & NNLO (QCD) + NLO (EW)~\cite{bkg_twiki_cs}   & $1358$   & $700$ & $277$\\  
 & $b\bar{b}jj$                    & LO                                          & $65776$  & $18716$ & $1315$\\ \cline{2-6}  
 & \multicolumn{2}{c|}{Total}                                                    & $156019$ & $41795$ & $31690$\\ \hline
\multicolumn{2}{|c|}{Signal ($hh \to b\bar{b}\tau_h\tau_h$)} & NNLO~\cite{hhtwiki}  & $1096$ & $981$ & $760$\\\hline 
\multicolumn{2}{|c|}{\multirow{2}{*}{Significance}} & $0\%$ $\sigma_{sys\_un}$ & $2.77$ & $4.78$ & $4.25$\\ \cline{3-6}
\multicolumn{2}{|c|}{} & $2\%~(5\%)$ $\sigma_{sys\_un}$ & $0.35~(0.14)$ & $1.13~(0.46)$ & $1.15~(0.47)$ \\ \hline  
\end{tabular}}
\caption{\it\it The signal and background yields at the HE-LHC are shown along with the signal significance in the $b\bar{b}\tau^{+}\tau^{-}$ channel from the analysis using BDTD, XGBoost and DNN classifiers.}
\label{bb2tau:tab1}
\end{table}
\end{center}

\begin{figure}[htb]
\centering
\includegraphics[scale=0.37]{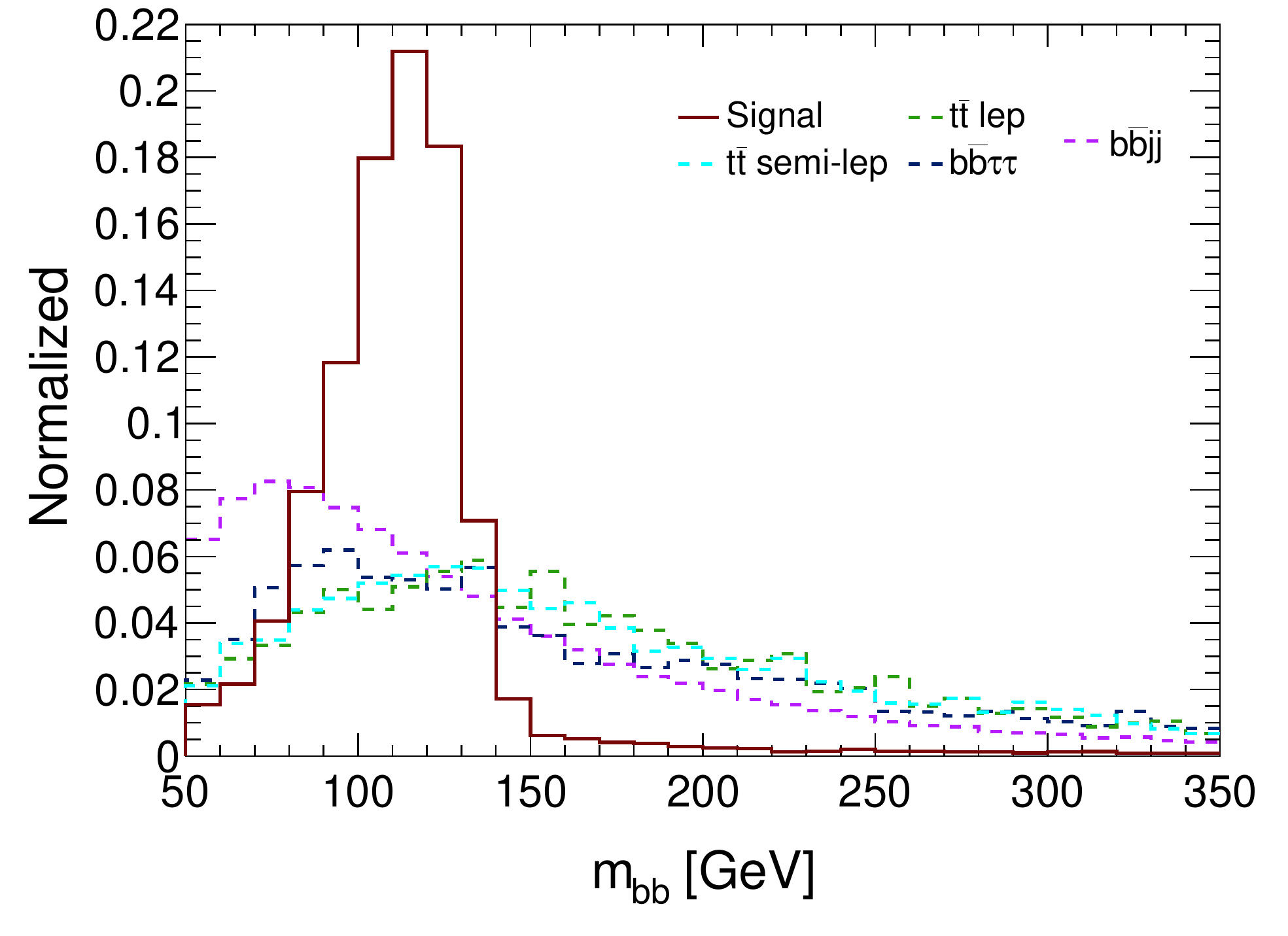}
\includegraphics[scale=0.37]{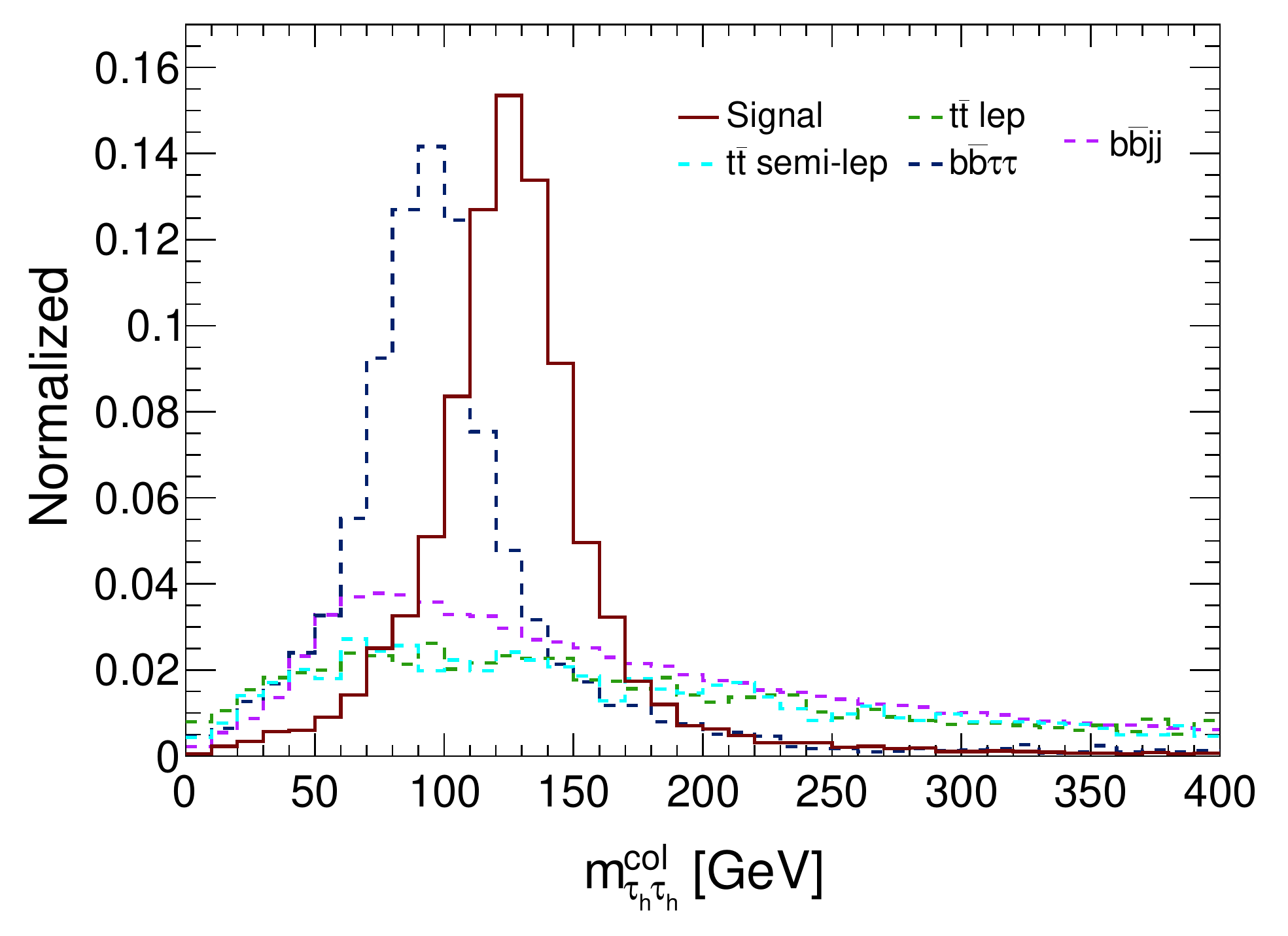}\\
\includegraphics[scale=0.37]{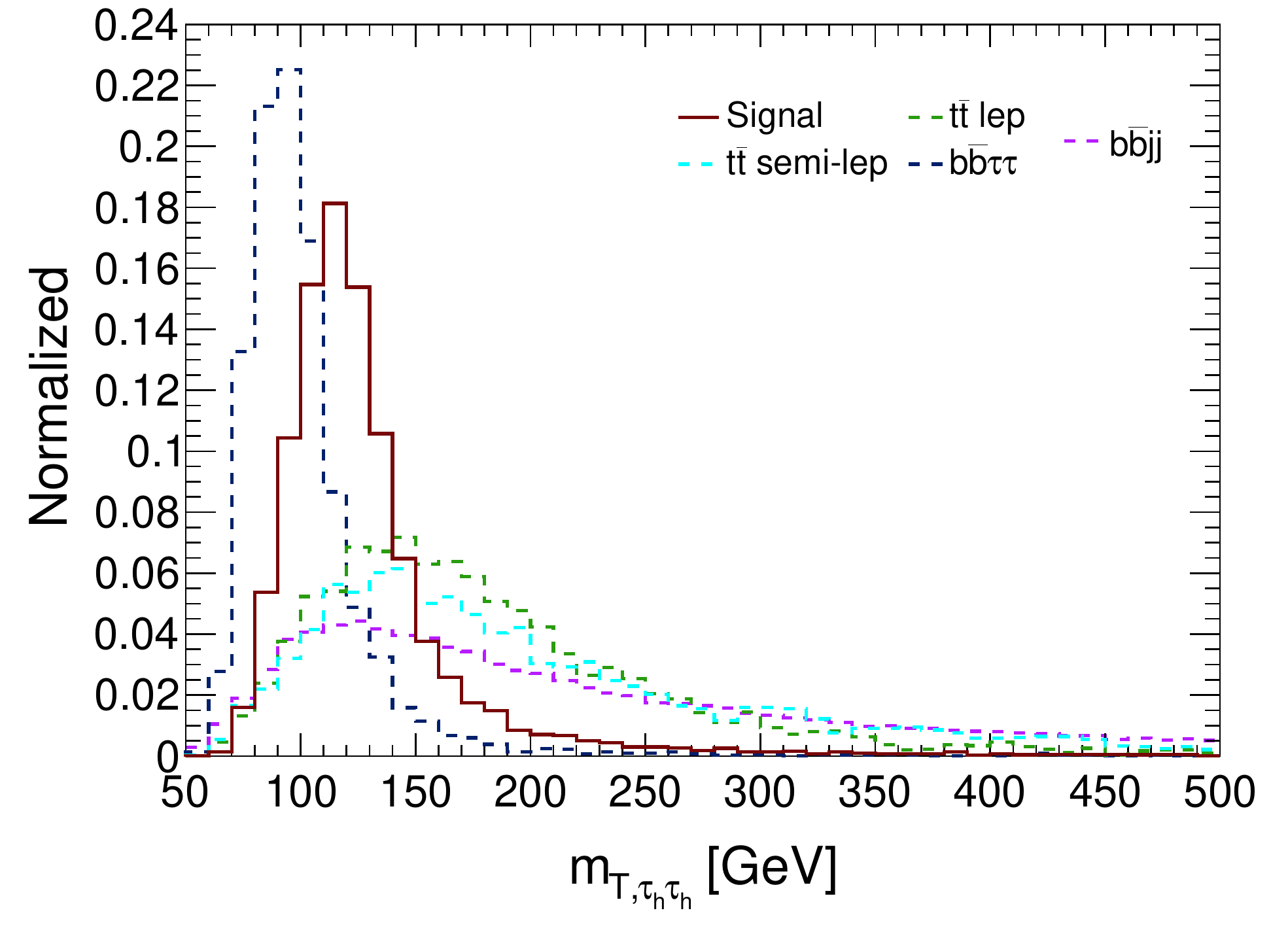}
\includegraphics[scale=0.37]{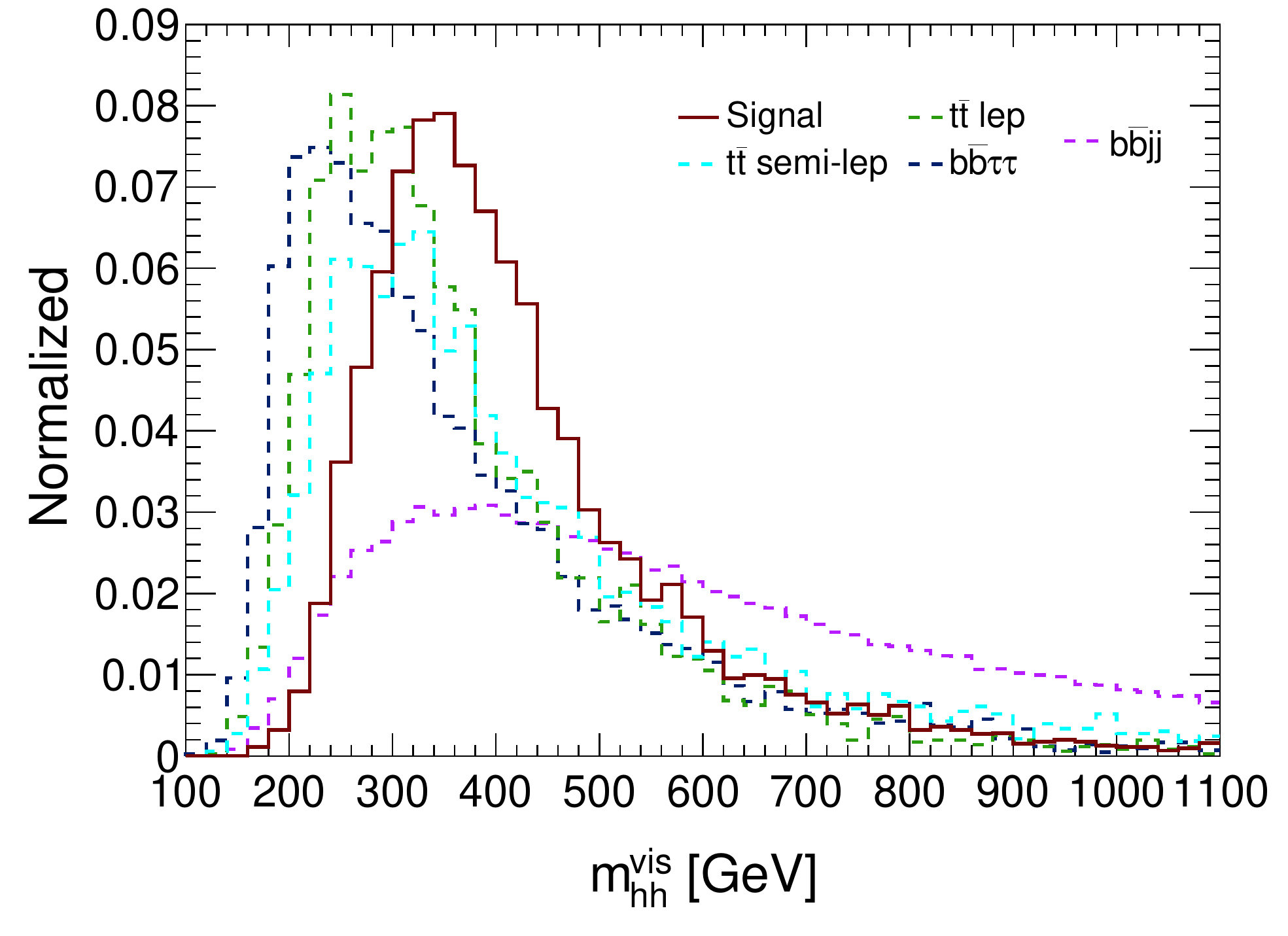}
\caption{\it Normalized distributions of $m_{bb}$, $m_{\tau_h\tau_h}^{col}$, $m_{T,\tau_h\tau_h}$ and $m_{hh}^{\textrm{vis}}$ for the $b\bar{b}\tau_h\tau_h$ signal and the dominant backgrounds after the acceptance cuts and the generation level cuts.}
\label{bb2tau:fig1}
\end{figure}

\begin{figure}
\centering
\includegraphics[scale=0.42]{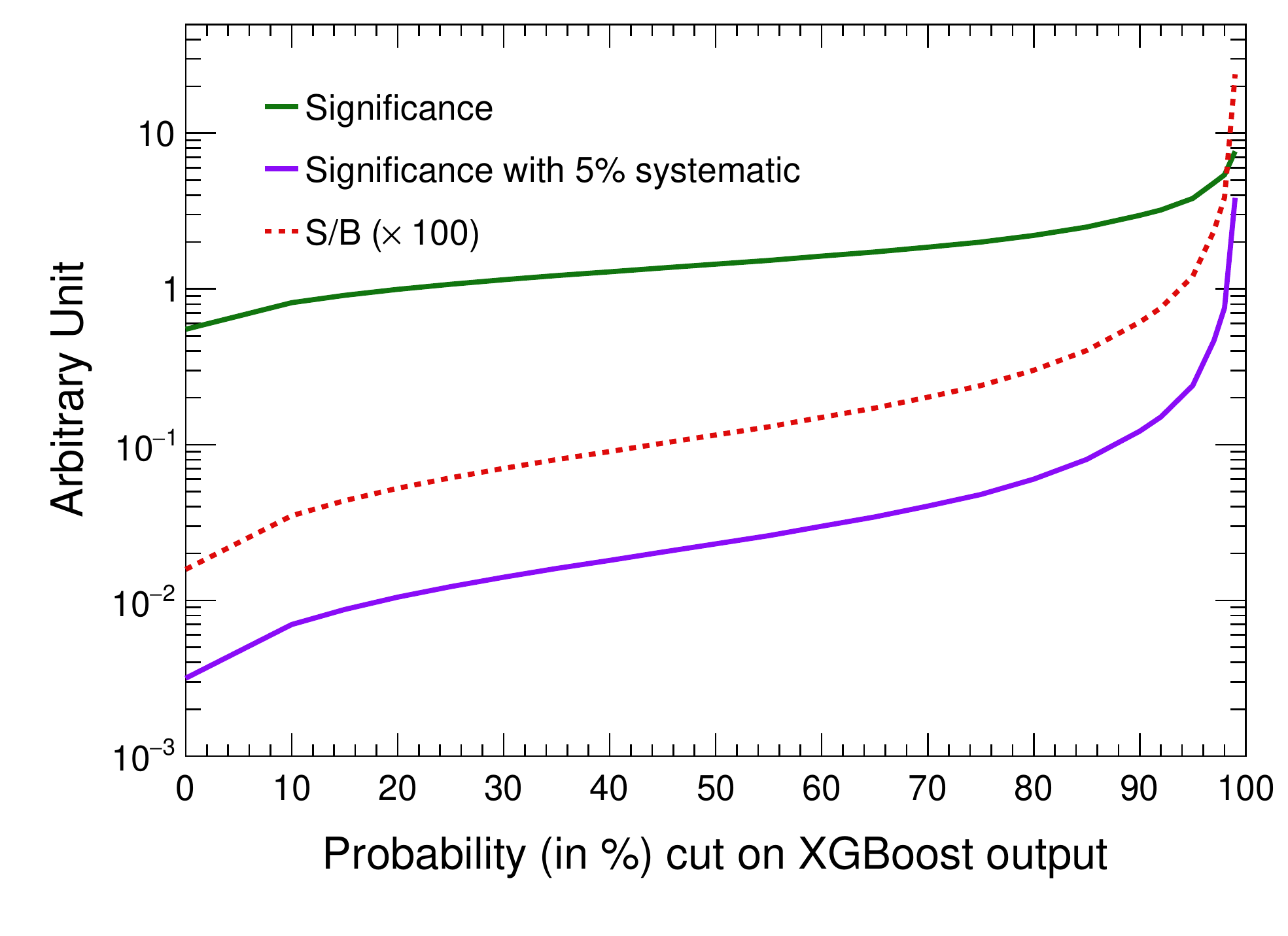}
\caption{\it The variation of significance~(with~($5\%$) and without systematic uncertainty) and $S/B$ is shown as a function of the probability cut on the XGBoost output for the $b\bar{b}\tau^{+}\tau^{-}$ channel.}
\label{bb2tau:fig2}
\end{figure}

where, $\Delta\phi_{\tau_{hi}\met}$, $i=1,2$ is the difference between the azimuthal angle of the visible decay product of the $\tau$ and the $\met$, $m_{T,\tau_h\tau_h}$ is the transverse mass of the $\tau$ pair and is defined by $m_{T,\tau_h\tau_h}^2=(\sum\limits_i E_{T,i})^2-(\sum\limits_i \vec{p}_{T,i})^2$ where i runs over the visible decay products of the $\tau$ and the $\met$, $m_{T2}$ is the stransverse mass~\cite{Lester:1999tx,Barr:2003rg} and $m_{\textrm{eff}}$ is the scalar sum of transverse momentum of all the visible decay products and $\met$. The $m_{T2}$ variable characterizes the topology of the fully leptonic $t\bar{t}$ background where the parent particles of the invisible neutrinos in the final state are of the same mass. In the case of fully leptonic $t\bar{t}$ background, it is bounded from above by the top quark mass, however, for the $pp \to hh \to b\bar{b}\tau^{+}\tau^{-}$ signal, $m_{T2}$ remains unbounded from above. In the construction of $m_{T2}$, the visible components are composed of two $b$ $jets$ and the two $\tau$-tagged $jets$ in the final state, while the invisible component is composed of the total $\met$ of the system. The other variables have their usual meaning. The kinematic variables which are most efficient in discriminating the signal and the backgrounds in the BDTD optimization are: $m_{bb}$, $m_{\tau_h \tau_h}^{col}$, $m_{T,\tau_h\tau_h}$ and $m_{hh}^{vis}$, and we illustrate their normalized distribution for the signal and the most dominant backgrounds in Fig.~\ref{bb2tau:fig1}. The BDTD optimization results in a total background yield of $1.56 \times 10^{5}$ and a signal yield of $1096$ leading to a signal significance of $2.77$ assuming zero systematic uncertainty (see Table~\ref{bb2tau:tab1}) and thus falls within the potential exclusion reach at the HE-LHC. We must however mention that $S/B$ value of this channel is of the order of $10^{-2}$ and the signal significance drops considerably upon introducing systematic uncertainties (significance becomes $0.14$ with $\sigma_{sys\_un} = 5\%$).

The signal and background discrimination is also performed using XGBoost and DNN, and the respective signal and background yields are shown in Table~\ref{bb2tau:tab1}. The signal and background yields for XGBoost are computed with a probability cut on the XGBoost output at $97\%$. The XGBoost signal yield is almost comparable to the signal yield from BDTD, however, the background yield is roughly $\sim 70\%$ less than its BDTD counterpart. Correspondingly, the signal significance improves to $4.78$ ($0.46$ with $\sigma_{sys\_un} = 5\%$). In Fig.~\ref{bb2tau:fig2}, we illustrate the variation of the signal significance obtained from the analysis using XGBoost, for $0\%$~(solid green line) and $5\%$~(solid purple line) systematic uncertainty, as a function of the probability cut on the XGBoost output. The red-dashed line represents the $S/B$ value scaled with $10^{2}$. The analysis using DNN results in a signal significance of $4.25$ ($0.47$ with $\sigma_{sys\_un} = 5\%$). Similar to our results in the $b\bar{b}\gamma\gamma$ channel~(Sec.~\ref{sec:bbaa}), we obtain the highest signal significance from XGBoost followed by DNN and BDTD. However, we would like to mention that more efficient optimization of the DNN analysis may lead to improved signal-background discrimination leading to comparable results with XGBoost. 

\subsection{The $b\bar{b}WW^{*}$ channel}

The $b\bar{b}WW^{*}$ channel manifests into three different final states: the fully leptonic $b\bar{b}l^{+}l^{-}+\met$, the semi-leptonic $b\bar{b}l+jets+\met$ and the fully hadronic $b\bar{b}+jets+\met$. Among these three, the fully leptonic final state is relatively cleaner, has a smaller background, and therefore, is the focus of this study.

The leading and sub-leading contribution to the background comes from leptonically decaying $t\bar{t}$ and $llb\bar{b}$, respectively. Lesser contributions to the background arises from $t\bar{t}h$, $t\bar{t}Z$, $t\bar{t}W$ and $tW$. The $tW$ process has been generated by merging with an additional $jet$ and the $W$ boson has been decayed leptonically. Furthermore, a hard cut of $m_{bb} \geq 50~{\rm GeV}$ is imposed at the generation level of $t\bar{t}$, $llb\bar{b}$ and $tW$ processes. We have listed all such generation level cuts in Appendix~\ref{sec:appendixA}.

The event selection criteria requires the presence of exactly two $b$ $jets$ with $p_{T} > 30~{\rm GeV}$ and $|\eta| < 4.0$, and two oppositely charged isolated leptons with $p_{T} > 20~{\rm GeV}$ and $|\eta| < 4.0$. In addition, for the sake of consistency, the generation level hard cuts: $\Delta R_{b_{i}l_{j}} > 0.2$ ($i,j =1,2$) and $m_{bb} > 50~{\rm GeV}$, imposed on $t\bar{t}$, $llb\bar{b}$ and $tW$, are applied on the signal and all the backgrounds at the selection level. The signal and the background samples thus obtained at the selection level are then passed through the BDTD algorithm. The following variables are considered to perform the multivariate analysis:

\begin{equation}
\begin{split}
logT,~logH,~M_{T2}^{(b)},~M_{T2}^{(\ell)},~\sqrt{\hat s_{min}^{(\ell\ell)}},~\sqrt{\hat s_{min}^{(bb\ell\ell)}},~
p_{T,\ell_{1/2}},~\met,~m_{\ell \ell},~m_{b b},\\\Delta R_{\ell \ell},~\Delta R_{b b},~p_{T,bb},~p_{T,\ell \ell},~\Delta \phi_{bb\; \ell \ell}.
\label{bbww:eq1}
\end{split}
\end{equation}

\begin{figure}[!ht]
\centering
\includegraphics[scale=0.37]{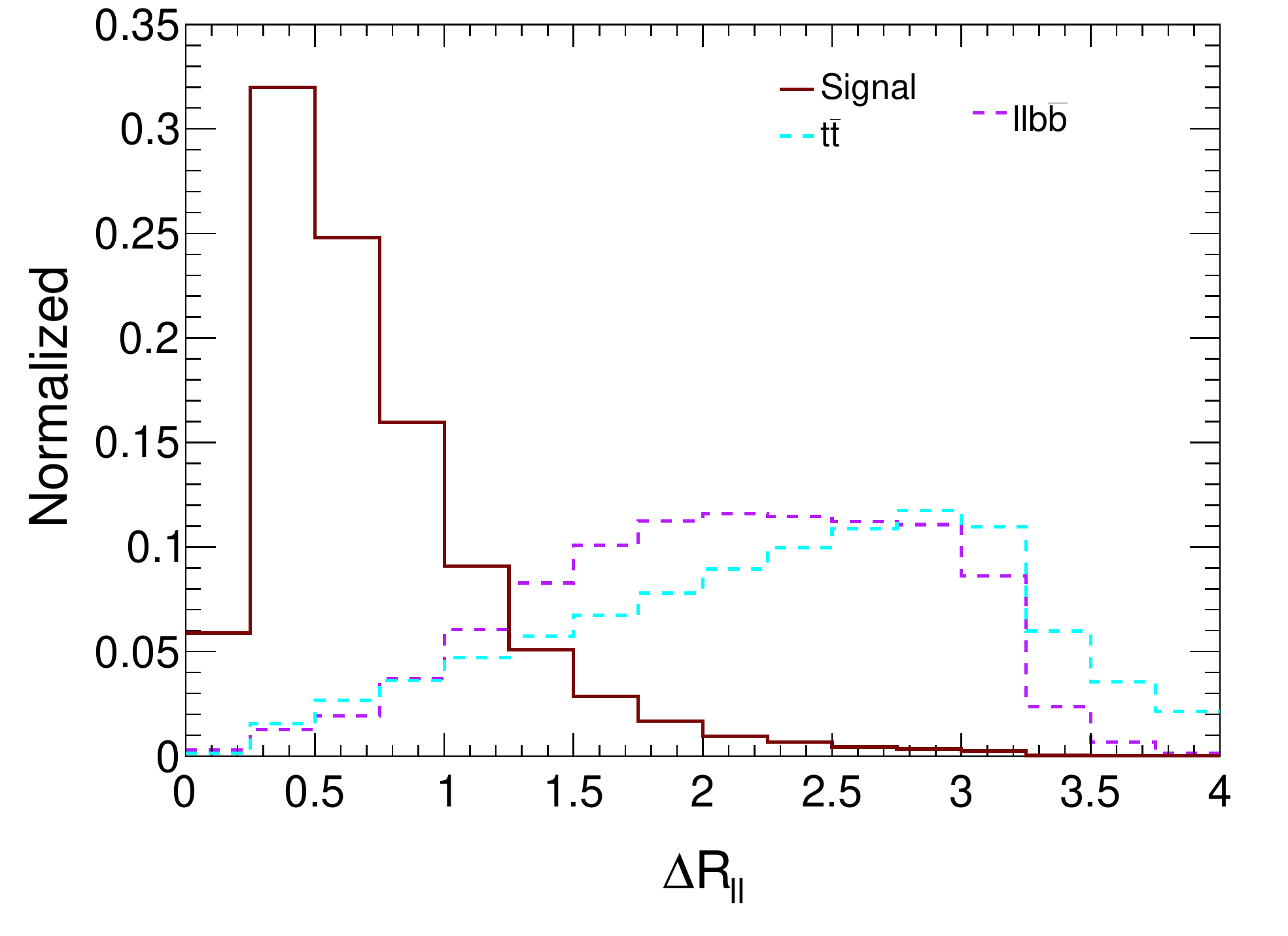}
\includegraphics[scale=0.37]{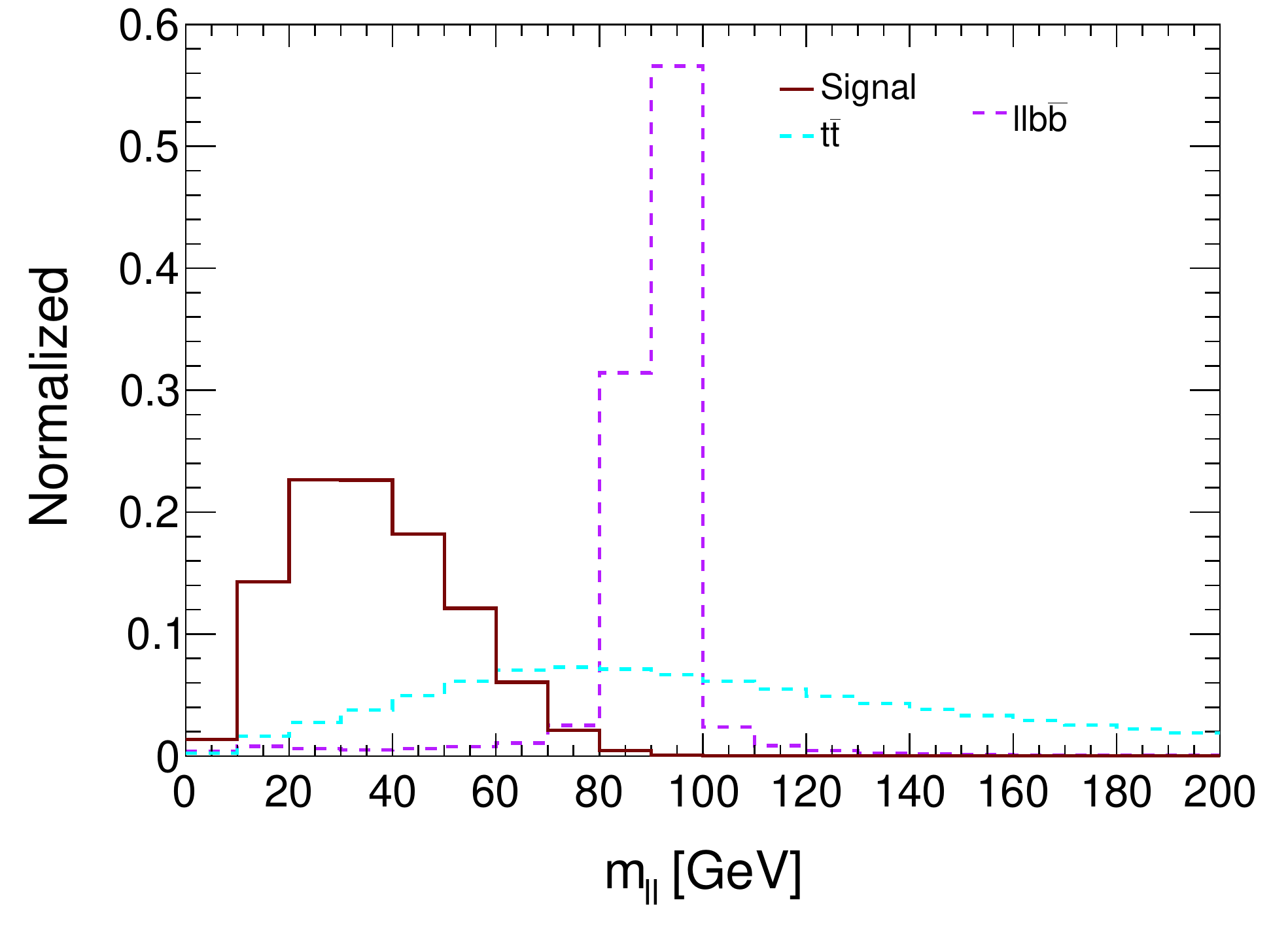}\\
\includegraphics[scale=0.37]{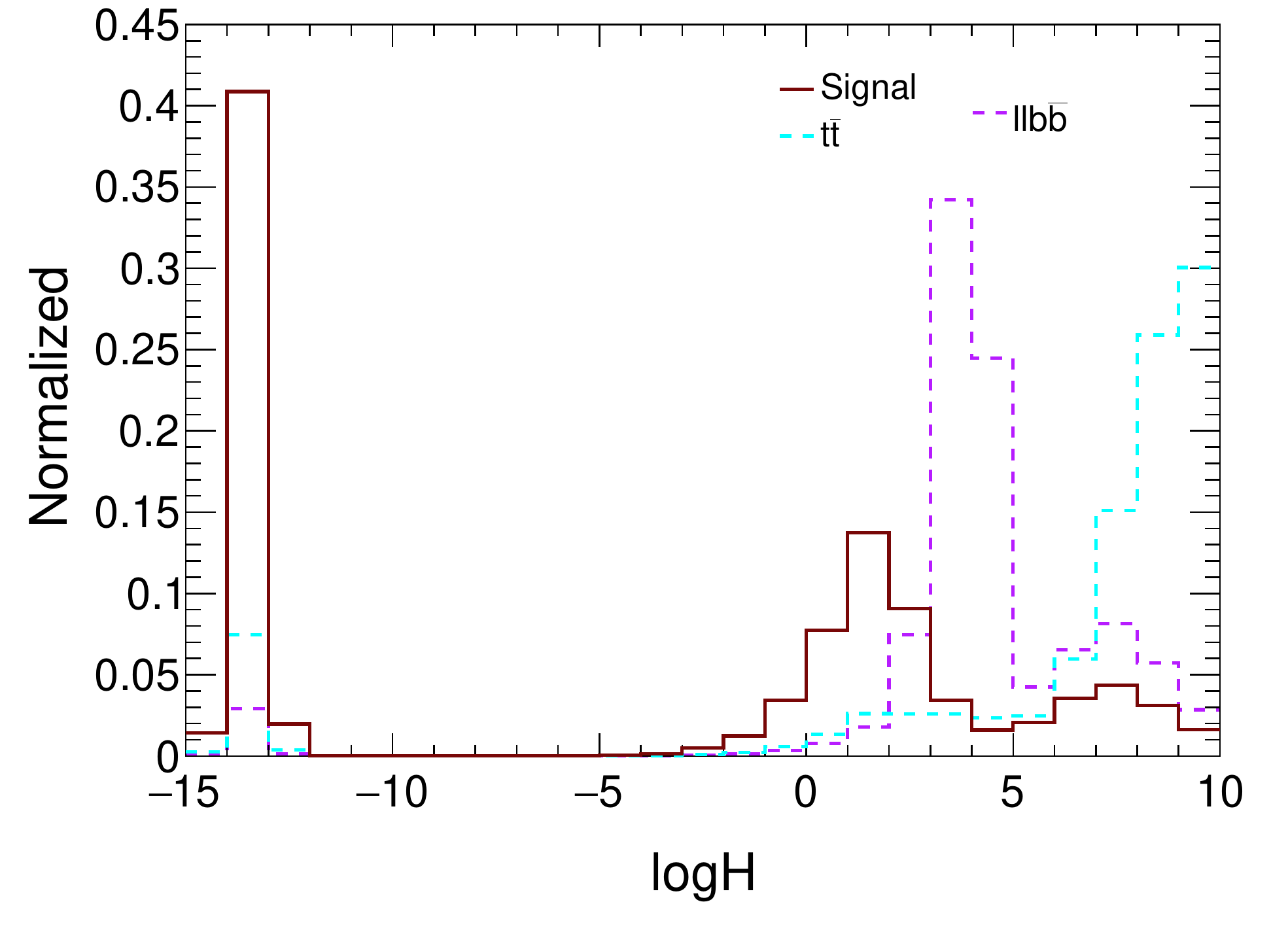}
\includegraphics[scale=0.37]{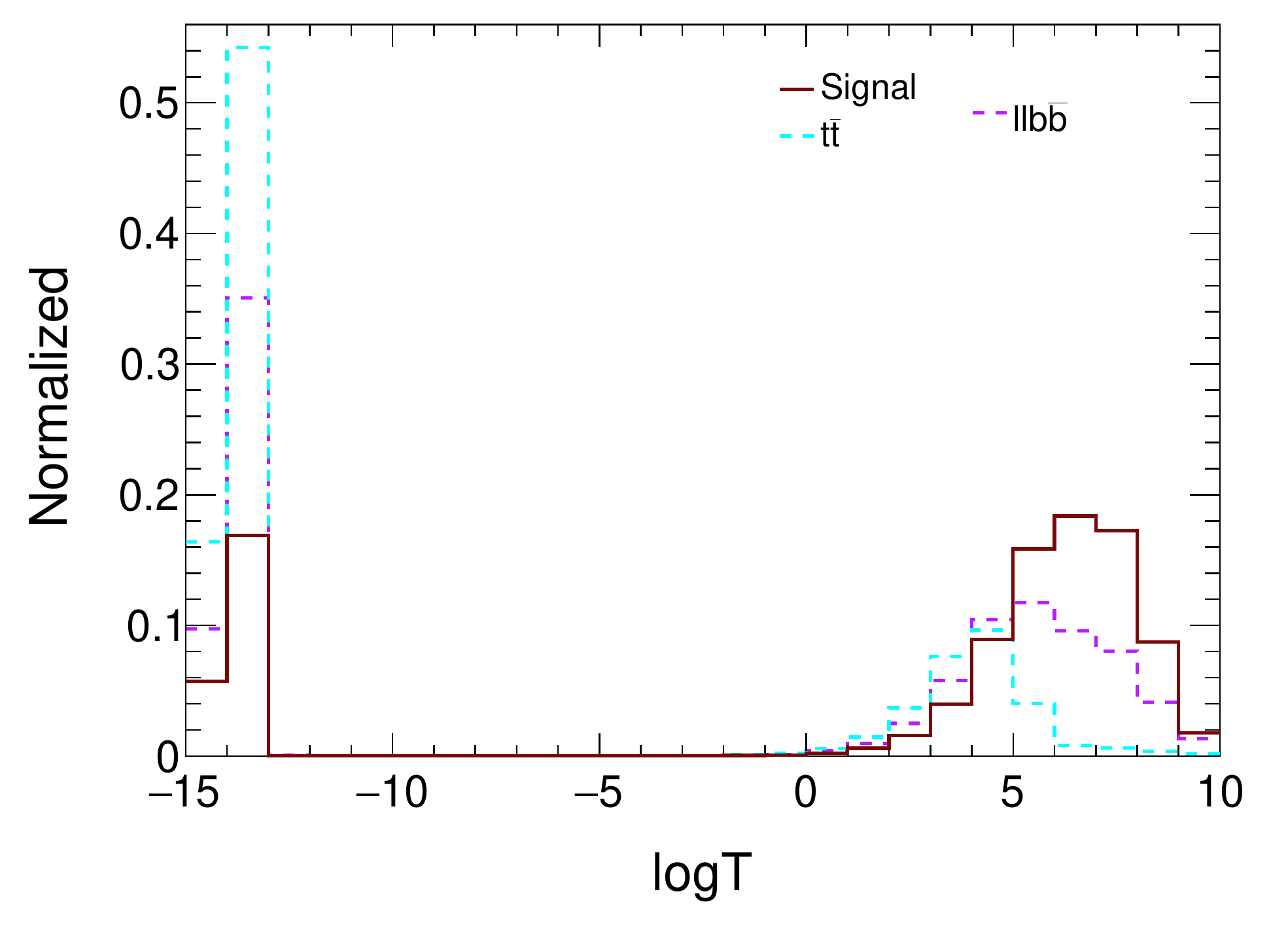}\\
\includegraphics[scale=0.37]{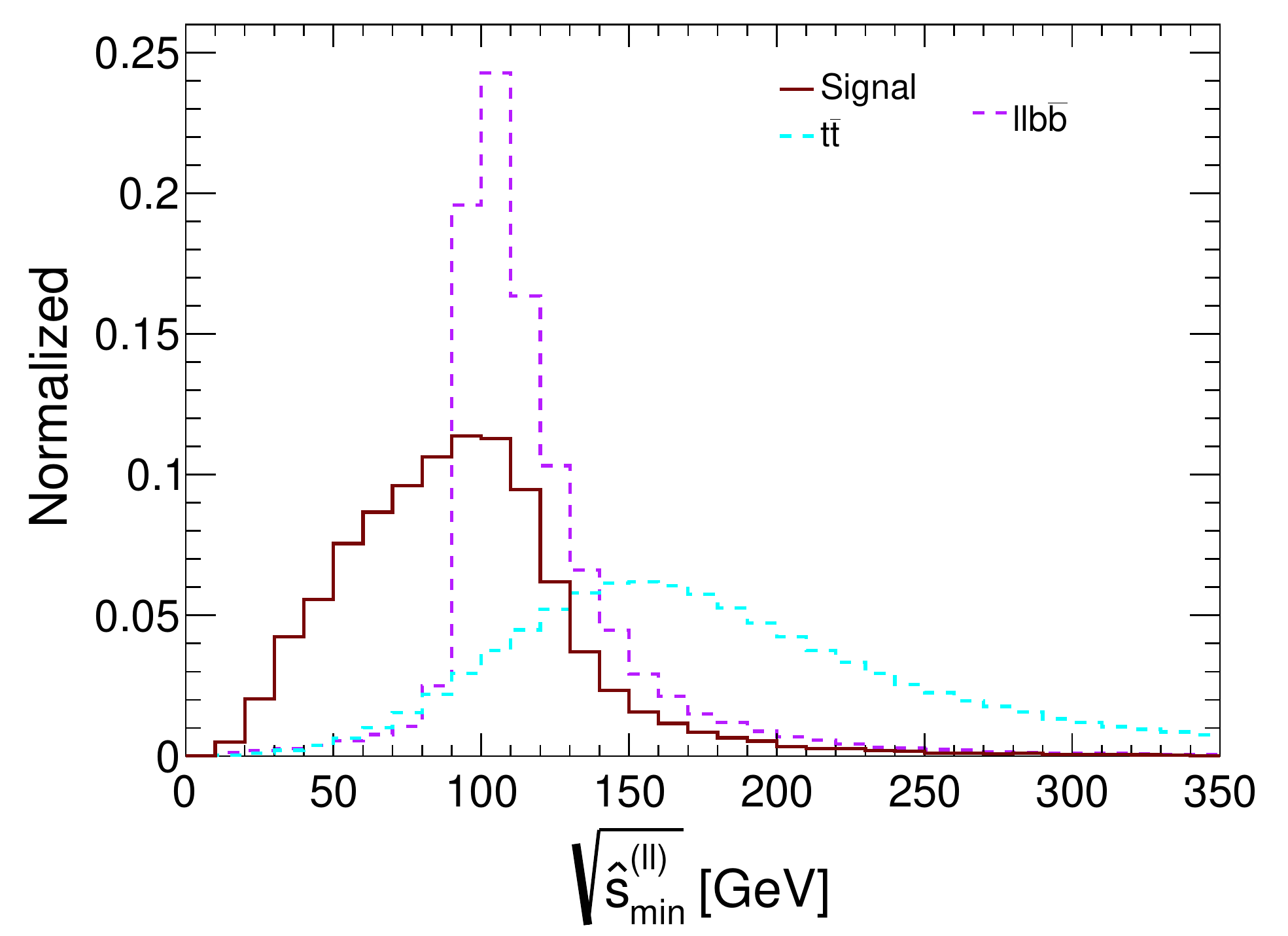}
\caption{\it Normalized distributions of $\Delta R_{\ell \ell}$, $m_{\ell \ell}$, $logH$, $logT$ and $\sqrt{\hat s_{min}^{(\ell\ell)}}$ for the $b\bar{b}WW^{*} \to 2b2\ell+\met$ signal and dominant backgrounds after the acceptance cuts and the generation level cuts.}
\label{bbww:fig1}
\end{figure}

Here, $T$ and $H$ refers to Topness and Higgsness~\cite{Graesser:2012qy,Kim:2018cxf}, respectively. The Topness variable is designed to characterize the fully-leptonic $t\bar{t}$ topology. The leptonic $t\bar{t}$ final state has two neutrinos whose three momenta are undetermined resulting in $6$ unknowns. Four on-shell mass conditions are provided by $m_{t}$, $m_{\bar{t}}$, $m_{W^{+}}$ and $m_{W^{-}}$. The three-momenta of the neutrino is fixed by minimizing $\chi_{ij}^{2}$ which is defined as~\cite{Kim:2018cxf}:

\begin{equation}
\chi^2_{ij} \equiv \min_{ \vec{\slashed{p}}_T = \vec{p}_{\nu T} + \vec{p}_{ \bar{\nu} T}}  \left [ \frac{\left ( m^2_{b_i \ell^+ \nu} - m^2_t \right )^2}{\sigma_t^4} + \frac{\left ( m^2_{\ell^+ \nu} - m^2_W \right )^2}{\sigma_W^4} + \frac{\left ( m^2_{b_j \ell^- \bar \nu} - m^2_t \right )^2}{\sigma_t^4} + \frac{\left ( m^2_{\ell^- \bar\nu} - m^2_W \right )^2}{\sigma_W^4} \right ]. 
\label{eq:chi2}
\end{equation}

Here, $i,j$ can take values of $1$ or $2$, corresponding to the leading~($b_{1}$) or sub-leading~($b_{2}$) $p_{T}$ ordered $b$ $jet$, respectively. Following Ref.~\cite{Kim:2018cxf}, we take $\sigma_{t} = \sigma_{W} = 5~{\rm GeV}$. The minimization of $\chi_{ij}^{2}$ is performed under the condition that the total missing transverse momentum of the system is the sum of the transverse momentum of the two neutrinos, $ \vec{\slashed{p}}_T = \vec{p}_{\nu T} + \vec{p}_{ \bar{\nu} T}$. There can be two different combinations of the $b$ $jet$, lepton and the neutrino: $b_{1}l^{+}\nu$ and $b_{2}l^{-}\bar{\nu}$, or $b_{2}l^{+}\nu$ and $b_{1}l^{-}\bar{\nu}$. The Topness is eventually defined by considering the minimum of the two $\chi_{ij}^{2}$ values, $T =  min \left ( \chi^2_{12} \, , \, \chi^2_{21} \right ) $. Similarly, the Higgsness variable targets the $h \to WW^{*}$ topology in the signal and is defined as~\cite{Kim:2018cxf}: 

\begin{gather}\label{eq:higgsness}
H \equiv \min \left [ \frac{\left ( m^2_{\ell^+\ell^-\nu \bar\nu} - m^2_h \right )^2}{\sigma_{h_\ell}^4} \right. + \frac{ \left ( m_{\nu  \bar\nu}^2 -  m_{\nu\bar\nu, peak}^2 \right )^2}{ \sigma^4_{\nu}} ~+  \\ \nonumber
 {min} \left (\frac{\left ( m^2_{\ell^+ \nu } - m^2_W \right )^2}{\sigma_W^4} + \frac{\left ( m^2_{\ell^- \bar \nu} - m^2_{W^*, peak} \right )^2}{\sigma_{W^*}^4} , \right. \left.  \left . \frac{\left ( m^2_{\ell^- \bar \nu} - m^2_W \right )^2}{\sigma_W^4} + \frac{\left ( m^2_{\ell^+ \nu} - m^2_{W^*, peak} \right )^2}{\sigma_{W^*}^4}  \right )  \right ]   \, . \nonumber
\end{gather}

Here, $m_{W^{*},peak}$ is the invariant mass of the off-shell $W$ boson produced from the decay of the Higgs boson. $m_{W^{*},peak}$ is defined as:
\begin{gather}
\frac{1}{\sqrt{3}} \sqrt{2(m_{h}^{2} + m_{W}^{2}) - \sqrt{m_{h}^{4} + 14 m_{h}^{2}m_{W}^{2} + m_{W}^{4}}}
\end{gather}
$m_{\nu\bar{\nu},peak}$ in Eqn.~\ref{eq:higgsness} corresponds to the location of the peak of the $\nu\bar{\nu}$ invariant mass distribution, and is allocated a value of $30~{\rm GeV}$ following Ref.~\cite{Kim:2018cxf}. We also fix $\sigma_{W^{*}}$, $\sigma_{h_{l}}$ and $\sigma_{\nu}$ at $5~{\rm GeV}$, 2~GeV and 10~GeV, respectively, following Ref.~\cite{Kim:2018cxf}. A combination of these two variables is found to be effective in discriminating the large $t\bar{t}$ background from the signal. The $b$ $jets$ and the decay products of the $W$ boson correspond to the visible and invisible components, respectively, of the $m_{T2}^{(b)}$ variable, while $~m_{T2}^{(\ell)}$ is computed by considering the two isolated final state leptons to form the visible component and the neutrinos to form the invisible component. These two variables have been taken from \cite{Konar:2008ei}. We also consider the minimum of the Mandelstam invariant mass variable: $\sqrt{\hat{s}_{min}}$~\cite{Konar:2008ei}, where $\sqrt{\hat{s}_{min}^{(ab)}}$ represents the minimum value of the centre of mass (c.o.m) energy required to produce the parton level parent particles of the final state particles $a$ and $b$. It is defined as: $$\hat{s}_{min}^{(ab)}=m_{ab}^2+2(\sqrt{{|p_T^{ab}|^2 + m_{ab}^2}~\slashed{p_T} - \vec{p}_T^{ab}.\vec{\slashed{p_T}}}$$ where $m_{ab}$ and $p_T^{ab}$ are the invariant mass and transverse momentum, respectively, of the visible system $ab$. In the present scenario, we consider two such variables: $\sqrt{\hat s_{min}^{(\ell\ell)}}$ and $\sqrt{\hat s_{min}^{(bb\ell\ell)}}$, where the former and the latter variables represent the minimum c.o.m. energy scale of the parton level $WW^{*}$ pair and the Higgs pair, respectively. Furthermore, in Eqn.~\ref{bbww:eq1}, $p_{T,\ell_{1/2}}$ represents the $p_{T}$ of the leading and sub-leading isolated leptons~(represented as $l_{1}$ and $l_{2}$, respectively), $m_{\ell \ell}$ represents the invariant mass of the $l_{1}l_{2}$ pair, $\Delta R_{\ell \ell}$ corresponds to the distance between $l_{1}$ and $l_{2}$ in the $\eta-\phi$ plane, $p_{T,\ell \ell}$ measures the $p_{T}$ of the di-lepton system and $\Delta \phi_{bb\; \ell \ell}$ computes the difference of azimuthal angles between the $b\bar{b}$ system and the di-lepton system.

The variables which are found to be most efficient in discriminating the signal and the background in the BDTD analysis are: $\Delta R_{\ell\ell}$, $m_{\ell\ell}$, $\log H$ and $\log T$. We illustrate the normalized distributions of these four variables and $\sqrt{\hat s_{min}^{(\ell\ell)}}$, for the signal and the dominant backgrounds: $t\bar{t}$ and $\ell\ell b\bar{b}$, in Fig.~\ref{bbww:fig1}. The signal and background yields obtained from the BDTD analysis have been listed in Table~\ref{bbww:tab1}. We obtain a signal significance of $1.42$ without assuming any systematic uncertainties. We must note that the signal significance value undergoes a significant reduction upon the introduction of systematic uncertainty due to the small $S/B$ which is shown in Table~\ref{bbww:tab1}.

\begin{center}
\begin{table}[htb!]
\centering
\scalebox{0.7}{%
\begin{tabular}{|c|c|c|c|c|c|}\hline
       & Process & Cross section order     & \multicolumn{3}{c|}{Event yield after the analysis with} \\ \cline{4-6}
 &         &           & BDTD & XGBoost & DNN\\ \hline\hline

\multirow{5}{*}{Background}   
 & $t\bar{t}$ lep             & NNLO~\cite{ttbarNNLO}                      & $481445$   & $305038$ & $242560$\\  
 & $\ell\ell b\bar{b}$        & LO                                         & $169857$   & $48365$ & $22441$\\
 & $t\bar{t}Z$                & NLO~\cite{Lazopoulos:2008de}               & $9544$     & $3364$ & $1930$\\  
 & $t\bar{t}h$                & NLO~\cite{bkg_twiki_cs}                    & $6285$     & $7636$ & $2960$\\  
 & $t\bar{t}W$                & LO                                         & $725$      & $775$ & $280$\\ 
 & $tW$                       & LO                                         & $5$        & $6.5$ & $3.7$\\ \cline{2-6} 
 & \multicolumn{2}{c|}{Total}                                              & $667861$   & $365185$ & $270175$\\ \hline
\multicolumn{2}{|c|}{Signal ($hh \to b\bar{b}WW^*\to b\bar{b} \ell \ell + \met$)} & NNLO~\cite{hhtwiki}  & $1162$ & $1661$ & $746$\\\hline 
\multicolumn{2}{|c|}{\multirow{2}{*}{Significance}} & $0\%$ $\sigma_{sys\_un}$ & $1.42$ & $2.75$ & $1.43$\\ \cline{3-6}
\multicolumn{2}{|c|}{} & $2\%~(5\%)$ $\sigma_{sys\_un}$ & $0.09~(0.03)$ & $0.23~(0.09)$ & $0.14~(0.06)$ \\ \hline
\end{tabular}}
\caption{\it \it The signal and background yields at the HE-LHC along with the signal significance for the $b\bar{b}WW^{*}$ channel from the analysis using BDTD, XGBoost and DNN classifiers.}
\label{bbww:tab1}
\end{table}
\end{center}

\begin{figure}
\centering
\includegraphics[scale=0.42]{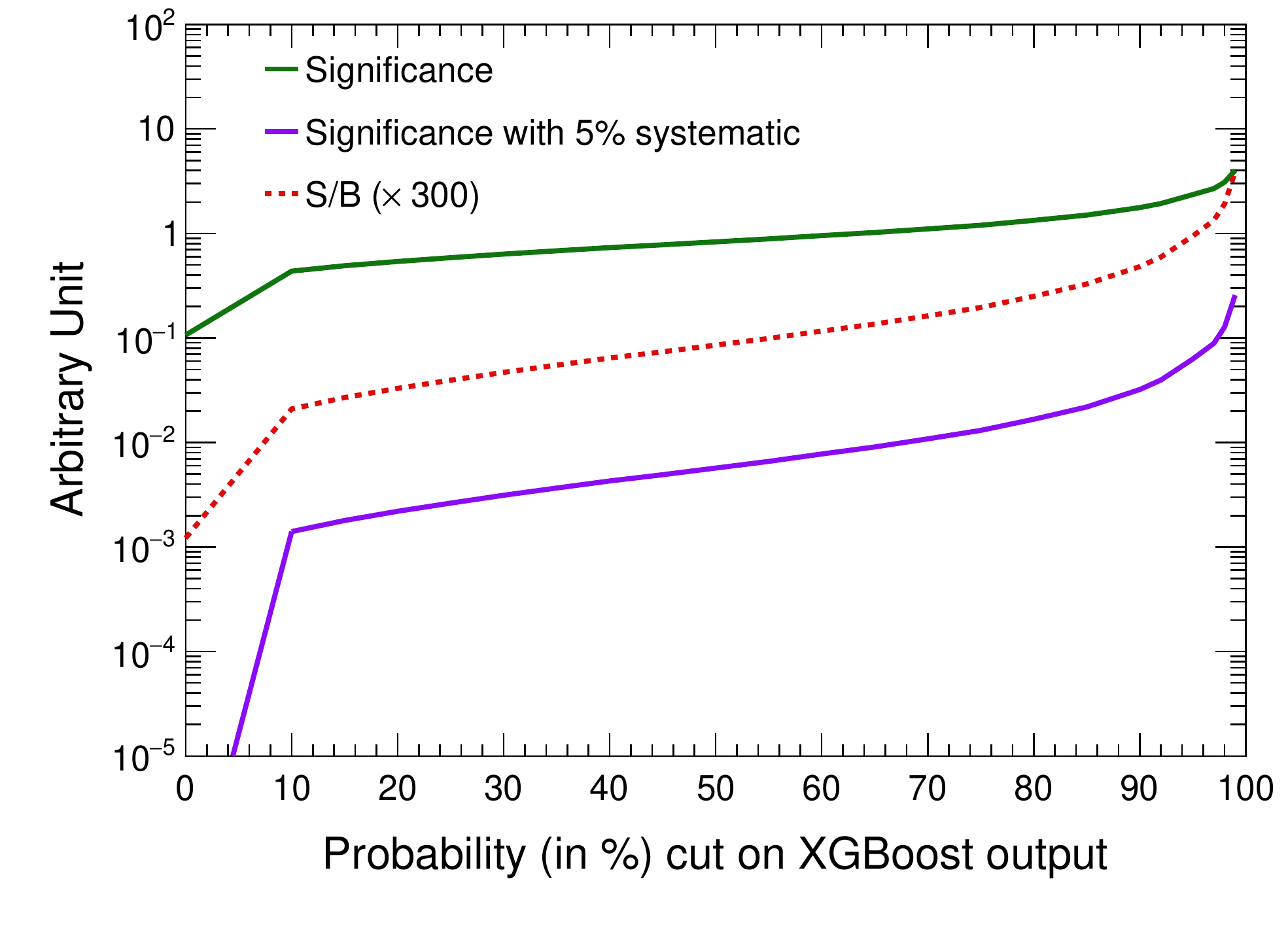}
\caption{\it The variation of significance~(with~($5\%$) and without systematic uncertainty) and $S/B$ is shown as a function of the probability cut on the XGBoost output for the $b\bar{b}WW^{*}$ final state.}
\label{bbww:fig2}
\end{figure}

Here again, we also use the XGBoost and DNN classifier to perform a detailed collider analysis. We obtain a signal yield of $1661$~($746$) and a total background yield of $3.6 \times 10^{5}$~($2.7 \times 10^{5}$) resulting in a signal significance of $2.75$~(1.43) which is $\sim 1.94$~($\sim 1.01$) times higher~(smaller) than the corresponding value from the BDTD optimization~(see Table.~\ref{bbww:tab1}). We would like to mention that signal yields from the BDTD optimization and the XGBoost toolkit are roughly within $30\%$ of each other, however, the XGBoost classifier performs background rejection more efficiently. The XGBoost background yield is approximately $45\%$ smaller than its BDTD counterpart. The XGBoost signal significance, computed assuming zero and $5\%$ systematic uncertainty, are also illustrated in Fig.~\ref{bbww:fig2} as a function of the probability cut applied on the XGBoost output, in solid green and solid purple lines, respectively. The red dashed line in Fig.~\ref{bbww:fig2} represents the respective variation in the value of $S/B$~(scaled with $300$).

\subsection{The $WW^{*}\gamma\gamma$ channel}

The production rate of $pp \to hh \to WW^{*}\gamma\gamma$ is smaller than the di-Higgs final states considered until now. However, the $WW^{*}\gamma\gamma$ channel draws an advantage from its relatively smaller backgrounds. In the present subsection, we focus only on the pure leptonic decay mode of $WW^{*}\gamma\gamma$ and defer the treatment of the semi-leptonic and hadronic decay modes to a future study. The pure leptonic decay mode results in the $WW^{*}\gamma\gamma \to l^{+}l^{-}\gamma\gamma+\met$ final state. In the analysis channels studied until now, we had performed the signal-background discrimination through the BDTD optimization, the XGBoost toolkit as well as the DNN framework. In all the previous channels, namely $b\bar{b}\gamma\gamma$, $b\bar{b}\tau^{+}\tau^{-}$ and $b\bar{b}WW^{*}$, we observed that the DNN analysis resulted in a slight weaker signal significance than the XGBoost classifier. Therefore, present subsection onwards, we perform the collider analysis only through the BDTD algorithm and the XGBoost toolkit. 

The dominant source of background is the $t\bar{t}h$ process. Sub-dominant contributions arise from $Zh$ and $ll\gamma\gamma$ processes. The later two backgrounds have been generated upon merging with two additional $jets$ and one additional $jet$, respectively. At the generation level, we impose a lower limit on the invariant mass of the di-lepton pair, $m_{ll} > 20~{\rm GeV}$, and a lower limit on the distance between the two final state leptons and the photons in the $\eta-\phi$ plane, $\Delta R_{ll/\gamma\gamma} > 2.0$. Additionally, the invariant mass of the photon pair, $m_{\gamma\gamma}$, is also restricted between 120~GeV$< m_{\gamma\gamma} <~$130~GeV at the generation level. The generation level cuts have been listed in Appendix~\ref{sec:appendixA}.
\begin{figure}[!htb]
\centering
\includegraphics[scale=0.37]{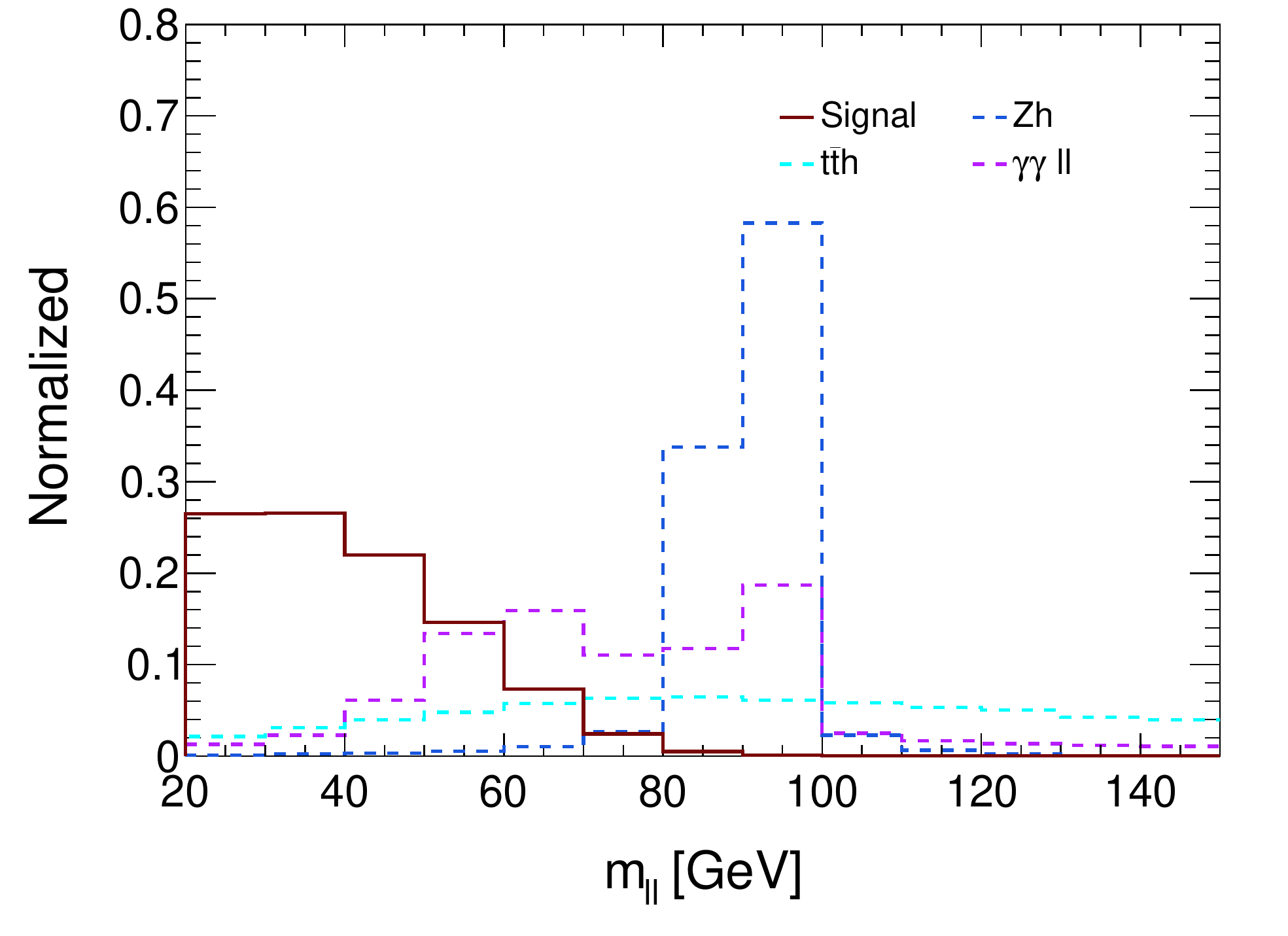}
\includegraphics[scale=0.37]{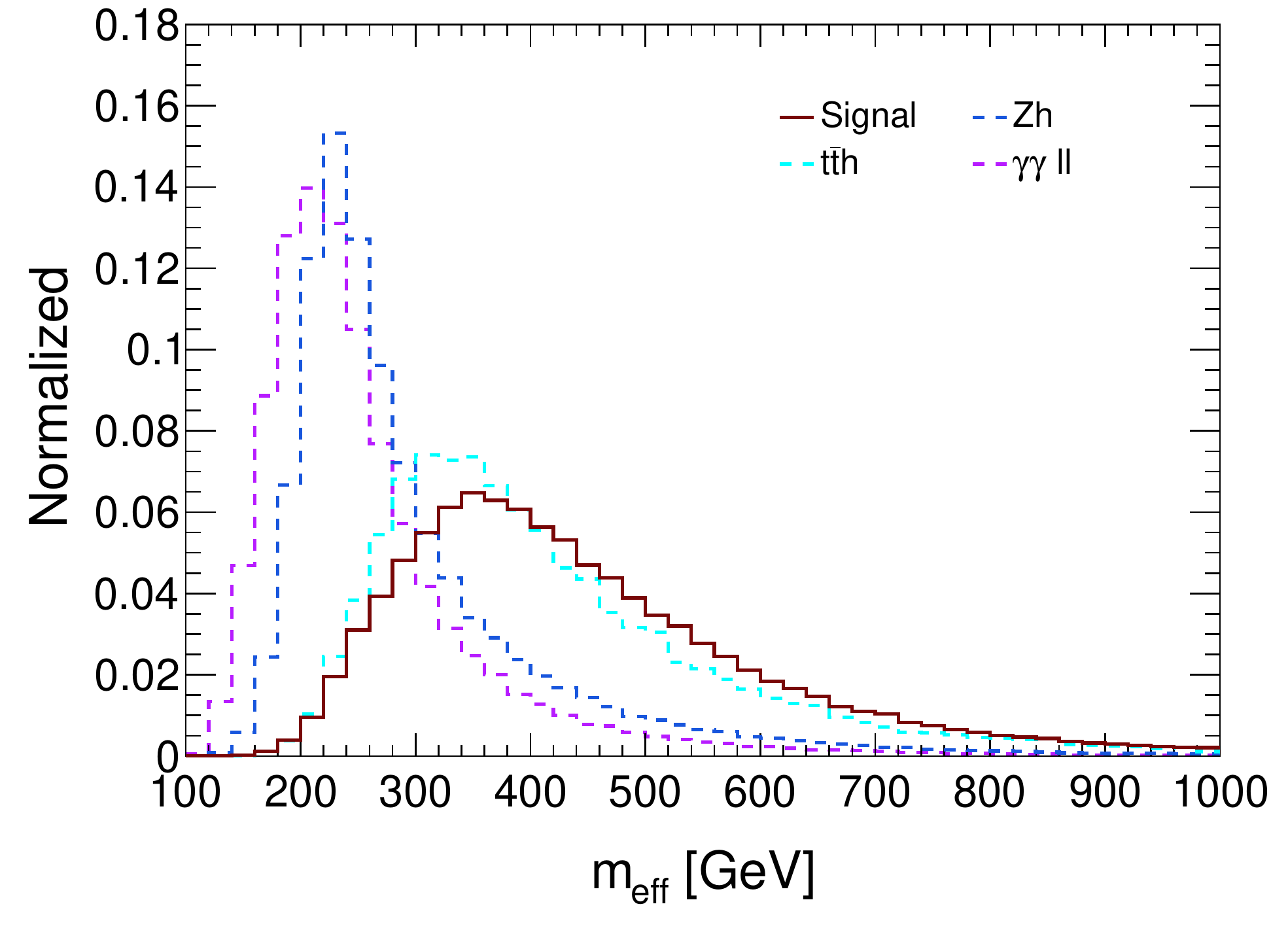}\\
\includegraphics[scale=0.37]{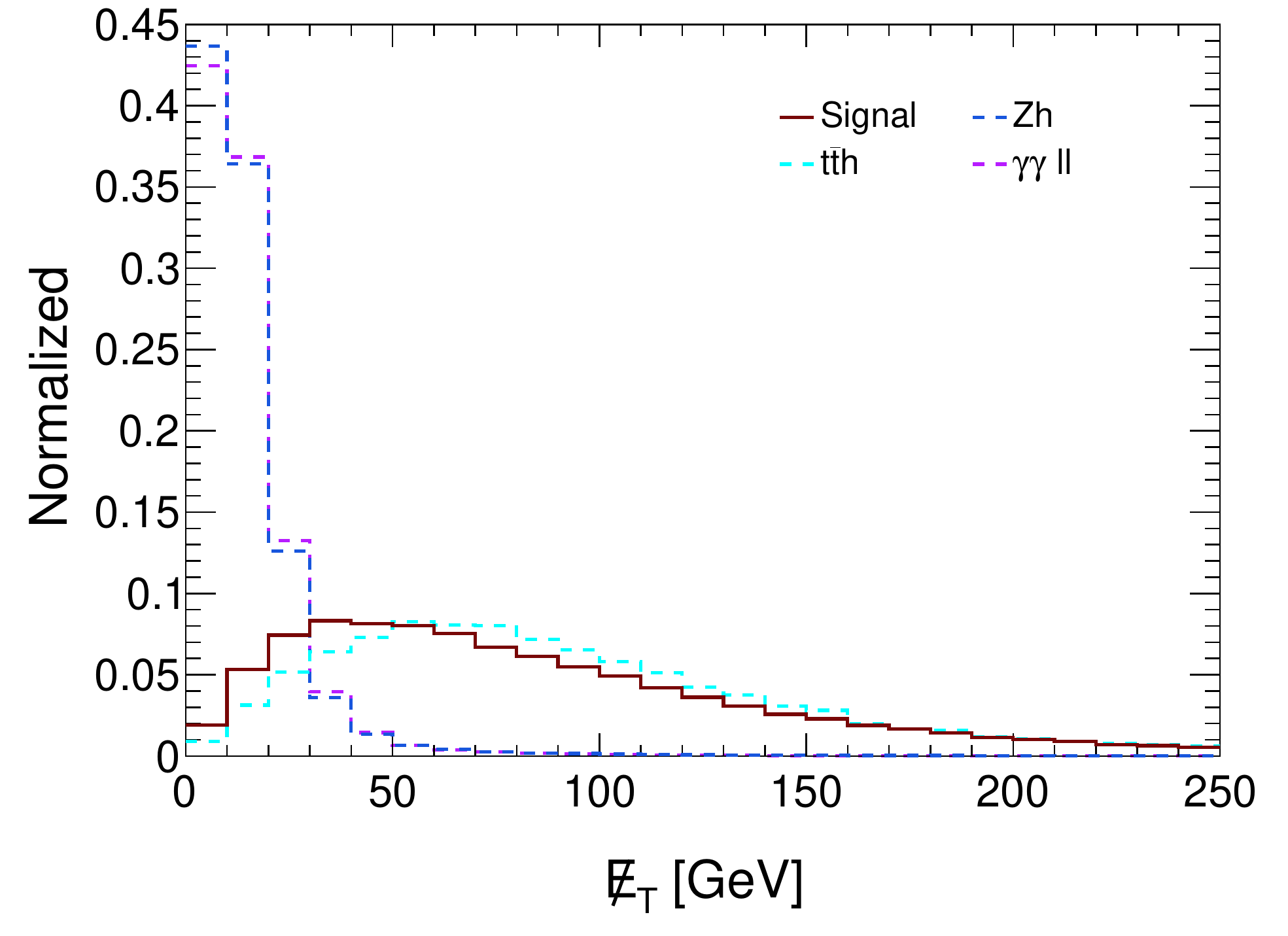}
\includegraphics[scale=0.37]{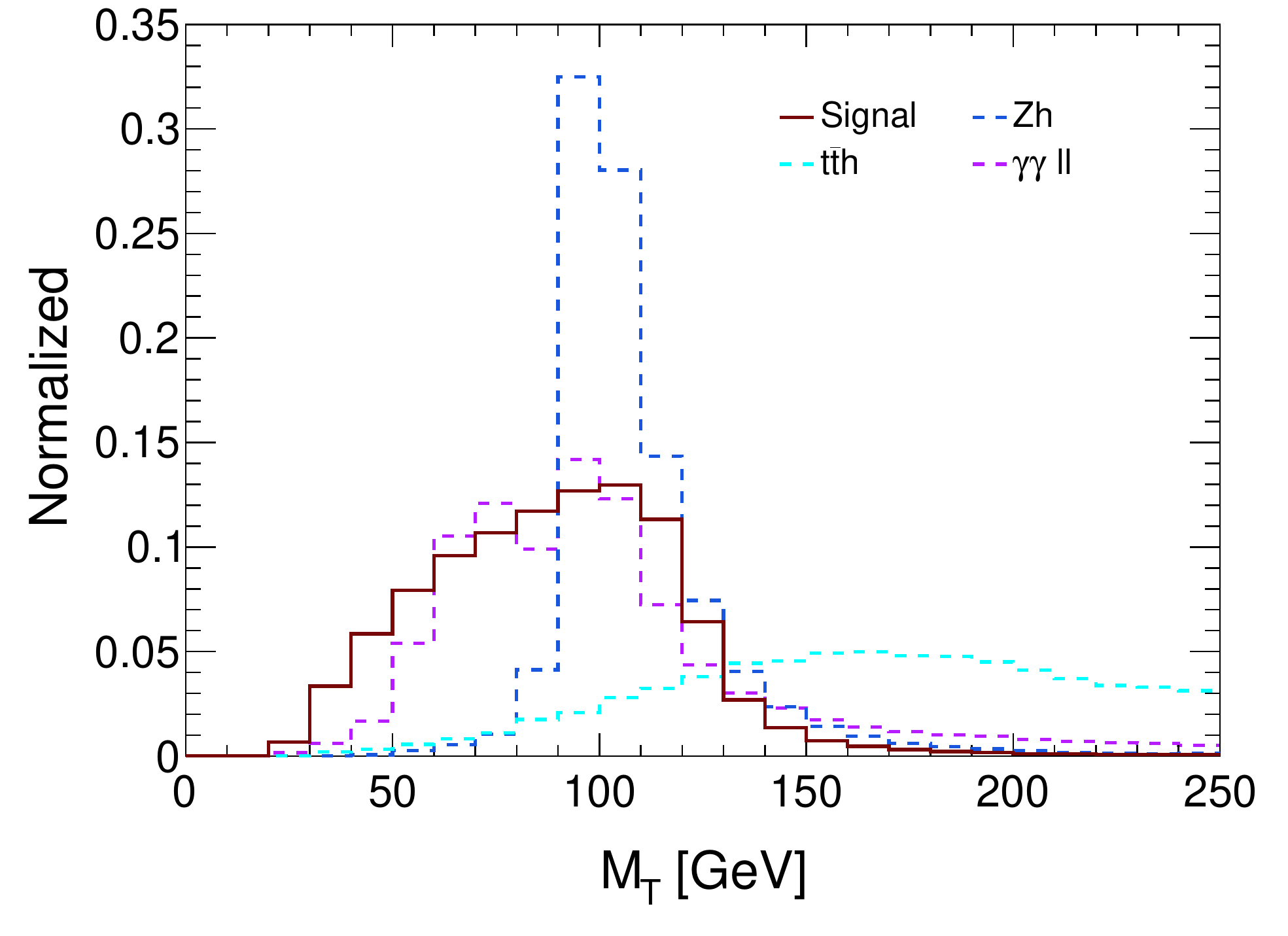}
\caption{\it Normalized distributions of $m_{\ell \ell}$, $m_{eff}$, $\met$ and $M_T$ for the $WW^{*}\gamma\gamma \to 2\ell 2\gamma + \met$ signal and the dominant backgrounds after the application of acceptance cuts and generation level cuts.}
\label{wwaa:fig1}
\end{figure}

An event is required to have two isolated opposite sign leptons (electrons or muons) with $p_{T} \geq 20~{\rm GeV}$ and two photons with $p_{T} \geq 30~{\rm GeV}$. The leptons and photons must lie within $|\eta| \leq 4.0$. To tackle the $t\bar{t}h$ background, a $b$ $jet$ veto is applied. In addition to these acceptance cuts, the generation level cuts are also imposed. We also veto events containing $\tau$ $jets$ and demand $m_{\gamma\gamma}$ to lie within $\left[122:128\right]~{\rm GeV}$. We perform a multivariate analysis with the BDTD algorithm and the XGBoost toolkit using the following kinematic variables:

\begin{equation}
\begin{split}
p_{T,\gamma \gamma},~\Delta R_{\gamma \gamma},~m_{\ell \ell},~p_{T,\ell \ell},~\Delta R_{\ell \ell},~\Delta R_{\ell \ell \; \gamma \gamma},~M_T,~p_{T,hh},~m_{eff},~\Delta R_{\gamma_1 \ell_1},\\Cos\theta^{*},~Cos\theta_{\gamma_1 h},~\met 
\end{split}
\end{equation}

Here, $M_{T}$ refers to the transverse mass of the $h \to WW^{*} \to l^{+}l^{-}+\met$ system while $m_{eff}$ represents the sum of the visible transverse momentum ($H_{T}$) and $\met$ in the event. The variables which are most efficient in discriminating the signal and background events in the BDTD analysis are: $m_{ll}$, $m_{eff}$, $\met$ and $M_{T}$, and we illustrate their normalized distributions in Fig.~\ref{wwaa:fig1}.  

\begin{figure}
\centering
\includegraphics[scale=0.42]{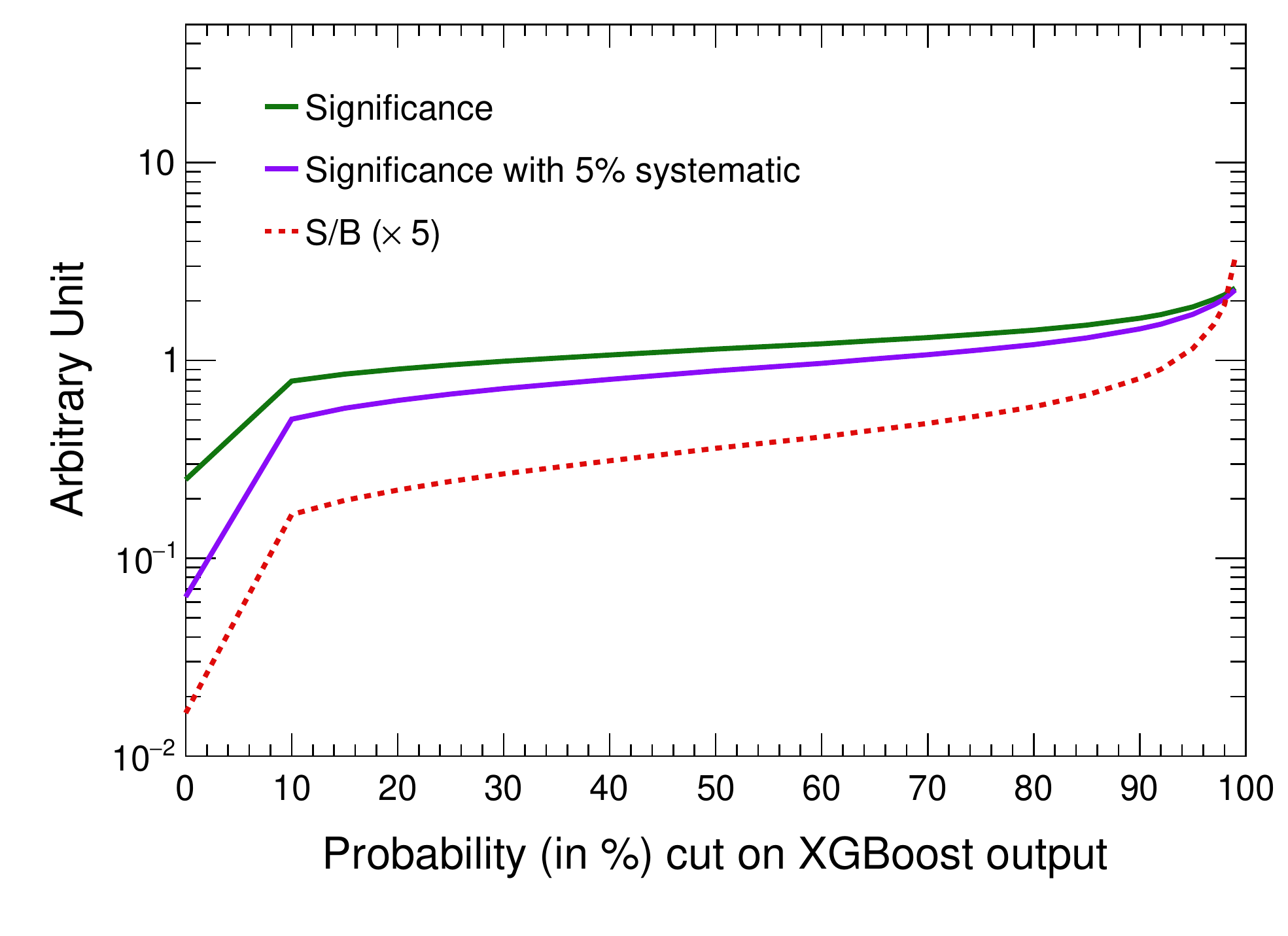}
\caption{\it The variation of significance~(with~($5\%$) and without systematic uncertainty) and $S/B$ is shown as a function of the probability cut on the XGBoost output for the $WW^{*}\gamma\gamma$ channel.}
\label{wwaa:fig2}
\end{figure}

The signal and background yields obtained from the BDTD and the XGBoost optimization (with probability cut at $97\%$) are listed in Table~\ref{wwaa:tab1} along with the respective signal significance values (assuming zero systematic uncertainty). The $S/B$ value for this channel is $\sim 0.2$~(from BDTD) and $\sim 0.3$~(from XGBoost), and are only behind the $S/B$ values from $b\bar{b}\gamma\gamma$ search channel. The signal yields from both the optimizations are roughly similar however the background yield from XGBoost is $\sim 38\%$ smaller. We obtain a signal significance of $1.64$ through BDTD analysis while the XGBoost framework leads to a signal significance of $2.05$. We follow the color code of Fig.~\ref{bbaa:fig3} and illustrate the signal significances computed at zero and $5\%$ systematic uncertainty as a function of the probability cut on the XGBoost output in Fig.~\ref{wwaa:fig2}. The relatively large $S/B$ value in the present channel is reflected by the red dashed line in Fig.~\ref{wwaa:fig2}.

\begin{center}
\begin{table}[htb!]
\centering
\scalebox{0.7}{%
\begin{tabular}{|c|c|c|c|c|}\hline
     & Process & Cross section order     & \multicolumn{2}{c|}{Event yield after the analysis with} \\ \cline{4-5}
 &         &           & BDTD & XGBoost \\ \hline\hline
 
\multirow{3}{*}{Background} 
        & $t\bar{t}h$                          & NLO (QCD + EW)~\cite{bkg_twiki_cs}        & $51$  & $33$\\
        & $Zh \; + $ jets                      & NNLO (QCD) + NLO (EW)~\cite{bkg_twiki_cs} & $7.7$   & $5.2$\\
        & $\ell \ell \gamma \gamma \; +$ jet   & LO                                        & $20$  & $11$\\ \cline{2-5}        
        & \multicolumn{2}{c|}{Total}   & $79$ & $49$\\ \hline
\multicolumn{2}{|c|}{Signal ($hh \to \gamma\gamma WW^*\to \gamma\gamma \ell\ell + \met$)} & NNLO~\cite{hhtwiki} & $15$ & $15$\\ \hline 
\multicolumn{2}{|c|}{\multirow{2}{*}{Significance}} & $0\%$ $\sigma_{sys\_un}$ & $1.64$ & $2.05$  \\ \cline{3-5}
\multicolumn{2}{|c|}{} & $2\%~(5\%)$ $\sigma_{sys\_un}$ & $1.61~(1.5)$ & $2.02~(1.92)$ \\ \hline
\end{tabular}}
\caption{\it The signal and background yields at the HE-LHC along with the signal significance for the $WW^{*}\gamma\gamma$ channel from the analysis using BDTD and XGBoost classifiers.}
\label{wwaa:tab1}
\end{table}
\end{center}

\subsection{The $b\bar{b}ZZ^{*}$ channel}
In this subsection, we study the potential reach of the largely unexplored $b\bar{b}ZZ^{*}$ channel. We study two different final states emerging from the leptonic decay of the $Z/Z^{*}$ bosons. In the first category, we consider the scenario where both the $Z$ bosons decay into electrons or muons (collectively represented as $l^{\prime}$) resulting in $2b4l^{\prime}$ final state while in the second category we consider the scenario where one of the $Z$ bosons decay into a pair of electrons while the other decays into muons leading to $2b2e2\mu$ final state, \textit{viz.}
$$pp\to hh\to b\bar{b}ZZ^{*}\to b\bar{b}~4l^{\prime},~l^{\prime}=e^\pm~or~\mu^\pm$$
$$pp\to hh\to b\bar{b}ZZ^{*}\to b\bar{b}~e^+e^- \mu^+\mu^-$$

The most dominant contribution to the background comes from $t\bar{t}h$. We must note that the inclusive production of $t\bar{t}h$ predominantly populates the hadronic final states over the leptonic states due to larger branching rates. Therefore, in order to generate a sufficiently large statistics of the specific final states emerging from $t\bar{t}h$ which are more relevant background sources to the $2b4l^{\prime}$ and $2b2e2\mu$ signal channels, the background generation is divided into six different categories. In the first three categories, the $h$ undergoes decay via $h \to ZZ^{*} \to 4l (l=e,\mu,\tau)$ while the top pairs are decayed leptonically, semi-leptonically and hadronically, respectively. In the fourth category, the $t\bar{t}$ pair is leptonically decayed while the $h$ undergoes decay via $h \to ZZ^{*} \to (Z/Z^{*} \to 2l)(Z^{*}/Z \to b\bar{b}/jj/\nu\bar{\nu})$. In the last two categories, $h$ undergoes decay into $WW^{*}$ and $\tau^{+}\tau^{-}$, respectively, while the top pair undergoes leptonic decay. In order to compute the overall background contribution from $t\bar{t}h$, the weighted sum of the six categories is considered. Sub-leading contribution to the background arises from $Zh$, $t\bar{t}Z$ and Higgs production in association with $b\bar{b}$ through gluon fusion ($ggF-b\bar{b}h$). Numerous fake backgrounds also contribute, $viz.$ $ggF-c\bar{c}h$, $Wh$, $Whc$ and higgs production through vector boson fusion ($VBF-hjj$). These processes contribute to the overall background yield when the $c$ $jets$ or the light $jets$ get faked as $b$ $jets$. Hard cuts have been applied at the generation level of these backgrounds and those have been listed in Appendix~\ref{sec:appendixA}. 

Before presenting the results from our collider analysis in the $2b4l^{\prime}$ and $2b2e2\mu$ channels, we briefly discuss the parton level kinematics of the final state leptons at $\sqrt{s} = 27~{\rm TeV}$. We illustrate the normalized distribution of the transverse momenta of the four leptons, $l_{1,2,3,4}$, with $l_{1}$ being the highest and $l_{4}$ being the lowest $p_{T}$ lepton. We would like to emphasize that the transverse momentum carried by the final state leptons is relatively small even at a $\sqrt{s}=27~{\rm TeV}$ proton-proton collider. The $p_{T,l_{1}}$ distribution for the signal peaks at a slightly larger value ($\sim 65~{\rm GeV}$) compared to the backgrounds, all of which peak roughly below $50~{\rm GeV}$. The peak of the $p_{T,l_{2}}$ and $p_{T,l_{3}}$ distributions for the signal and the backgrounds fall roughly within a similar range, while the peak of $p_{T,l_{4}}$ for the signal and the backgrounds is located below $\sim 5~{\rm GeV}$. These observations are important while choosing the trigger cuts at the event selection level.

In the following subsections, we study the projected potential of observing the non-resonant di-Higgs signal in the $2b4l^{\prime}$ and $2b2e2\mu$ channel at the HE-LHC, and detail the results from the BDTD optimization and the analysis using XGBoost toolkit. 

\begin{figure}[htb!]
\centering
\includegraphics[scale=0.37]{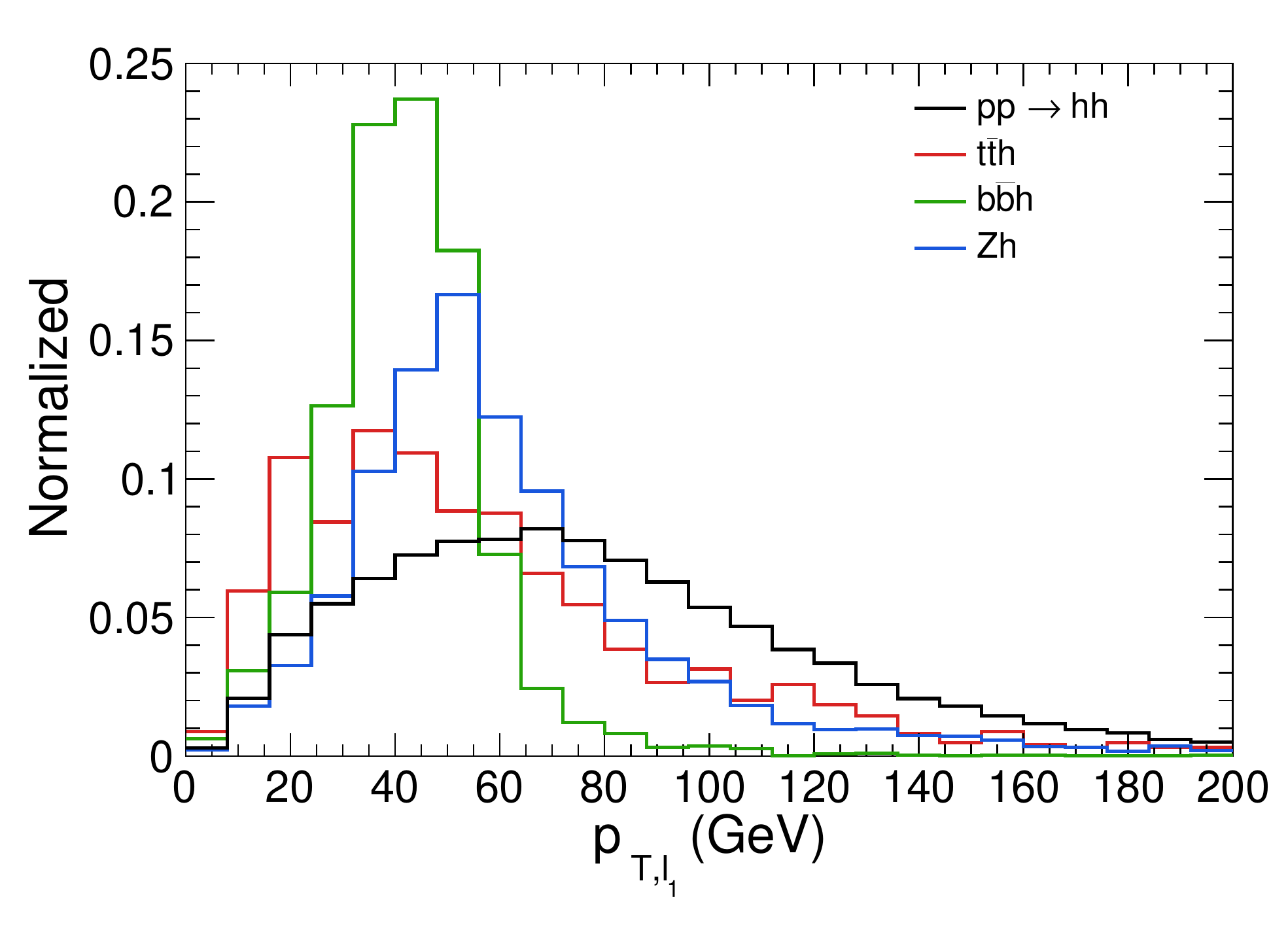}\includegraphics[scale=0.37]{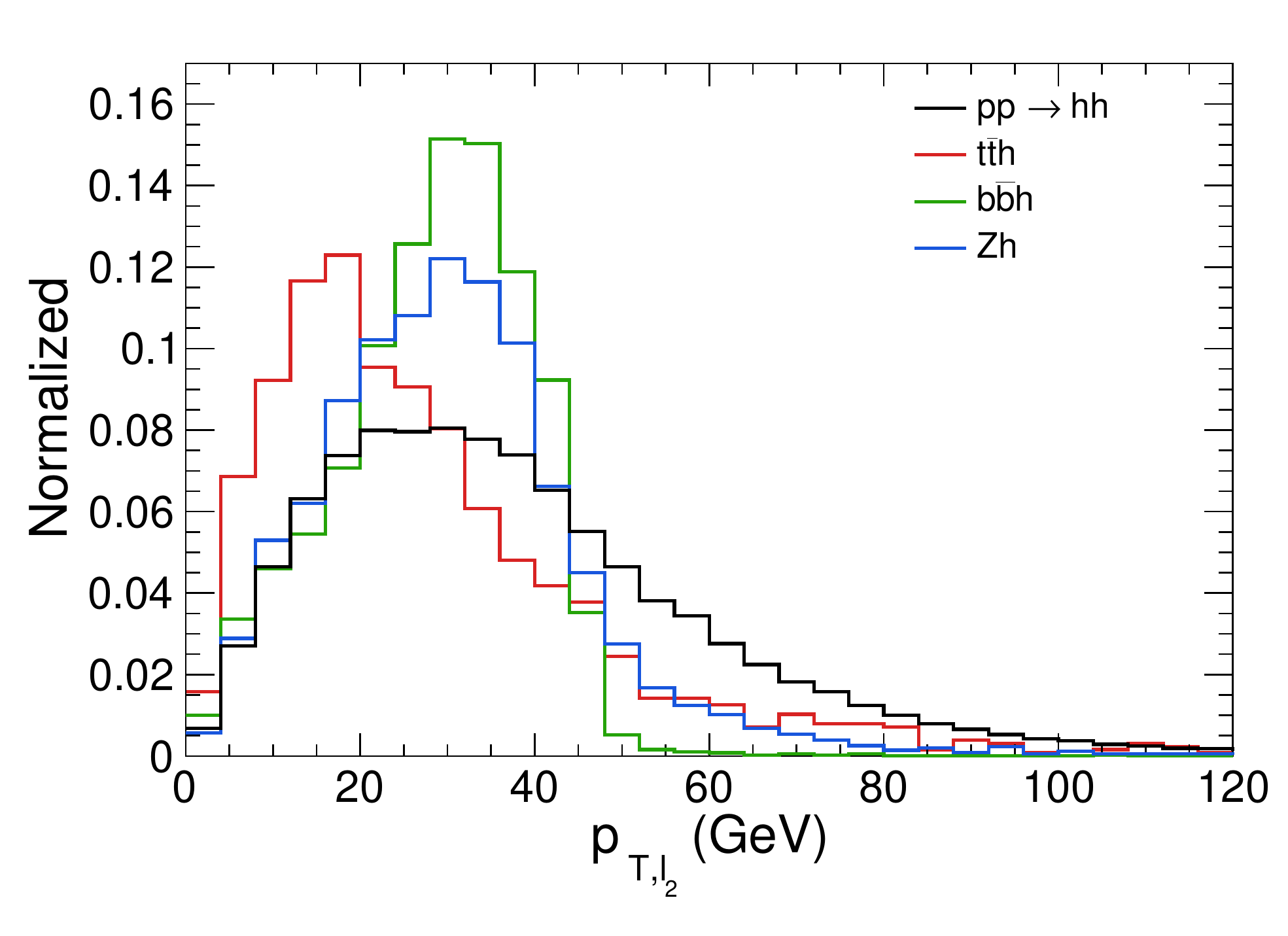}\\
\includegraphics[scale=0.37]{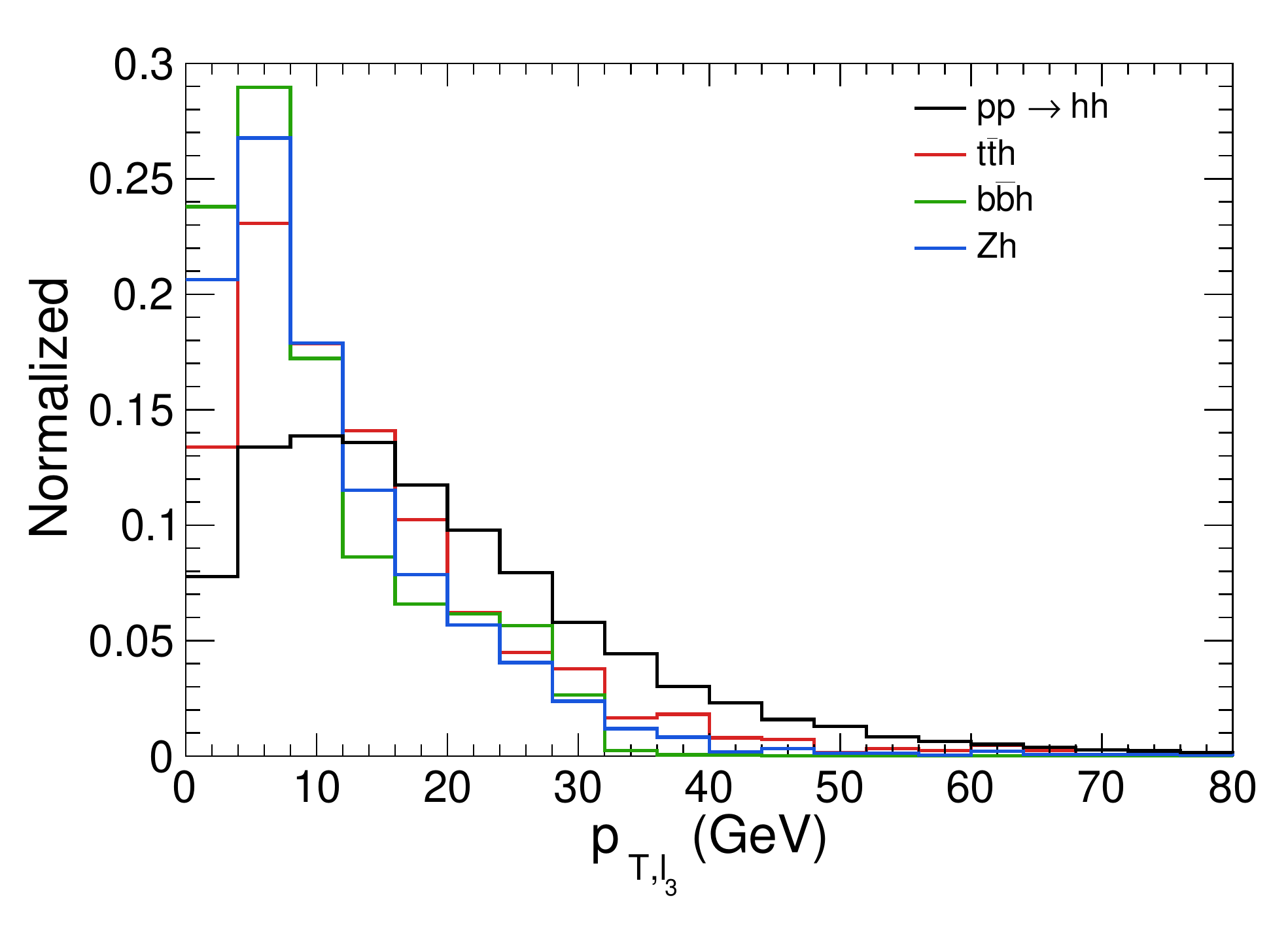}\includegraphics[scale=0.37]{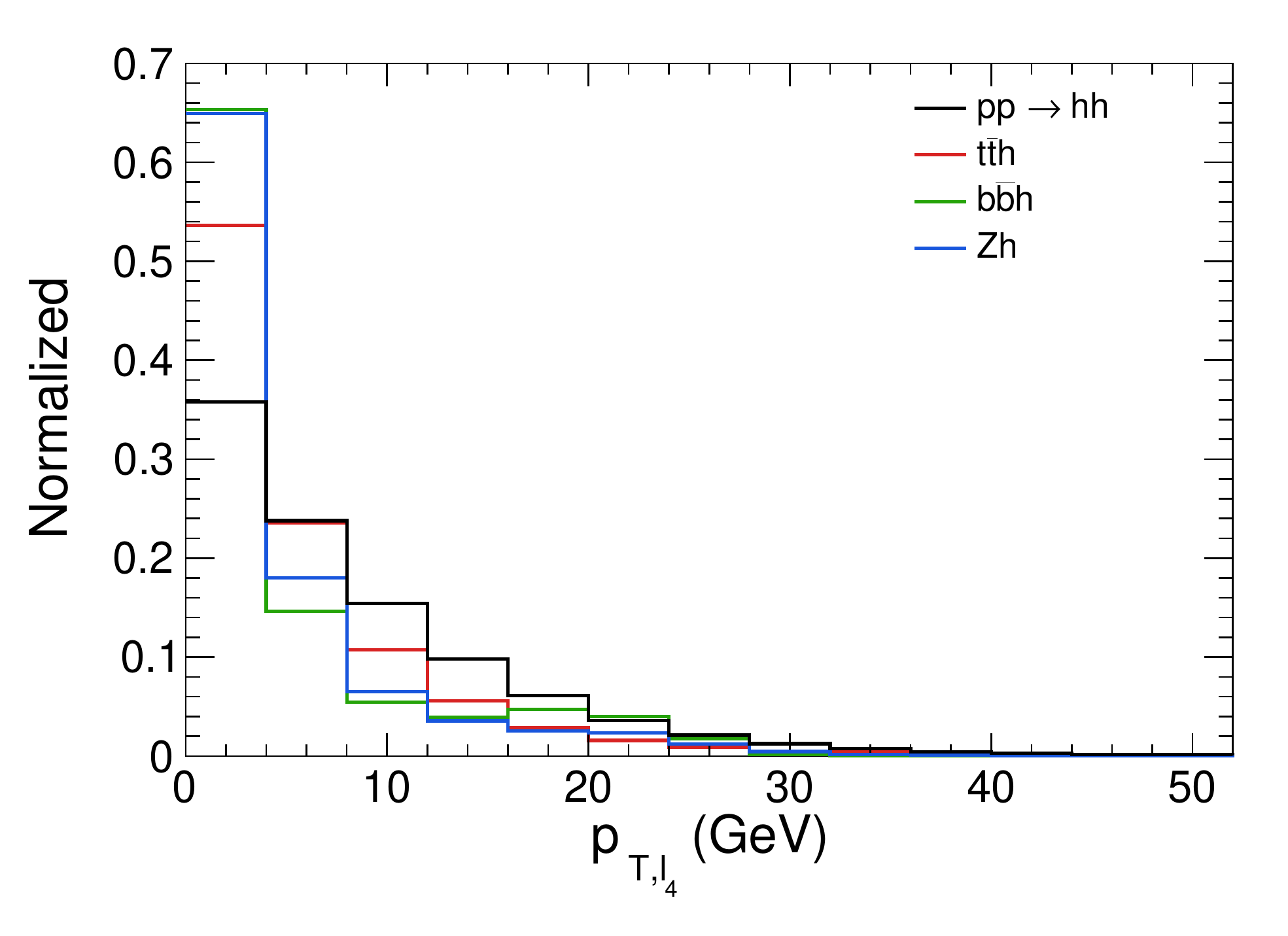}
\caption{\it Normalized distribution of the transverse momentum, $p_{T}$, of the four leptons produced via: $pp \to hh \to b\bar{b}ZZ^{*} \to b\bar{b}(ZZ^{*} \to 4l)$, at the parton level.}
\label{fig:ptbl}
\end{figure}

\subsubsection{The $2b4l^{\prime}$ channel}
\label{bbZZ:2b4l_sec}

The selected events are required to have exactly two $b$ tagged $jets$ with $p_{T} > 20~{\rm GeV}$ and $|\eta| < 4.0$, and four isolated $l^{\prime}$'s carrying $p_{T} > 5 ~{\rm GeV}$ within $|\eta| < 4.0$. The invariant mass of the four leptons is also restricted between 120~GeV and 130~GeV. In addition, $\Delta R_{bb}$ is required to be greater than $0.2$ and $m_{bb}$ must be $> 50~{\rm GeV}$. We must note that it is possible to construct four different combinations of same flavor opposite sign (SFOS) $l^{\prime}$ pairs. Thus, the $Z$ and $Z^{*}$ can be reconstructed in two different ways. We represent these two possibilities with superscript $1$ and $2$. If we represent the two $Z$ bosons as $Z_{1}$ and $Z_{2}$, then the two different SFOS pairs would result in either ($Z_{1}^{1}, Z_{2}^{1}$) or ($Z_{1}^{2}, Z_{2}^{2}$). We utilize kinematic variables constructed from both, $Z^{1}$ and $Z^{2}$, in performing the collider analysis. First, we perform a multivariate analysis with BDTD algorithm and subsequently, we also perform signal optimization with XGBoost with the aim to maximize the signal significance. The following kinematic variables are used as inputs to the analysis techniques:
      
\begin{figure}
\centering
\includegraphics[scale=0.37]{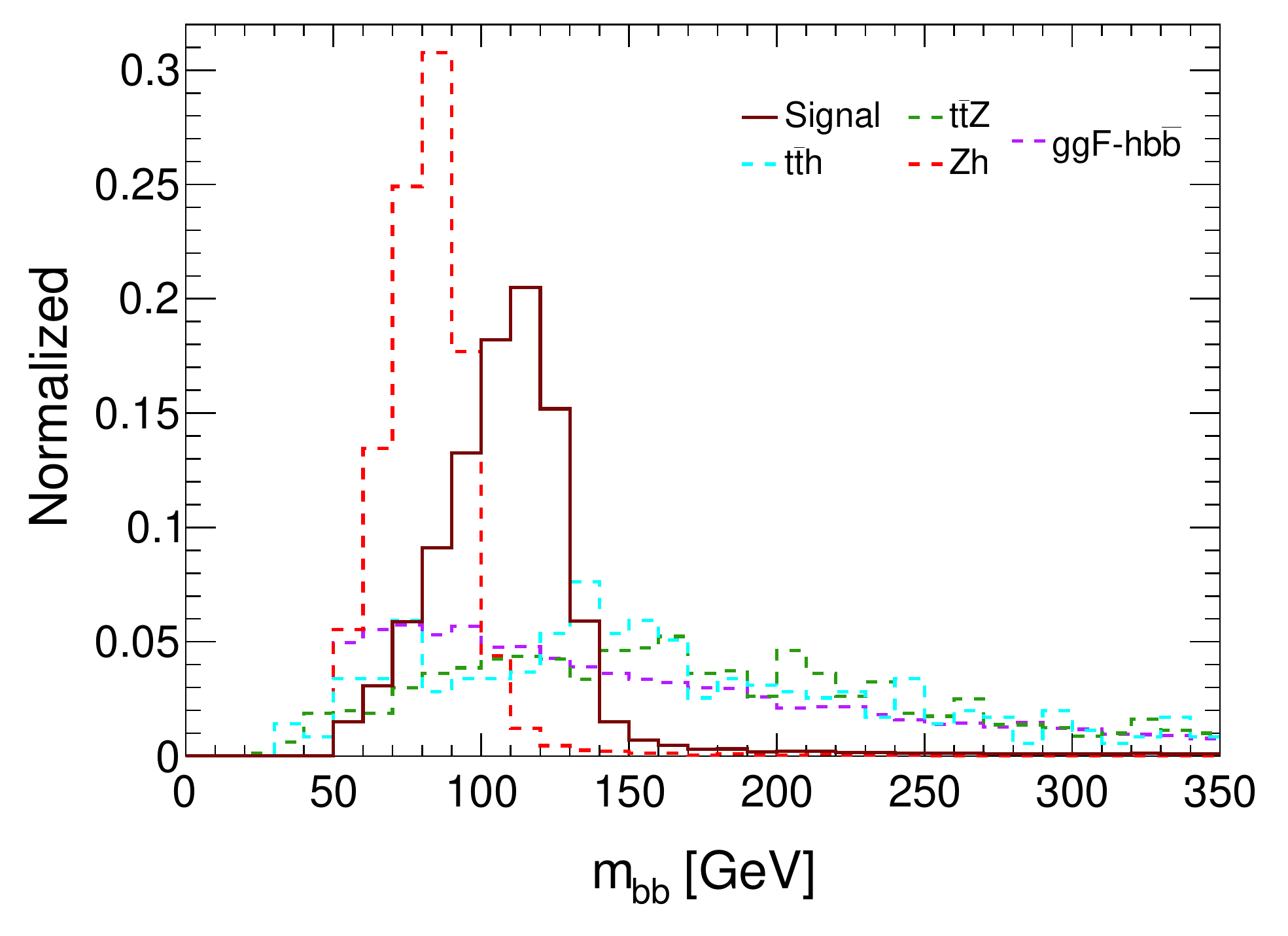}
\includegraphics[scale=0.37]{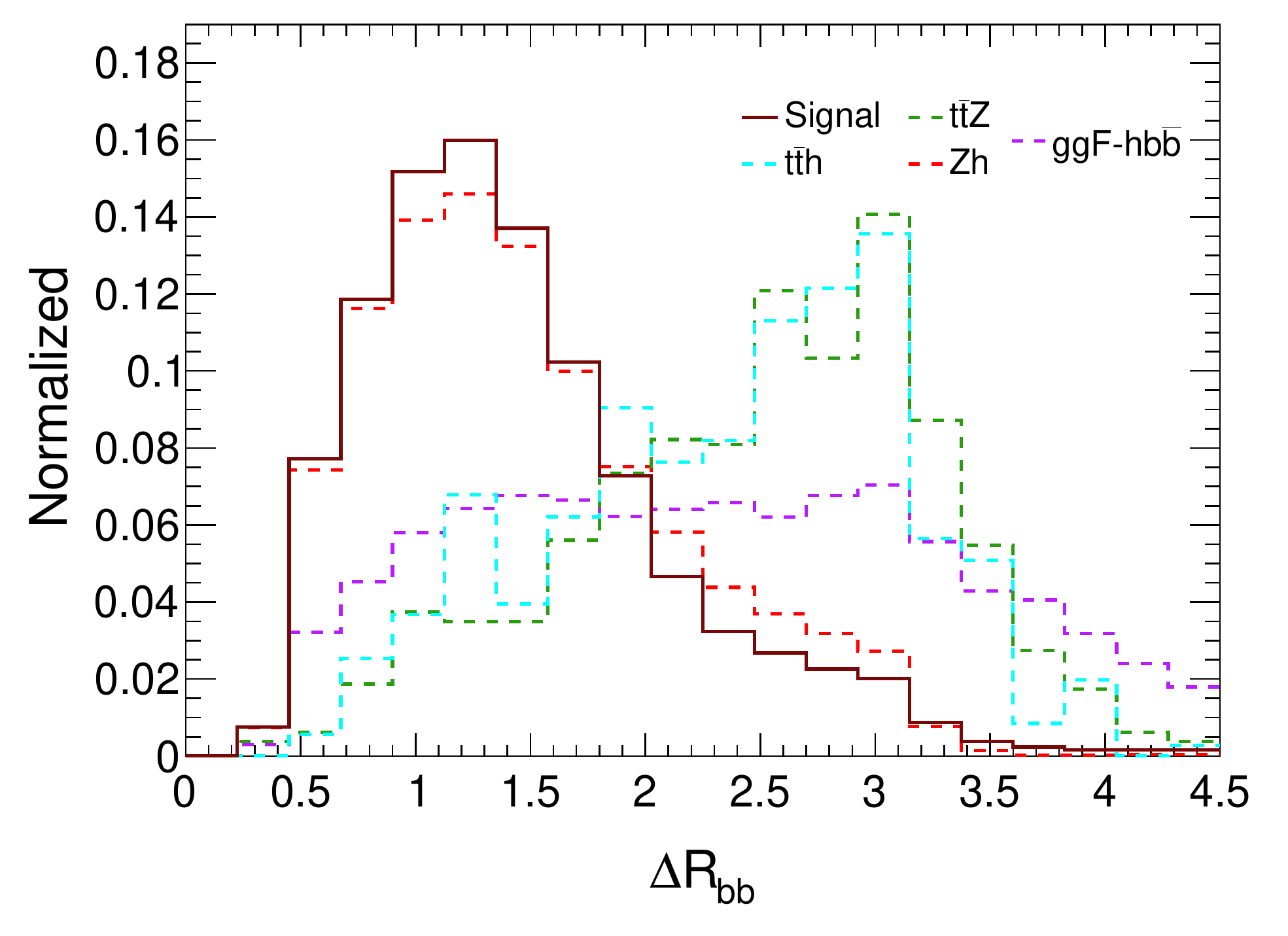}\\
\includegraphics[scale=0.37]{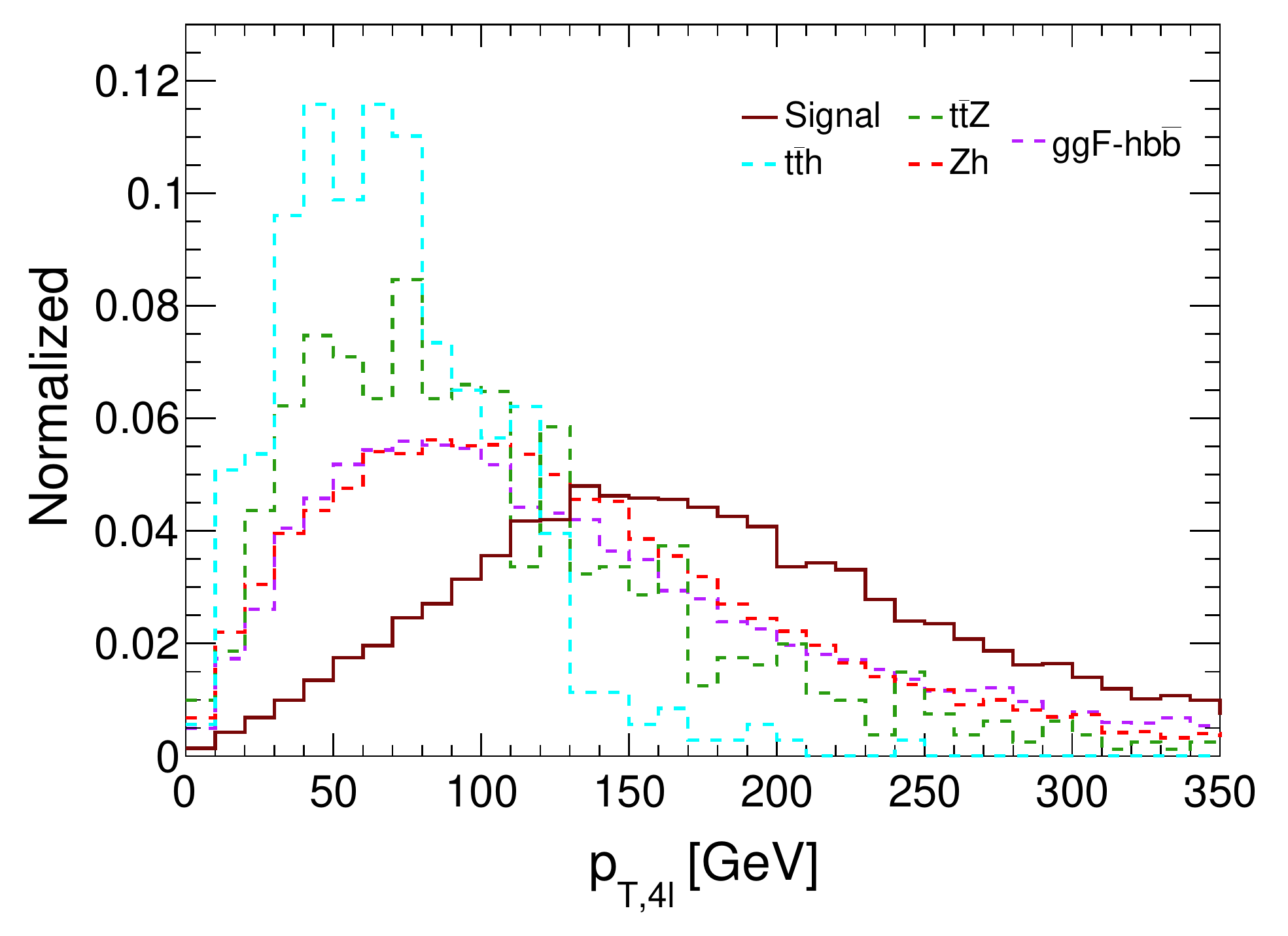}
\includegraphics[scale=0.37]{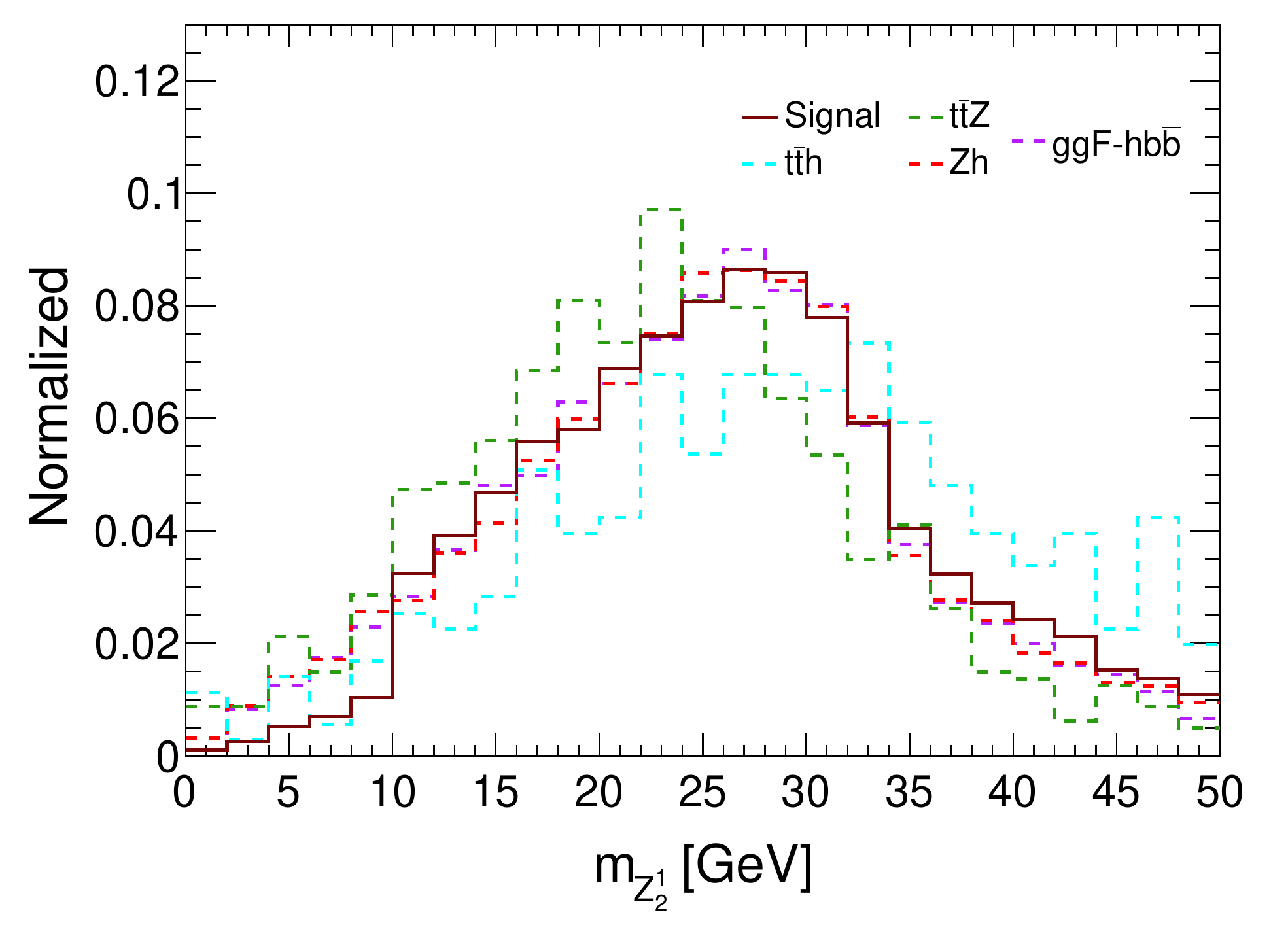}
\caption{\it Normalized distributions of $m_{bb}$, $\Delta R_{bb}$, $p_{T,4\ell}$ and $m_{Z_2}^{1}$ for the $2b4l^{\prime}$ signal and the dominant backgrounds after the acceptance cuts and the generation level cuts.}
\label{bbzz:fig1}
\end{figure}
      
\begin{equation}
\begin{split}
p_{T,bb},~\Delta R_{bb},~m_{bb},~p_{T,4\ell},~m_{Z_i}^{1,2},~\Delta R_{ZZ}^{1,2},~m_{T2},~m_{\textrm{eff}},~\Delta R_{b_1\ell_{1}},~m_{hh},~\Delta R_{hh},~\met \nonumber
\end{split}
\end{equation}

Here, $m_{Z_i}^{1}$ and $m_{Z_i}^{2}$, with $i=1,2$, corresponds to the invariant mass of the four reconstructed $Z$ bosons as discussed previously, $\Delta R^{1}_{ZZ}$ represents the $\Delta R$ value between $Z_{1}^{1}$ and $Z_{2}^{1}$, and $\Delta R^{2}_{ZZ}$ represents the $\Delta R$ value between $Z_{1}^{2}$ and $Z_{2}^{2}$. The other variables have their usual meanings. $m_{bb}$, $\Delta R_{bb}$, $p_{T,4l}$ and $m_{Z_{2}^{1}}$ are found to be the variables which are most efficient in discriminating the signal and the background through BDTD optimization. We illustrate their normalized distribution in Fig.~\ref{bbzz:fig1}. 
\begin{center}
\begin{table}[htb!]
\centering
\scalebox{0.7}{%
\begin{tabular}{|c|c|c|c|c|}\hline
       & Process & Cross section order     & \multicolumn{2}{c|}{Event yield after the analysis with} \\ \cline{4-5}
 &         &           & BDTD & XGBoost (probability cut $>95\%$) \\ \hline\hline

\multirow{10}{*}{Background} 
 & $t\bar{t}h$              		& NLO~\cite{bkg_twiki_cs}                     & $16.5$ & $32$\\  
 & $t\bar{t}Z$             	    	& NLO~\cite{Lazopoulos:2008de}                & $1.38$ & $4.9$\\  
 & $Zh$                            & NNLO (QCD) + NLO (EW)~\cite{bkg_twiki_cs}   & $7$ & $8.5$\\  
 & $Wh$                            & NNLO (QCD) + NLO (EW)~\cite{bkg_twiki_cs}   & $3\times 10^{-3}$ & $3\times 10^{-3}$\\
 & $Whc$                           & LO                                          & $6\times 10^{-4}$ & $9\times 10^{-4}$\\ 
 & ggF-$hb\bar{b}$                 & LO                                          & $2.5$ & $3.5$\\ 
 & ggF-$hc\bar{c}$                 & LO                                          & $6\times 10^{-3}$ & $8\times 10^{-3}$\\ 
 & VBF-$hjj$                       & NNLO (QCD) + NLO (EW)~\cite{bkg_twiki_cs}   & $2.7\times 10^{-4}$ & $4\times 10^{-4}$\\ \cline{2-5}  
 & \multicolumn{2}{c|}{Total}                                                    & $\sim 27$ & $49$\\ \hline
\multicolumn{2}{|c|}{Signal ($hh \to b\bar{b}ZZ^*\to 2b4lep$)} & NNLO~\cite{hhtwiki}  & $5$ & $6.7$\\\hline 
\multicolumn{2}{|c|}{\multirow{2}{*}{Significance}} & $0\%$ $\sigma_{sys\_un}$ &  $0.93$ & $0.94$\\ \cline{3-5}
\multicolumn{2}{|c|}{} & $2\%~(5\%)$ $\sigma_{sys\_un}$ & $0.93~(0.9)$ & $0.93~(0.88)$  \\ \hline
\end{tabular}}
\caption{\it The signal and background yields at the HE-LHC along with the signal significance for the $b\bar{b}ZZ^{*} \to 2b4l^{\prime}$ channel from the analysis using the BDTD and the XGBoost classifiers.}
\label{bbzz:tab1}
\end{table}
\end{center}

\begin{figure}
\centering
\includegraphics[scale=0.42]{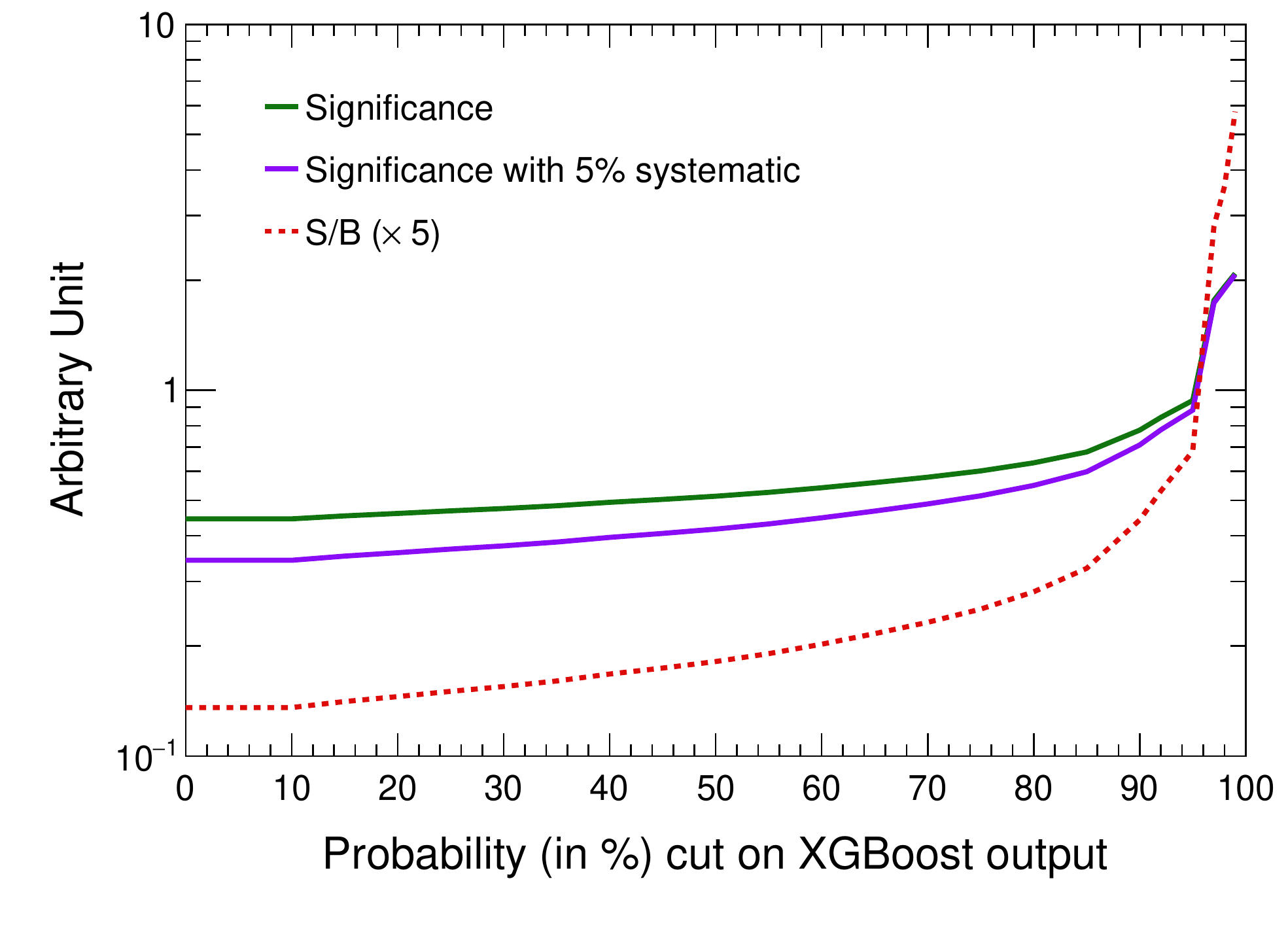}
\caption{\it The variation of significance~(with~($5\%$) and without systematic uncertainty) and $S/B$ is shown as a function of the probability cut on the XGBoost output for the $b\bar{b}ZZ^{*}\to b\bar{b}~4l^{\prime}$ channel.}
\label{bbzz:fig2}
\end{figure}

The signal and background yields obtained from the BDTD analysis and the XGBoost analysis are presented in Table~\ref{bbzz:tab1}. The corresponding signal significance is also listed. We observe a background yield of $27$ and a signal yield of $5$ from the BDTD analysis resulting in a signal significance of $0.93$. We must also note that this channel registers an impressive S/B value of $0.19$. The analysis using XGBoost gives a background yield of $49$ and a signal yield of $6.7$ leading to a signal significance of $0.94$. Here, we have imposed the probability cut on the XGBoost output at $95\%$. Imposing the probability cut at $97\%$ leads to an improved signal significance of $1.8$. We show the variation of significance and $S/B$ as a function of the probability cut on the XGBoost output in Fig.~\ref{bbzz:fig2} following the color code of Fig.~\ref{bbaa:fig3}.

\subsubsection{The $2b2e2\mu$ channel}

The event selection criteria requires the presence of exactly two $b$ $jets$ with $p_{T} > 20~{\rm GeV}$ and $|\eta| < 4.0$, two isolated electrons and two isolated muons with $p_{} > 5~{\rm GeV}$ and $|\eta| < 4.0$. Similar to the selection cuts prescribed in Sec.~\ref{bbZZ:2b4l_sec}, we apply $\Delta R_{bb} > 0.2$ and $m_{bb} > 50~{\rm GeV}$ . Furthermore, we also demand that the invariant mass of the $2e2\mu$ system ($m_{2e2\mu}$) must lie within $120~{\rm GeV}$ and $130~{\rm GeV}$. In the present scenario, the two electrons and the muons are combined, respectively, to reconstruct the $Z$ bosons. Among the two $Z$ bosons, the one with invariant mass closest to $m_{Z}$ will be referred to as $Z_{1}$ while the other $Z$ boson will be referred as $Z_{2}$.

\begin{figure}
\centering
\includegraphics[scale=0.37]{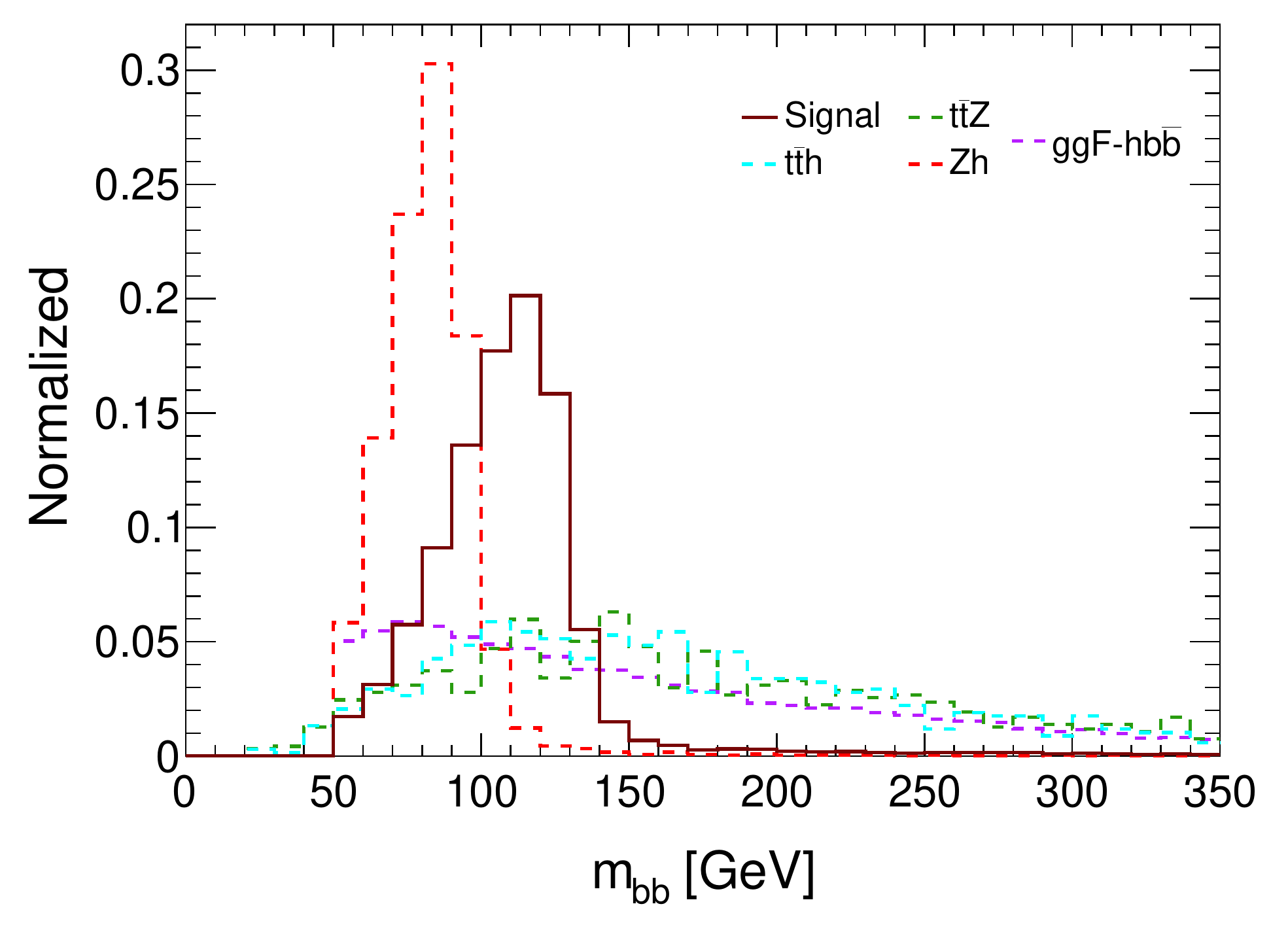}
\includegraphics[scale=0.37]{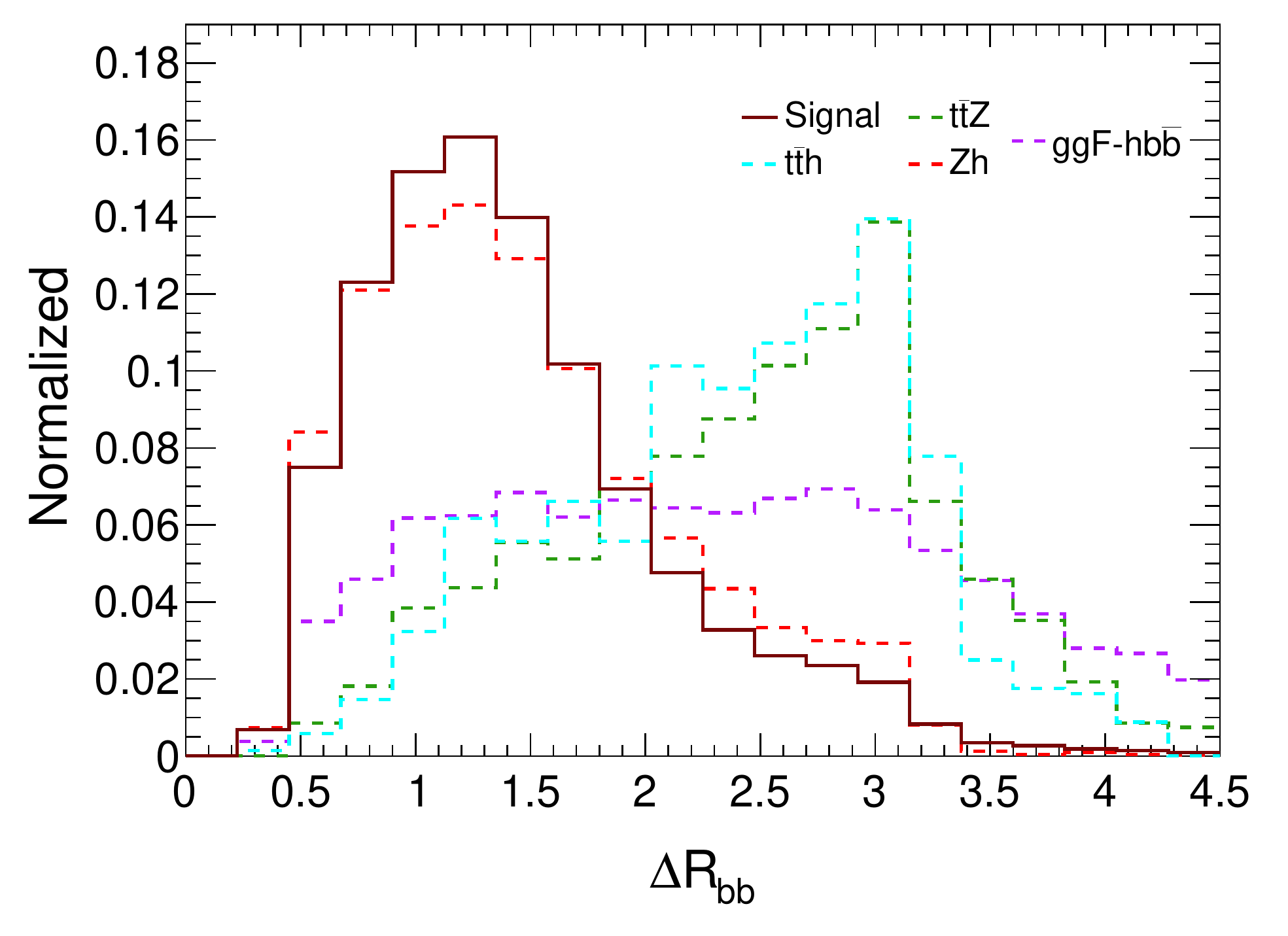}\\
\includegraphics[scale=0.37]{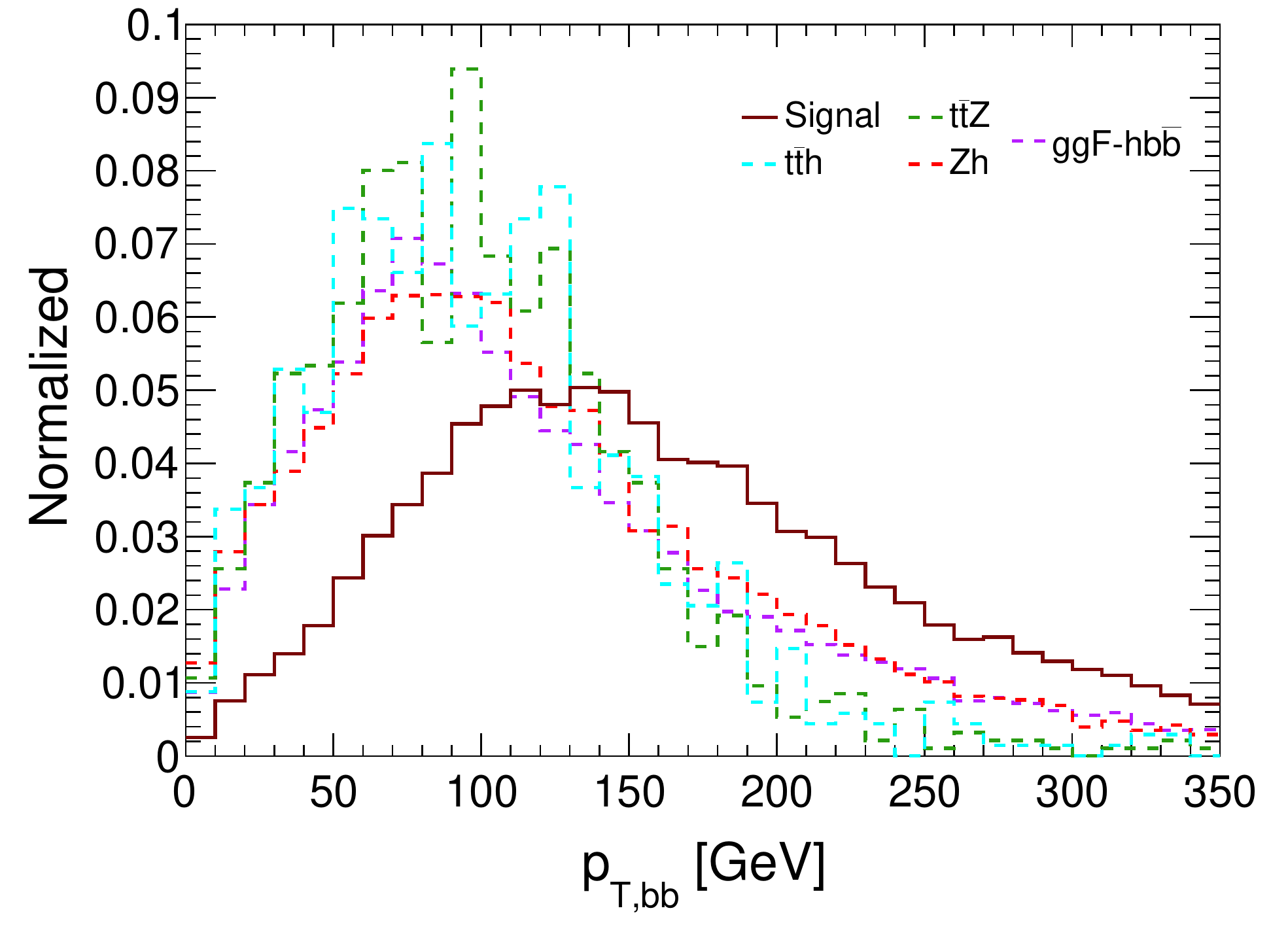}
\includegraphics[scale=0.37]{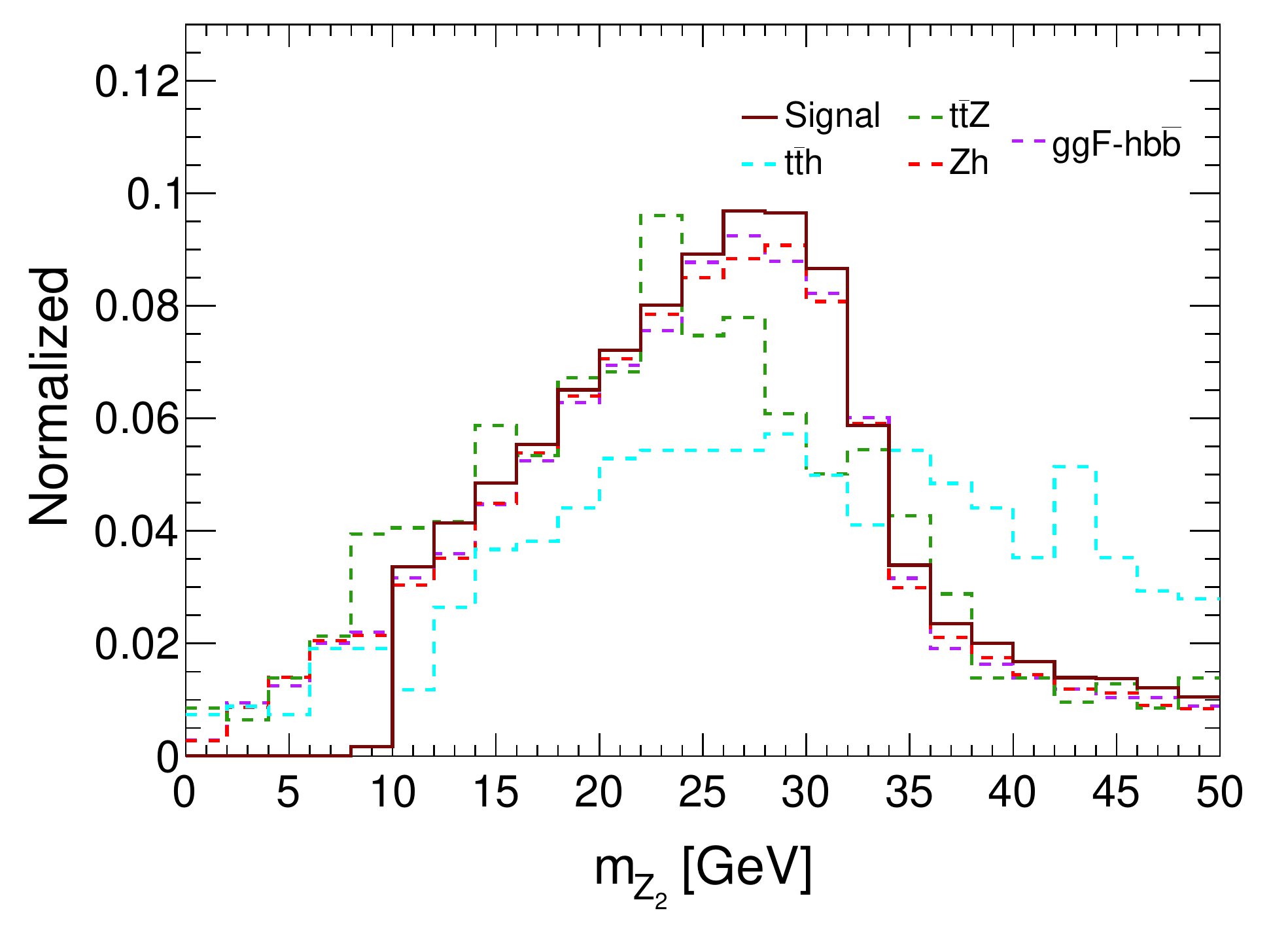}
\caption{\it Normalized distributions of $m_{bb}$, $\Delta R_{bb}$, $p_{T,bb}$ and $m_{Z_2}$ for the $2b 2e 2\mu$ signal and the dominant backgrounds after the acceptance cuts and the generation level cuts.}
\label{bbzz:fig3}
\end{figure}

The signal and background events which pass the selection cuts are subjected to multivariate analysis using the BDTD algorithm and the XGBoost package. In this regard, we consider the following $13$ kinematic variables:
\begin{equation}
\begin{split}
p_{T,bb},~\Delta R_{bb},~m_{bb},~p_{T,2e2\mu},~m_{Z_i},~\Delta R_{ZZ},~m_{T2},~m_{\textrm{eff}},~\Delta R_{b_1\ell_{1}},~m_{hh},~\Delta R_{hh},~\met \nonumber
\end{split}
\end{equation}

\begin{figure}
\centering
\includegraphics[scale=0.42]{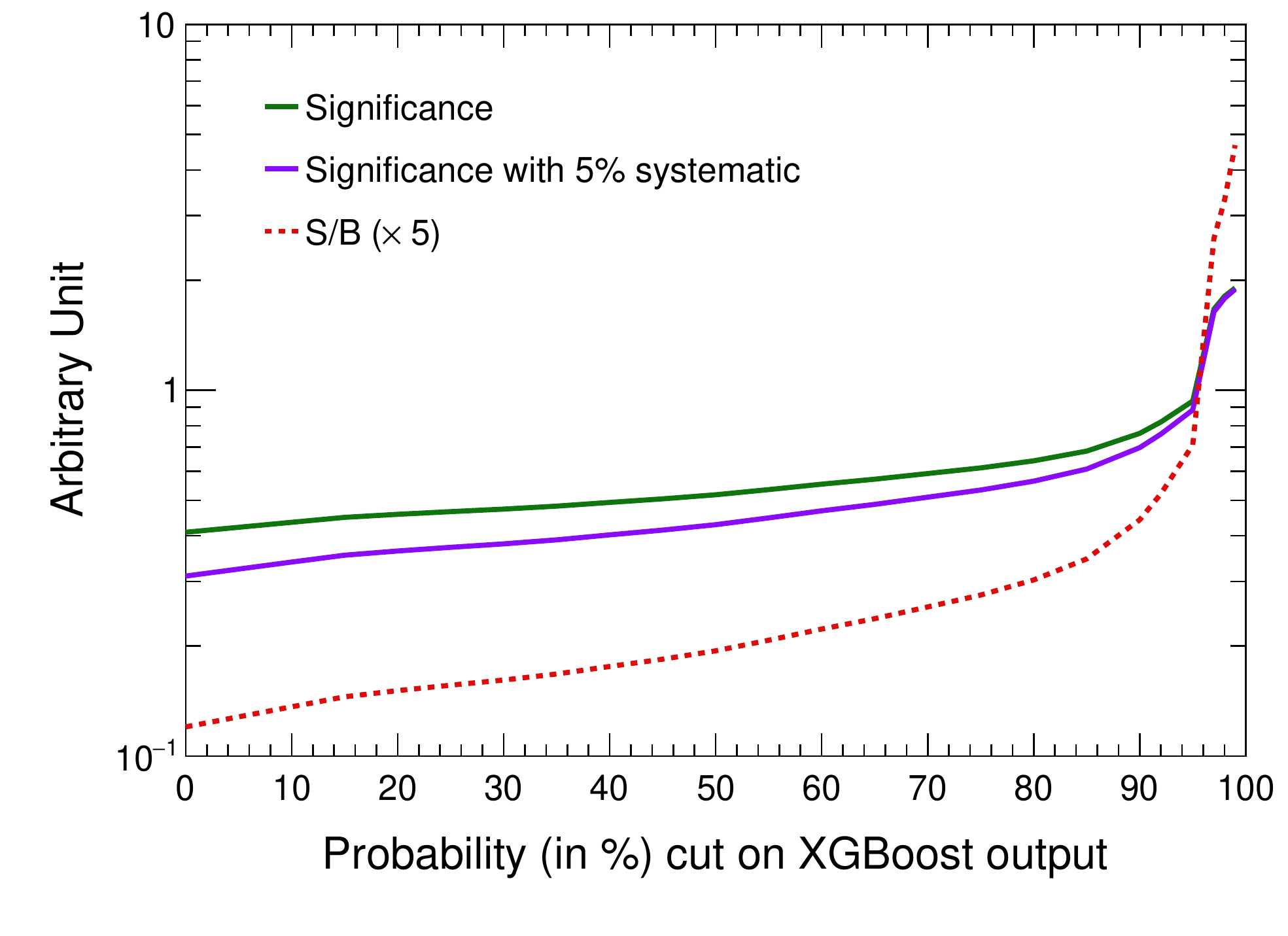}
\caption{\it The variation of significance~(with~($5\%$) and without systematic uncertainty) and $S/B$ is shown as a function of the probability cut on the XGBoost output for the $b\bar{b}ZZ^{*}\to b\bar{b}~2e2\mu$ channel.}
\label{bbzz:fig4}
\end{figure}

Here, $p_{T,2e2\mu}$ is the transverse momentum of the $2e2\mu$ system while the other variables have their usual meaning. The kinematic variables which discriminate between the signal and the backgrounds with maximal efficiency are: $m_{bb}$, $\Delta R_{bb}$, $p_{T,bb}$ and $m_{Z_{2}}$. The normalized distribution of these four variables is illustrated in Fig.~\ref{bbzz:fig3}. The multivariate analysis using BDTD algorithm gives a signal significance of $0.86$ and a $S/B$ value of $0.15$. The corresponding signal and background yields are listed in Table~\ref{bbzz:tab2}. Among the two different final states of $hh \to b\bar{b}ZZ^{*}$ considered in this work, the $2b4l^{\prime}$ channel exhibits a slightly larger sensitivity compared to the $2b2e2\mu$ search channel. The signal and background yields from the XGBoost analysis are also listed in the same table. We obtain a slightly higher signal significance (0.93) from the XGBoost analysis. Similar to the previous subsection, the probability cut on the XGBoost output has been applied at $95\%$. For the sake of completeness, we also show the variation of significance and $S/B$ as a function of the probability cut in Fig.~\ref{bbzz:fig4}. The color code of Fig.~\ref{bbaa:fig3} has been followed here.

\begin{center}
\begin{table}[htb!]
\centering
\scalebox{0.7}{%
\begin{tabular}{|c|c|c|c|c|}\hline
       & Process & Cross section order     & \multicolumn{2}{c|}{Event yield after the analysis with} \\ \cline{4-5}
 &         &           & BDTD & XGBoost (probability cut $>95\%$)\\ \hline\hline

\multirow{10}{*}{Background} 
 & $t\bar{t}h$              		& NLO~\cite{bkg_twiki_cs}                     & $23$ & $28$\\  
 & $t\bar{t}Z$             	    	& NLO~\cite{Lazopoulos:2008de}                & $3.6$ & $6.4$\\  
 & $Zh$                            & NNLO (QCD) + NLO (EW)~\cite{bkg_twiki_cs}   & $6.9$ & $7.6$\\  
 & $Wh$                            & NNLO (QCD) + NLO (EW)~\cite{bkg_twiki_cs}   & $3\times 10^{-3}$ & $3\times 10^{-3}$\\
 & $Whc$                           & LO                                          & $7\times 10^{-4}$ & $9\times 10^{-4}$\\ 
 & ggF-$hb\bar{b}$                 & LO                                          & $2.8$ & $3.5$\\ 
 & ggF-$hc\bar{c}$                 & LO                                          & $7\times 10^{-3}$ & $8\times 10^{-3}$\\ 
 & VBF-$hjj$                       & NNLO (QCD) + NLO (EW)~\cite{bkg_twiki_cs}   & $3.5\times 10^{-4}$ & $4\times 10^{-4}$\\ \cline{2-5}  
 & \multicolumn{2}{c|}{Total}                                                    & $36$ & $45$\\ \hline
\multicolumn{2}{|c|}{Signal ($hh \to b\bar{b}ZZ^*\to 2b 2e 2\mu$)} & NNLO~\cite{hhtwiki}  & $5.3$ & $6.4$\\\hline 
\multicolumn{2}{|c|}{\multirow{2}{*}{Significance}} & $0\%$ $\sigma_{sys\_un}$ & $0.86$ & $0.93$ \\ \cline{3-5}
\multicolumn{2}{|c|}{} & $2\%~(5\%)$ $\sigma_{sys\_un}$ & $0.86~(0.82)$ & $0.92~(0.88)$  \\ \hline
\end{tabular}}
\caption{\it The signal and background yields at the HE-LHC along with the signal significance for the $b\bar{b}ZZ^{*} \to 2b2e2\mu$ channel from the analysis using the BDTD and the XGBoost classifiers.}
\label{bbzz:tab2}
\end{table}
\end{center}

\subsection{The $b\bar{b}\mu\mu$ channel}
\label{sec:bbmumu}

The di-Higgs final state often ignored in terms of clarity is the $b\bar{b}\mu\mu$ channel. The production rate of this channel is even lower than the $b\bar{b}\gamma\gamma$ final state due to the smaller branching ratio of $h \to \mu^{+}\mu^{-} \simeq 2.18 \times 10^{-4}$. The dominant background sources are: $t\bar{t}$ and the QCD-QED $b\bar{b}\mu\mu$. Sub-dominant contribution to the background comes from $Zh$, $t\bar{t}h$ and $b\bar{b}h$. Additionally, the fake backgrounds, $viz.$ $c\bar{c}\mu\mu$ and $jj\mu\mu$ also contribute when the $c$ $jets$ or the light $jets$ get mistagged as $b$ $jets$, respectively. We must mention that the $t\bar{t}$ background has been generated by decaying both the $W$ bosons into the $\mu\nu_{\mu}$ pair. Contributions can also arise when the $W$ bosons decay into the $\tau\nu_{\tau}$, however, we do not consider this contribution in our analysis due to its negligible rate. We also impose hard cuts at the generation level of the backgrounds and those have been listed in Appendix~\ref{sec:appendixA}.

Event selection demands exactly two $b$ $jets$ with $p_{T} > 30~{\rm GeV}$ and two isolated muons with $p_{T} > 20~{\rm GeV}$. The final state $b$ $jets$ and the leptons are required to lie within $|\eta| < 4.0$. For consistency, we impose the generation level cuts: $\Delta R_{b_{i}\mu_{j}} > 0.2$~($i,j=1,2$), $m_{\mu\mu} > 100~{\rm GeV}$ and $m_{bb} > 50~{\rm GeV}$, at the selection level. The selected signal and background events are then subjected to a multivariate analysis using the BDTD algorithm and the XGBoost toolkit. The multivariate analysis is performed using the following kinematic variables:

\begin{figure}[!htb]
\centering
\includegraphics[scale=0.37]{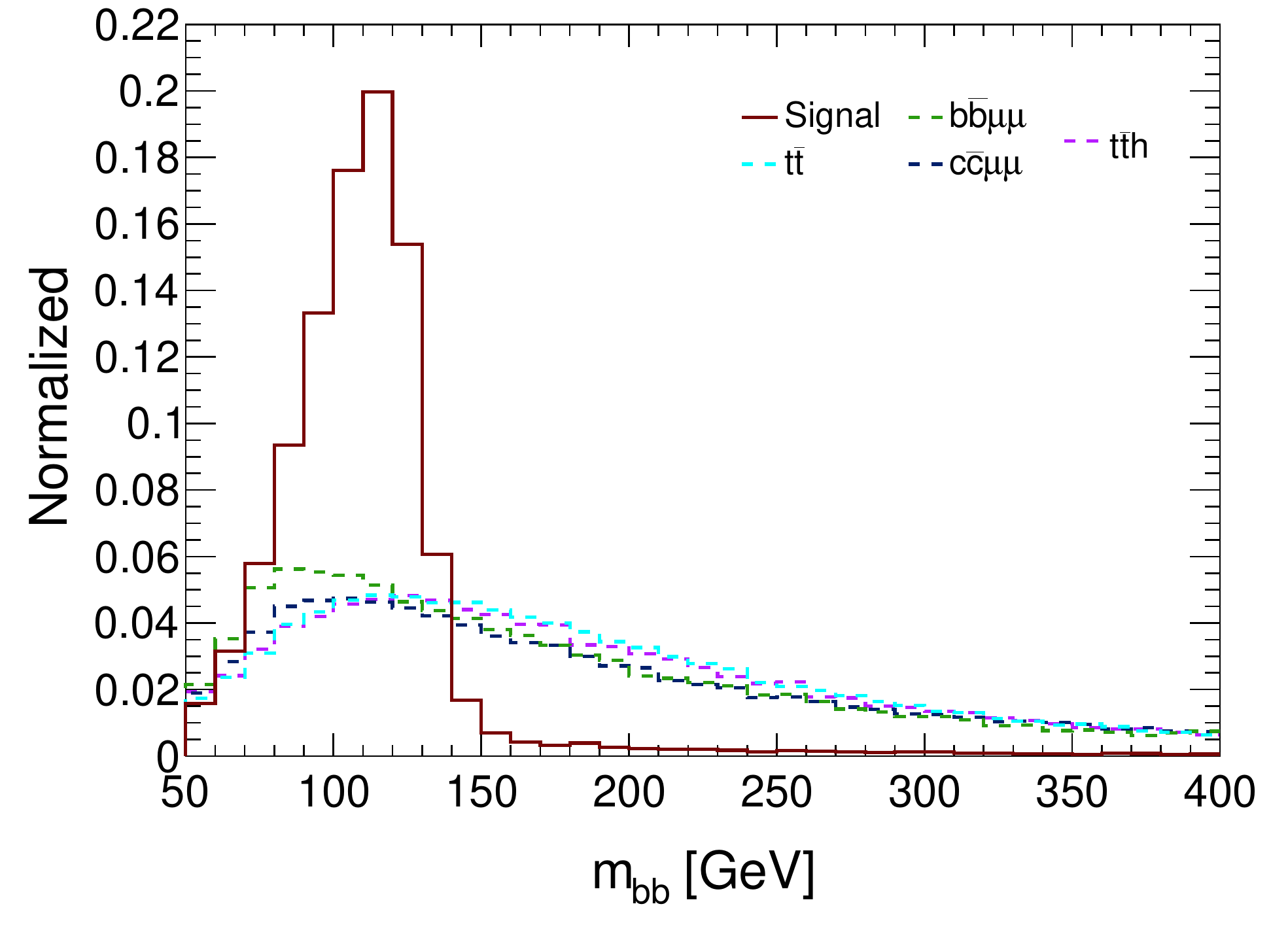}
\includegraphics[scale=0.37]{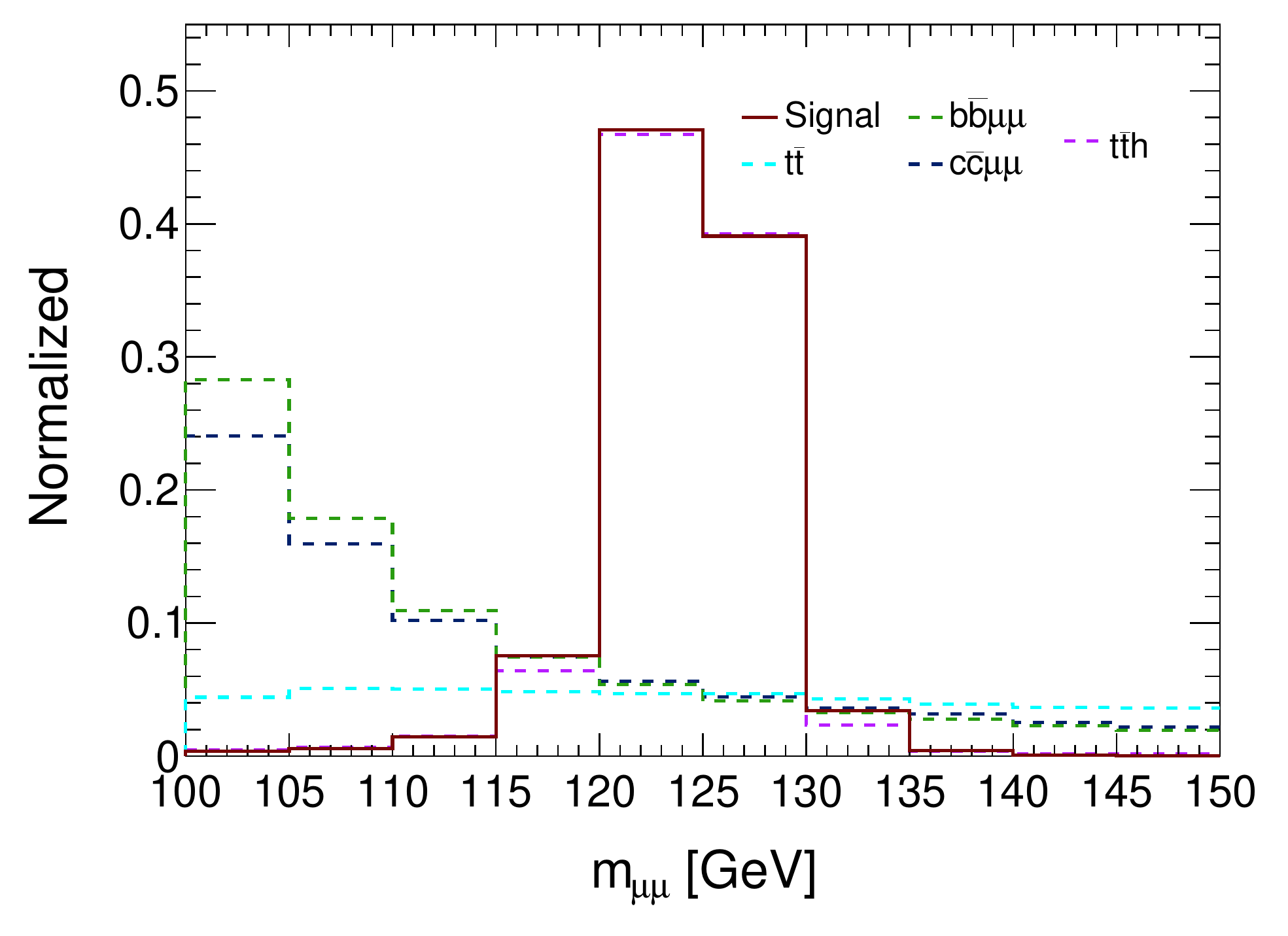}\\
\includegraphics[scale=0.37]{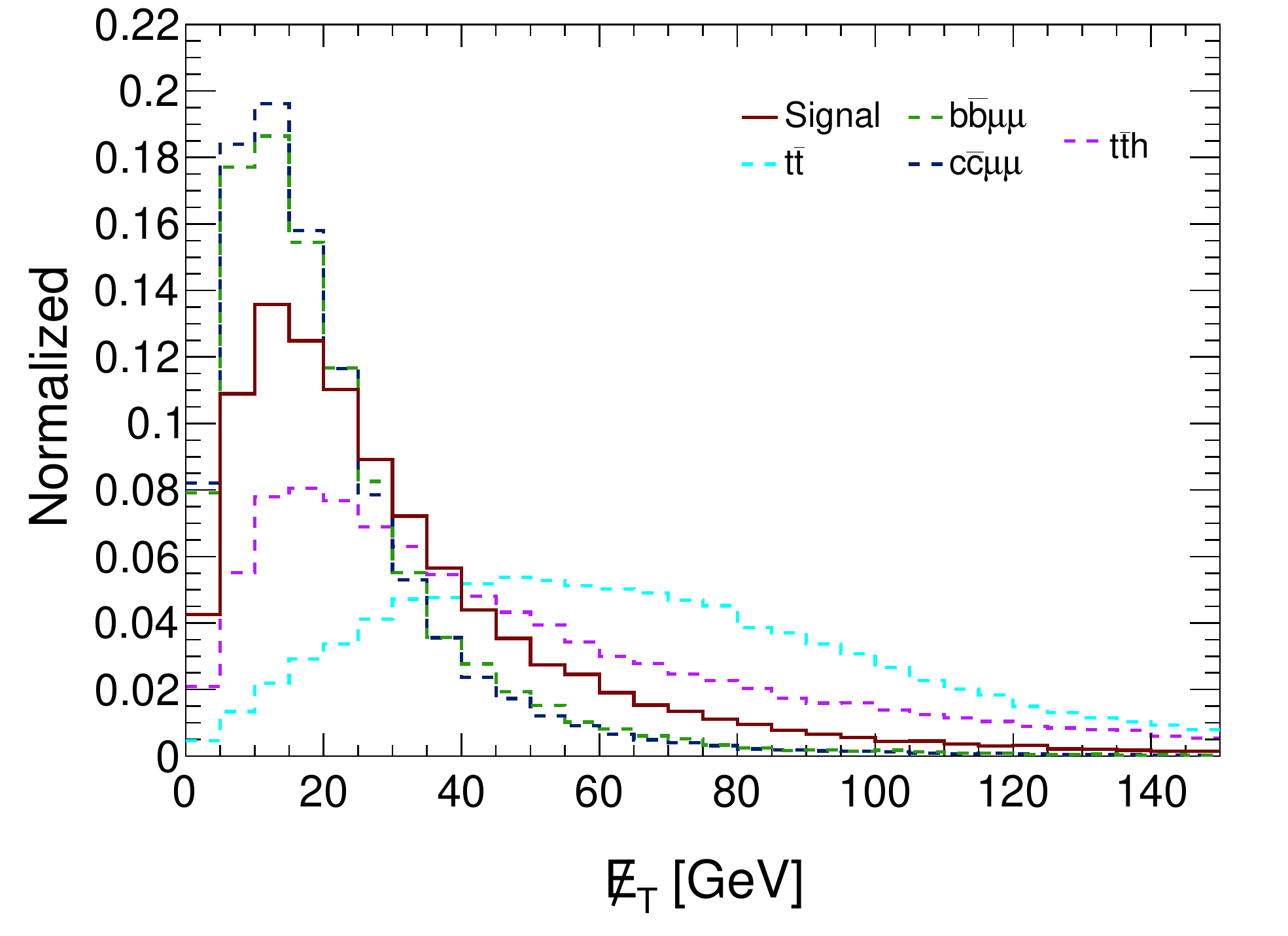}
\includegraphics[scale=0.37]{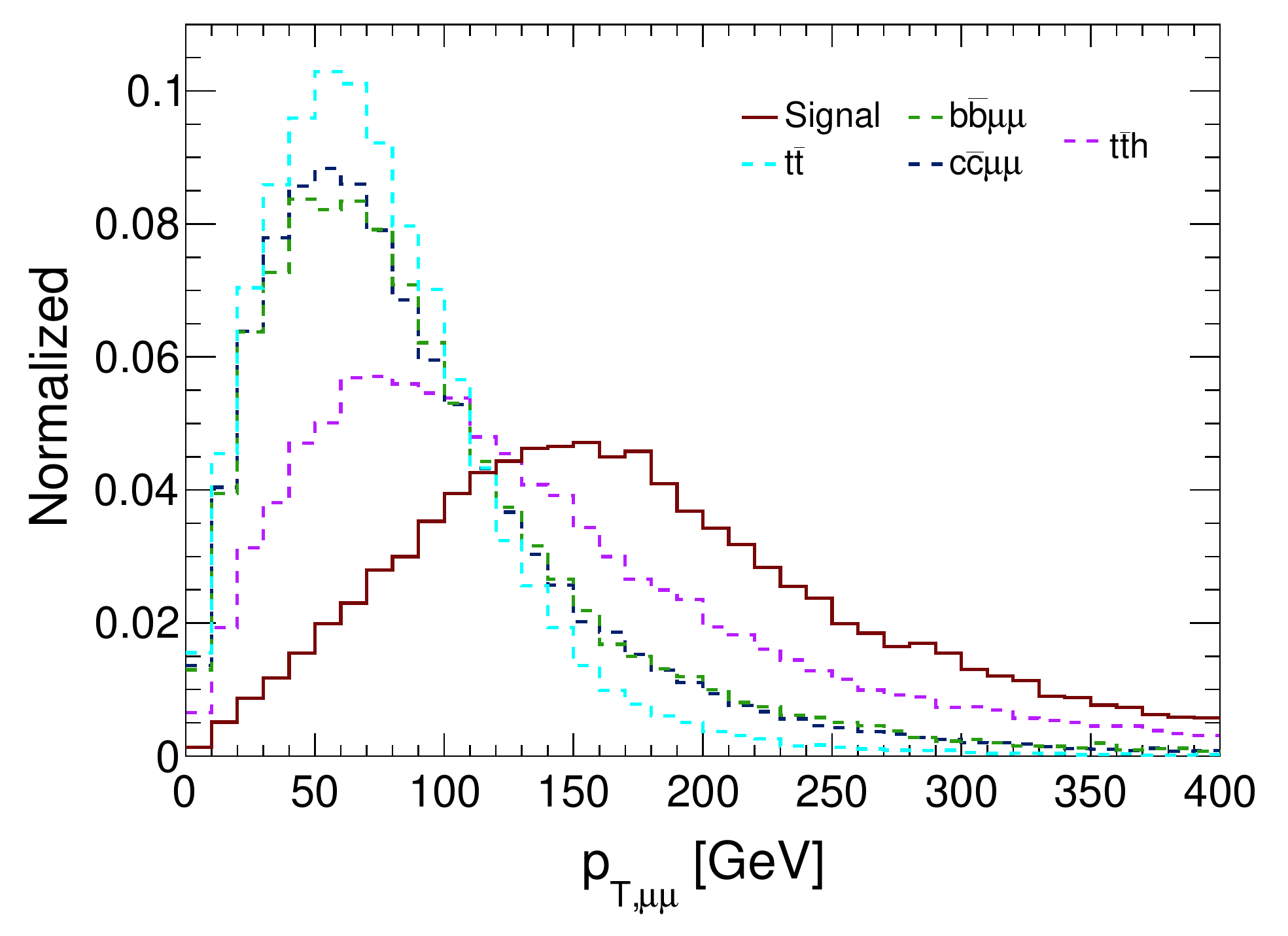}
\caption{\it Normalized distributions of $m_{bb}$, $m_{\mu\mu}$, $\met$ and $p_{T,\mu\mu}$ for the $b\bar{b}\mu^{+}\mu^{-}$ signal and the dominant backgrounds after the acceptance cuts and the generation level cuts.}
\label{bb2mu:fig1}
\end{figure}

\begin{equation}
p_{T,\mu\mu},~\Delta R_{\mu\mu},~m_{\mu\mu},~p_{T,bb},~\Delta R_{bb},~m_{bb},~H_T,~p_{T,hh},~m_{hh},~\Delta R_{hh},~\met, \nonumber
\end{equation}
where, $H_{T}$ is the sum of the visible transverse momenta~(sum of the $p_{T}$ of the two $b$ $jets$ and the two final state muons) and all the variables have their usual meaning. The kinematic variables which are most effective in discriminating the signal from the backgrounds in the BDTD analysis are: $m_{bb}$, $m_{\mu\mu}$, $\met$ and $p_{T, \mu\mu}$. We illustrate the normalized distribution of these variables in Fig.~\ref{bb2mu:fig1}. The signal and background yields along with the signal significance from the BDTD analysis have been listed in Table~\ref{bbmumu:tab1}. Although, the $b\bar{b}\mu\mu$ channel offers a clean signal, it manifests a low value of signal significance due to huge contamination from the $t\bar{t}$ and $b\bar{b}\mu\mu$ backgrounds which are difficult to suppress. We also utilize XGBoost to discriminate the signal and the background events and perform a detailed collider analysis. The corresponding signal and background yields along with the signal significance are listed in Table~\ref{bbmumu:tab1}. The XGBoost analysis results in a signal significance of 0.42 (with probability cut on XGBoost output at $97\%$) which is $2$ times the signal significance obtained from the BDTD analysis, but still remains short of being within the exclusion reach of HE-LHC. In Fig.~\ref{bbmumu:fig2}, we show significance and $S/B$ $vs.$ probability cut on the XGBoost output following the color code from Fig.~\ref{bbaa:fig3}. At this point, we conclude our discussion on the future prospects of observing the non-resonant di-Higgs signature at the HE-LHC. In the next section, we study the ramifications of varying $\lambda_{h}$ on the optimized di-Higgs search strategies and the difference in the kinematics of the di-Higgs final states emerging from different values of $\lambda_{h}$.

\begin{center}
\begin{table}[htb!]
\centering
\scalebox{0.7}{%
\begin{tabular}{|c|c|c|c|c|}\hline
 & Process & Cross section order     & \multicolumn{2}{c|}{Event yield after the analysis with} \\ \cline{4-5}
 &         &           & BDTD & XGBoost \\ \hline\hline

\multirow{9}{*}{Background} 
 & $t\bar{t}~(\mu\mu)$             & NNLO~\cite{ttbarNNLO}                       & $30113$ & $24285$\\ 
 & $b\bar{b}\mu\mu$      		    & LO                                          & $32985$ & $11671$\\ 
 & $c\bar{c}\mu\mu$              	& LO                                          & $135$   & $46$\\  
 & $jj\mu\mu$             	    	& LO                                          & $14$    & $4.1$\\  
 & $Zh$             	        	& NNLO (QCD) + NLO (EW)~\cite{bkg_twiki_cs}   & $24$    & $38$\\  
 & $b\bar{b}h$                     & LO                                          & $0.8$     & $2.2$\\
 & $t\bar{t}h$                     & NLO~\cite{bkg_twiki_cs}                     & $110$   & $243$\\ \cline{2-5} 
 & \multicolumn{2}{c|}{Total}                                                    & $63382$ & $36289$\\ \hline
\multicolumn{2}{|c|}{Signal ($hh \to b\bar{b}\mu\mu$)} & NNLO~\cite{hhtwiki}     & $50$    & $80$ \\\hline 
\multicolumn{2}{|c|}{\multirow{2}{*}{Significance}} & $0\%$ $\sigma_{sys\_un}$ & $0.2$ & $0.42$ \\ \cline{3-5}
\multicolumn{2}{|c|}{} & $2\%~(5\%)$ $\sigma_{sys\_un}$ & $0.04~(0.02)$ & $0.11~(0.04)$ \\ \hline
\end{tabular}}
\caption{\it The signal and background yields at the HE-LHC along with the signal significance for the $b\bar{b}\mu^{+}\mu^{-}$ channel from the analysis using the BDTD and the XGBoost classifiers.}
\label{bbmumu:tab1}
\end{table}
\end{center}

\begin{figure}
\centering
\includegraphics[scale=0.42]{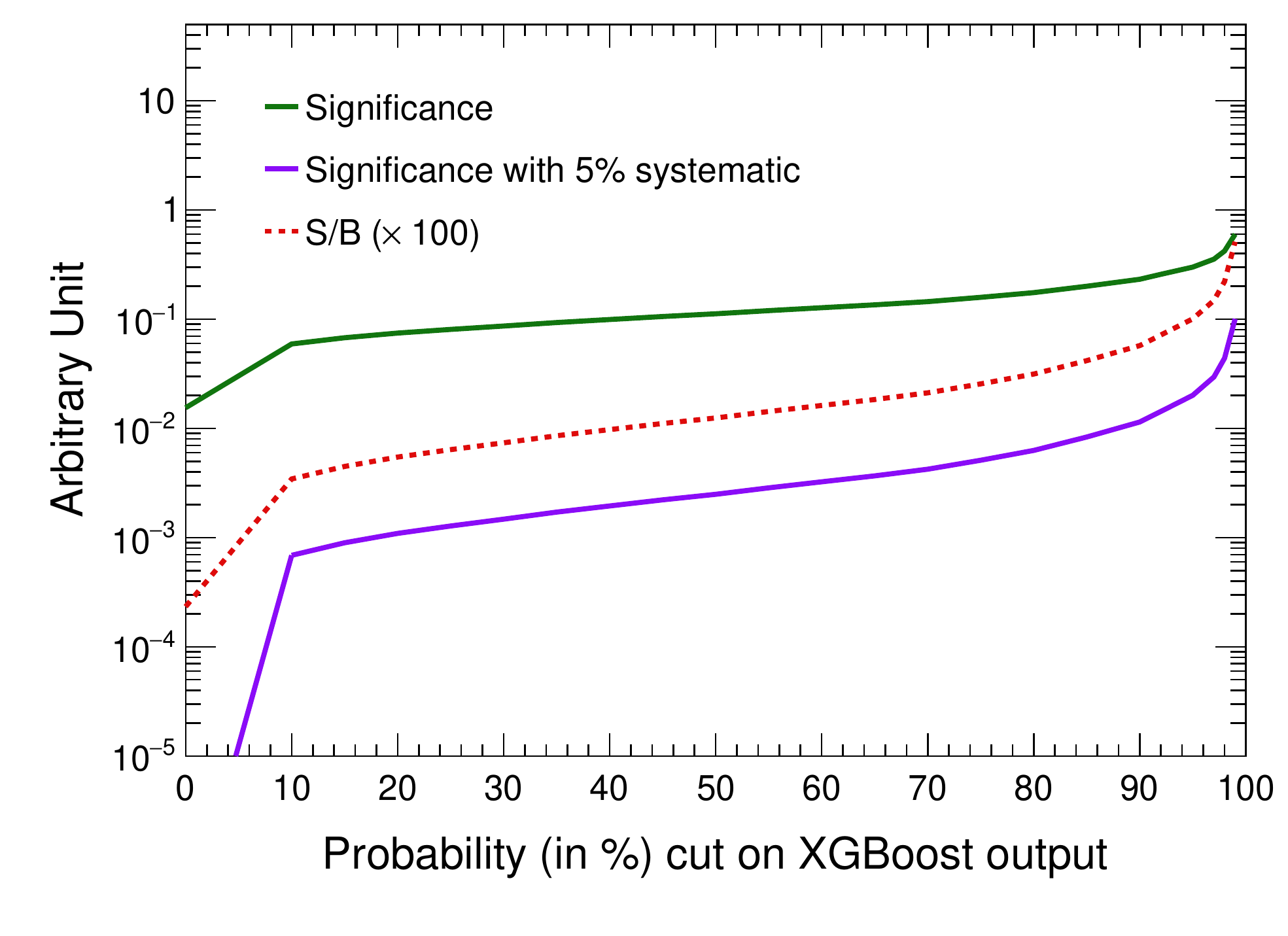}
\caption{\it The variation of significance~(with~($5\%$) and without systematic uncertainty) and $S/B$ is shown as a function of the probability cut on the XGBoost output for the $b\bar{b}\mu^+\mu^-$ channel.}
\label{bbmumu:fig2}
\end{figure}


\section{Higgs self-coupling measurement}
\label{sec:Cons-selfC}

As discussed in Sec.~\ref{intro}, the LO di-Higgs production cross-section in the SM is an outcome of the destructive interference between the triangle ($\lambda_{h}$ dependent) and the box diagrams. We must also note that the SM value of $\lambda_{h}$ is a small number. Hence, the magnitude of interference between the two diagrams is highly sensitive to even small changes in $\lambda_{h}$. This strongly necessitates a detailed study of the implications of varying $\lambda_{h}$ on the projected reach of non-resonant di-Higgs searches and the underlying changes in the kinematic distributions, which is precisely the goal of this section.

Before delving into the details of this section, we briefly summarize some of the existing results on the projected capability of constraining $k_{\lambda}$ at the HL-LHC and the HE-LHC through direct and indirect measurements. The ATLAS detector is projected to be capable of constraining $k_{\lambda}$ within $0.8  \leq k_{\lambda} \leq 7.7$ at $95\%$ CL at the HL-LHC through a direct search of non-resonant di-Higgs production in the $b\bar{b}\gamma\gamma$ channel~\cite{ATL-PHYS-PUB-2017-001}, while indirect probes of $\lambda_{h}$ at the HL-LHC have projected an exclusion range of $0.1 \leq k_{\lambda} \leq 2.3$ at $68\%$ CL~\cite{DiVita:2017eyz}. The projected performance of the HE-LHC in constraining $\lambda_{h}$ through a direct search in the $b\bar{b}\gamma\gamma$ channel has also been studied in Refs.~\cite{Homiller:2018dgu,Goncalves:2018qas}. Ref.~\cite{Homiller:2018dgu} and \cite{Goncalves:2018qas} have reported a projected sensitivity in the range of $0.6\leq \lambda_{h} \leq 1.46$ at $1\sigma$ and $k_{\lambda} = 1 \pm 30\%$ at $95\%$ CL, respectively. With these estimates in mind, and assuming a conservative approach, we study a wider range of $k_{\lambda}$ and consider $7$ different values of the same, $viz.$ $k_{\lambda} = $~4,~3,~2,~1,~-0.5,~-1 and -2.

\begin{center}
\begin{table}[htb!]
\centering
\scalebox{0.67}{
\begin{tabular}{|c|c|c|c|c|c|c|c|c|c|c|}\hline
$k_{\lambda}$ & \makecell{Signal cross-\\ section (fb)} & \multicolumn{2}{c|}{Efficiency} & \multicolumn{2}{c|}{Signal yield} & \multicolumn{2}{c|}{Background yield} & \multicolumn{2}{c|}{Significance} \\\hline 

&& \multirow{2}{*}{BDTD} & \multirow{2}{*}{XGBoost} & \multirow{2}{*}{BDTD} & \multirow{2}{*}{XGBoost} & \multirow{2}{*}{BDTD} & \multirow{2}{*}{XGBoost} & BDTD & XGBoost\\\cline{9-10}
&&&&&&&& \makecell{with $0\%~(2\%,~5\%)$ \\ $\sigma_{sys\_un}$} & \makecell{with $0\%~(2\%,~5\%)$ \\ $\sigma_{sys\_un}$}\\\hline\hline

\multicolumn{10}{|c|}{A. $pp\to hh\to b\bar{b}\gamma\gamma$}\\\hline
                                      
$-2$      &  $2.12$    &   $0.068$ & $0.07$    &  $2162$  & $2226$    & \multirow{7}{*}{$3410$} & \multirow{7}{*}{$1806$} &  $34~(21,10)$ & $45~(32,17)$\\ \cline{1-6} \cline{9-10}                                      
$-1$      &  $1.35$    &   $0.076$ & $0.078$   &  $1539$    & $1579$  &                           & &  $25~(15,7.6)$ & $33~(24,13)$\\   \cline{1-6} \cline{9-10}
$-0.5$    &  $1.04$    &   $0.081$ & $0.082$   &  $1264$  & $1279$  &                           & &  $21~(13,6.3)$ & $27~(20,11)$\\  \cline{1-6} \cline{9-10}
$1$       &  $0.37$    &   $0.103$ & $0.1$     &  $572$  & $555$     &                           & &  $9.5~(6.1,3)$ & $12.5~(9.3,5.1)$\\   \cline{1-6} \cline{9-10}
$2$       &  $0.188$   &   $0.113$ & $0.108$   &  $319$  & $305$  &                           & &  $5.4~(3.5,1.7)$ & $7~(5.3,3)$\\   \cline{1-6} \cline{9-10}
$3$       &  $0.185$   &   $0.066$ & $0.064$   &  $183$  & $178$   &                           & &  $3.1~(2,1)$ & $4.1~(3.1,1.7)$\\   \cline{1-6} \cline{9-10}
$4$       &  $0.38$    &   $0.032$ & $0.035$   &  $182$   & $199$   &                           & &  $3.1~(2,1)$ & $4.6~(3.5,1.9)$\\   \cline{1-6} \cline{9-10}
\hline\hline

\multicolumn{10}{|c|}{B. $pp\to hh\to b\bar{b}\tau\tau$}\\\hline
                                      
$-2$      &  $58.7$    &   $0.0045$  & $0.004$  &  $3962$  & $3522$ & \multirow{7}{*}{$155260$} & \multirow{7}{*}{$41795$} & $10~(1.3,0.5)$ & $17~(4,1.6)$\\ \cline{1-6} \cline{9-10}                                   
$-1$      &  $37.3$   &   $0.0052$  & $0.0046$  &  $2912$ & $2576$    &                         &  &  $7.4~(0.9,0.4)$ & $12.5~(3,1.2)$\\   \cline{1-6} \cline{9-10}
$-0.5$    &  $28.7$   &   $0.0055$  & $0.0047$  &  $2364$ & $2020$  &                          & &  $6~(0.8,0.3)$ & $9.8~(2.3,0.9)$\\  \cline{1-6} \cline{9-10}
$1$       &  $10.2$   &   $0.0071$  & $0.0064$  &  $1088$ & $981$  &                           &&  $2.8~(0.3,0.1)$ & $4.8~(1.1,0.5)$\\  \cline{1-6} \cline{9-10}
$2$       &  $5.19$    &   $0.0077$  & $0.0071$  &  $599$  & $553$  &                          & &  $1.5~(0.2,0.08)$ & $2.7~(0.6,0.3)$\\  \cline{1-6} \cline{9-10} 
$3$       &  $5.12$    &   $0.0047$  & $0.0045$  &  $361$  & $346$  &                           &&  $0.92~(0.1,0.05)$ & $1.7~(0.4,0.2)$\\   \cline{1-6} \cline{9-10}
$4$       &  $10.4$   &   $0.0023$  & $0.0021$  &  $358$  & $327$  &                           &&  $0.91~(0.1,0.05)$ & $1.6~(0.4,0.2)$\\   \cline{1-6} \cline{9-10}
\hline\hline

\multicolumn{10}{|c|}{C. $pp\to hh\to b\bar{b}WW^{*}$}\\\hline
                                      
$-2$      &  $21.2$    &   $0.01$   & $0.016$  &  $3183$      & $5093$  & \multirow{7}{*}{$531600$} & \multirow{7}{*}{$365185$} &  $4.4~(0.3,0.1)$ & $8.4~(0.7,0.3)$\\ \cline{1-6} \cline{9-10}                                    
$-1$      &  $13.5$    &   $0.0125$ & $0.019$  &  $2531$   & $3847$   &                          & &  $3.5~(0.2,0.1)$ & $6.4~(0.5,0.2)$\\  \cline{1-6} \cline{9-10} 
$-0.5$    &  $10.4$    &   $0.013$  & $0.021$  &  $2020$    & $3263$  &                           &&  $2.8~(0.2,0.08)$ & $5.4~(0.4,0.2)$\\  \cline{1-6} \cline{9-10}
$1$       &  $3.69$     &   $0.02$   & $0.03$   &  $1107$      & $1661$  &                           &&  $1.5~(0.1,0.04)$ & $2.7~(0.2,0.09)$\\   \cline{1-6} \cline{9-10}
$2$       &  $1.87$     &   $0.026$  & $0.037$  &  $729$     & $1038$  &                          & &  $1~(0.07,0.03)$ & $1.7~(0.14,0.06)$\\   \cline{1-6} \cline{9-10}
$3$       &  $1.85$     &   $0.016$  & $0.022$  &  $444$       & $611$  &                          & &  $0.61~(0.04,0.02)$ & $1~(0.08,0.03)$\\   \cline{1-6} \cline{9-10}
$4$       &  $3.75$     &   $0.0062$ & $0.009$  &  $349$    & $506$  &                         &  &  $0.48~(0.03,0.01)$ & $0.84~(0.07,0.03)$\\   \hline 
\end{tabular}
}

\caption{\it The cross-sections, signal efficiencies, signal yields, background yields, and the signal
significances at the HE-LHC, obtained from the BDTD and the XGBoost analysis are listed along with the respective values of $k_{\lambda}$ for $\sigma_{sys\_un} = 0\%,~2\%$ and $5\%$.}
\label{tab:lambda}
\end{table}
\end{center}

We generate the LO signal events for the new values of $k_{\lambda}$ in the \texttt{MG5\_aMC@NLO} framework by incorporating the UFO model file from \cite{hh}. We restrict our study of the Higgs self-coupling to the three most sensitive di-Higgs search channels from Sec.~\ref{Sec:non_res_dihiggs}, $viz.$  $b\bar{b}\gamma\gamma$, $b\bar{b}\tau^{+}\tau^{-}$ and $b\bar{b}WW^{*}$. The newly generated signal events corresponding to the different values of $k_{\lambda}$ are passed through the BDTD and XGBoost frameworks which were optimized for the SM di-Higgs signal ($k_{\lambda} = 1$). The results for the $b\bar{b}\gamma\gamma$, $b\bar{b}\tau^{+}\tau^{-}$ and the $b\bar{b}WW^{*}$ channels are listed in Table~\ref{tab:lambda}. We have shown the production cross-section of the respective final states at different $k_{\lambda}$ values along with the signal efficiency\footnote{Signal efficiency has been defined as the ratio of the number of signal events which pass the respective analysis frameworks to the total number of generated event samples.} and the signal yields at the HE-LHC obtained from both, the BDTD and the \texttt{XGBoost} frameworks optimized for the SM hypothesis. The respective total background yields are also listed along with the signal significances computed assuming zero systematic uncertainties. Among the chosen values of $k_{\lambda}$, the non-resonant double Higgs production cross-section is the smallest for $k_{\lambda}=3$, and continues to increase on its either side. The leading order squared amplitude for double Higgs production can be factorized into three different contributions: the triangle diagram ($\propto \lambda_{h}^{2}$), box diagram (independent of $\lambda_{h}$) and the interference between these two diagrams ($\propto \lambda_{h}$). For positive values of $\lambda_{h}$, the interference term contributes negatively, resulting in a destructive interference between the box and triangle diagrams~\cite{Agrawal:2019bpm}. Therefore, as we increase the values of $k_{\lambda}$ away from the SM hypothesis, we initially observe a reduction in the di-Higgs production rate due to a larger destructive interference, $viz.$ $k_{\lambda}=~2,~3$. Upon moving to an even larger value of $\lambda_{h}$ ($k_{\lambda}=4$), the contribution from the triangle loop which is proportional to $\lambda_{h}$ squared, starts compensating the reduction from the negative interference term resulting in an increase in the di-Higgs production rate. As we move towards the negative values of $k_{\lambda}$, a dramatic increase is observed in the di-Higgs cross-section since the interference term begins to contribute constructively.

\begin{figure}[!htb]
\centering
\includegraphics[scale=0.44]{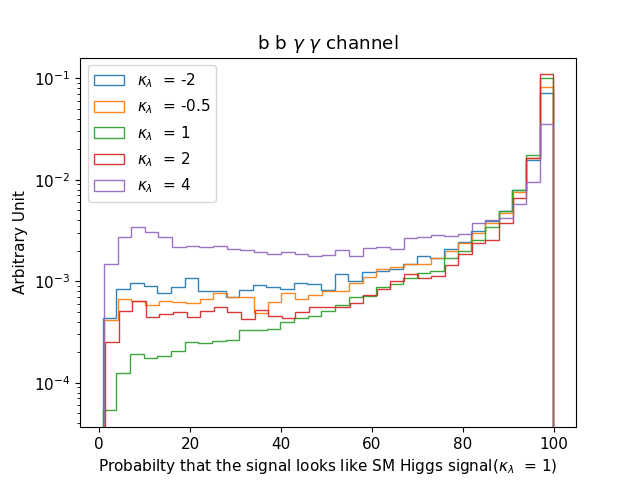}
\includegraphics[scale=0.44]{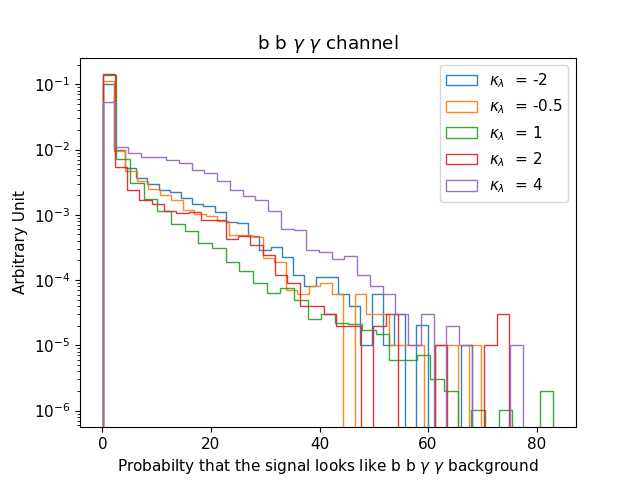}\\
\includegraphics[scale=0.44]{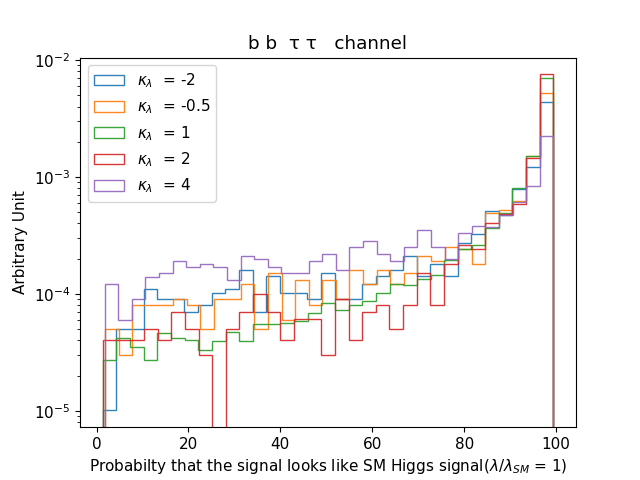}
\includegraphics[scale=0.44]{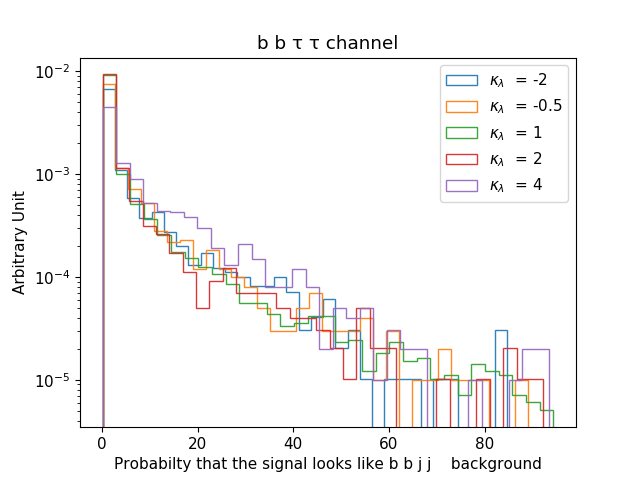}\\
\includegraphics[scale=0.44]{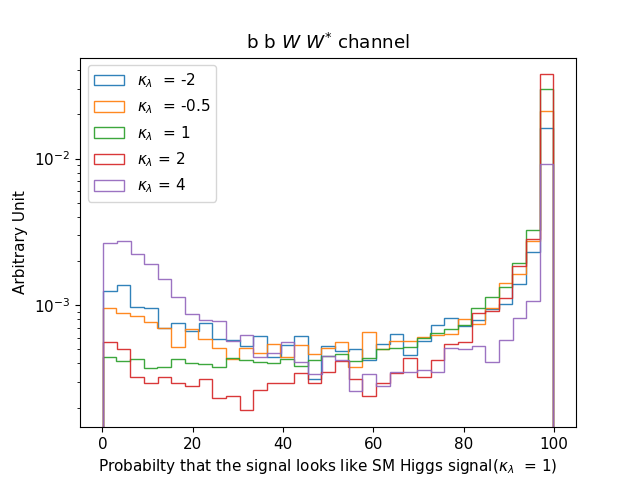}
\includegraphics[scale=0.44]{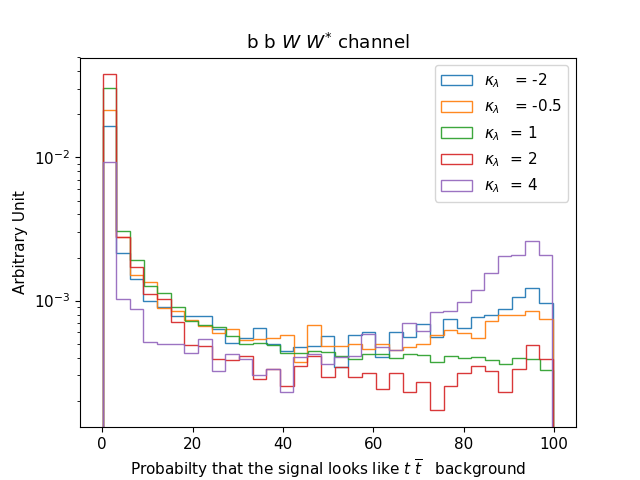}
\caption{\it Normalized distributions illustrating the probability that the signal events in the $b\bar{b}\gamma\gamma$ (top), $b\bar{b}\tau\tau$ (center) and $b\bar{b}WW^{*}$ (bottom) channel look like the respective SM di-Higgs signal (shown in the left panels) and respective dominant SM backgrounds~(shown in the right panels).}
\label{selfC:fig2}
\end{figure}

In all the di-Higgs search channels considered in this section, we observe a roughly uniform trend in the variation of signal efficiency as a function of $k_{\lambda}$. Both, the BDTD and the XGBoost classifiers~(both trained for the SM hypothesis), exhibit the highest signal efficiency for $k_{\lambda}=2$ with a gradual decrease towards higher as well as smaller values of $k_{\lambda}$. For example, in the case of $b\bar{b}\gamma\gamma$ channel, the BDTD~(XGBoost) optimization result in a signal efficiency of 0.068~(0.07), 0.103~(0.1), 0.113~(0.108) and 0.032~(0.035) at $k_{\lambda}=-2,~1,~2$ and $4$, respectively. We would like to clarify again that we have not performed the signal significance optimization for the respective $k_{\lambda}$ signals, and have used the BDTD and the XGBoost classifiers optimized for the case of $k_{\lambda} = 1$. The signals corresponding to the different values of $k_{\lambda}$ have been simply passed through the classifiers optimized for the case of $k_{\lambda}=1$. We must note that even with the classifiers optimized for the case of $k_{\lambda}=1$, the signal efficiency is highest for $k_{\lambda}=2$. This occurs despite the $k_{\lambda}=2$ signal having a relatively smaller cross-section compared to most of its counterparts. In order to attain a better viewpoint, we illustrate the normalized distribution of the probability of new physics signal events corresponding to the different values of $k_{\lambda}$ to look like the respective SM di-Higgs signal~($k_{\lambda}=1$) in the left panel of Fig.~\ref{selfC:fig2}. The probability distribution of the NP signals to look like the respective dominant SM backgrounds: QCD-QED $b\bar{b}\gamma\gamma$~(for the $b\bar{b}\gamma\gamma$ channel), $b\bar{b}jj$~(for the $b\bar{b}\tau^{+}\tau^{-}$ channel) and $t\bar{t}$~(for the $b\bar{b}WW^{*}$ channel), respectively, is illustrated in the right panels of Fig.~\ref{selfC:fig2}. The XGBoost toolkit was used to derive these probability distributions. It can be observed from the distributions in the left panel of Fig.~\ref{selfC:fig2} that the $k_{\lambda}=2$ scenario (shown in red color) has the largest number of events in the highest probability bin for all three signal channels. The $k_{\lambda}$ signals in decreasing order of signal efficiency at the highest probability bin can be noted from the left panels in Fig.~\ref{selfC:fig2} to be: $k_{\lambda}= 2$, 1~(green color), -0.5~(orange color), -2~(cyan color) and 4~(purple color). This sequence holds true for all the three di-Higgs final states considered in the present section. It must be noted that the above mentioned ordering also stands at the lowest probability bin in the background-like probability distribution of the $k_{\lambda}$ signals~(right panel of Fig.~\ref{selfC:fig2}). Furthermore, in the lower right panel of Fig.~\ref{selfC:fig2}, where we illustrate the normalized distribution of the probability that the various $k_{\lambda}$ signals in the fully-leptonic $b\bar{b}WW^{*}$ channel look like the dominant $t\bar{t}$ background, we observe a second peak towards the higher end. This indicates that the $b\bar{b}WW^{*}$ signal corresponding to $k_{\lambda} = 4$~(purple color), $k_{\lambda}=-2$~(cyan color) and $k_{\lambda}=-0.5$~(orange color), also has a relatively high probability to resemble the dominant $t\bar{t}$ background, making the signal-background discrimination more challenging. This trend is also reflected in the low signal efficiency for these $k_{\lambda}$ signals. 

\begin{figure}[!htb]
\centering
\includegraphics[scale=0.37]{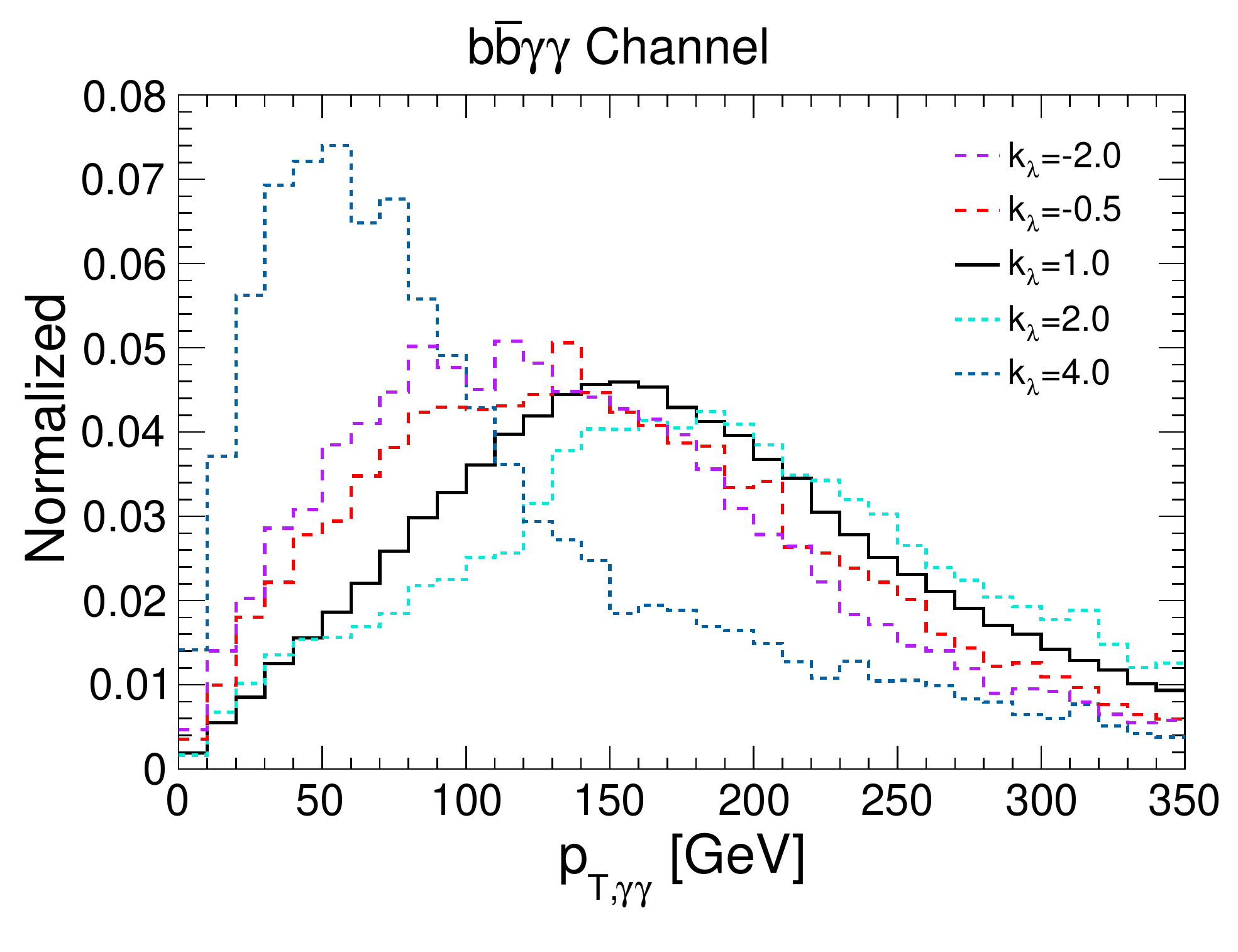}
\includegraphics[scale=0.37]{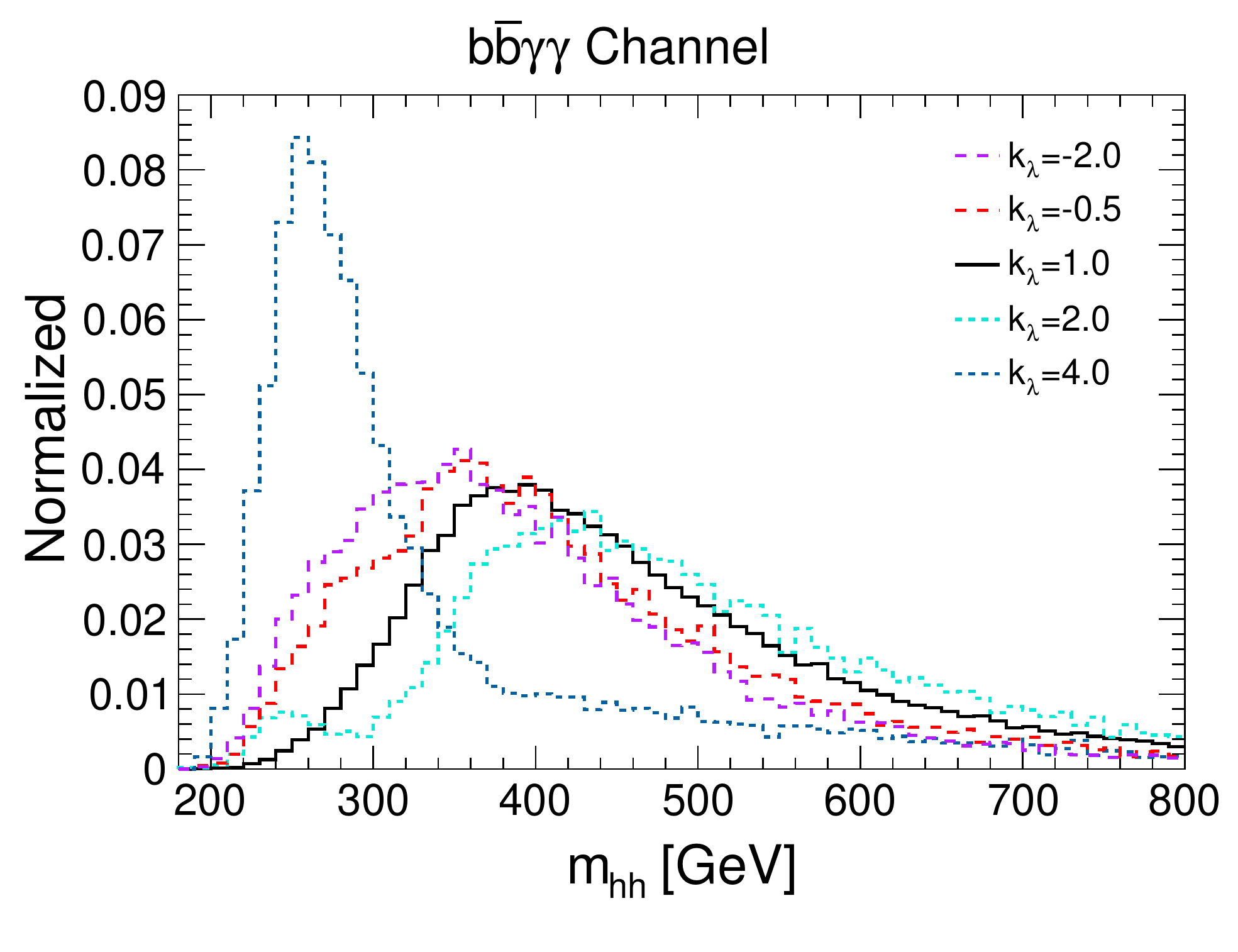}\\
\includegraphics[scale=0.37]{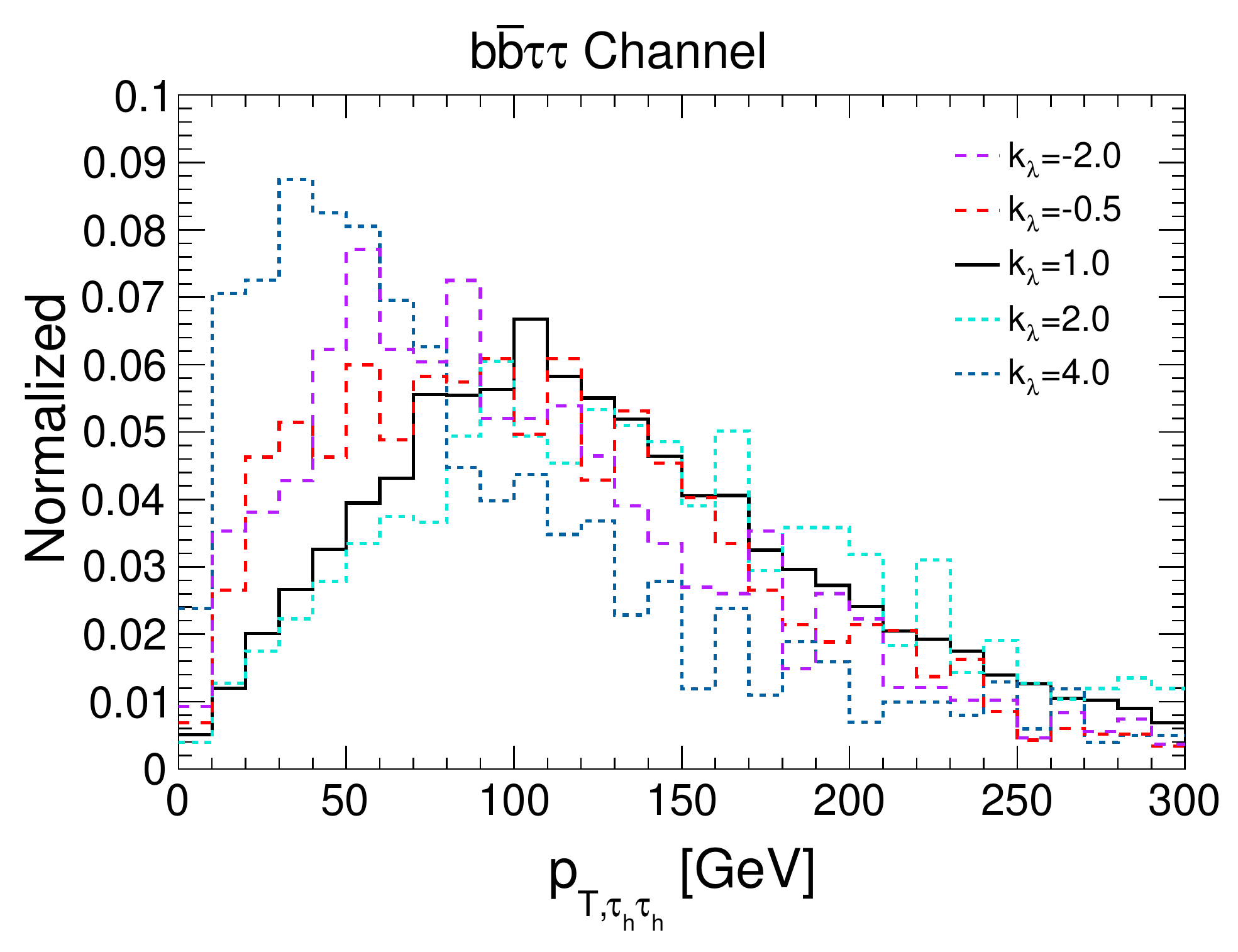}
\includegraphics[scale=0.37]{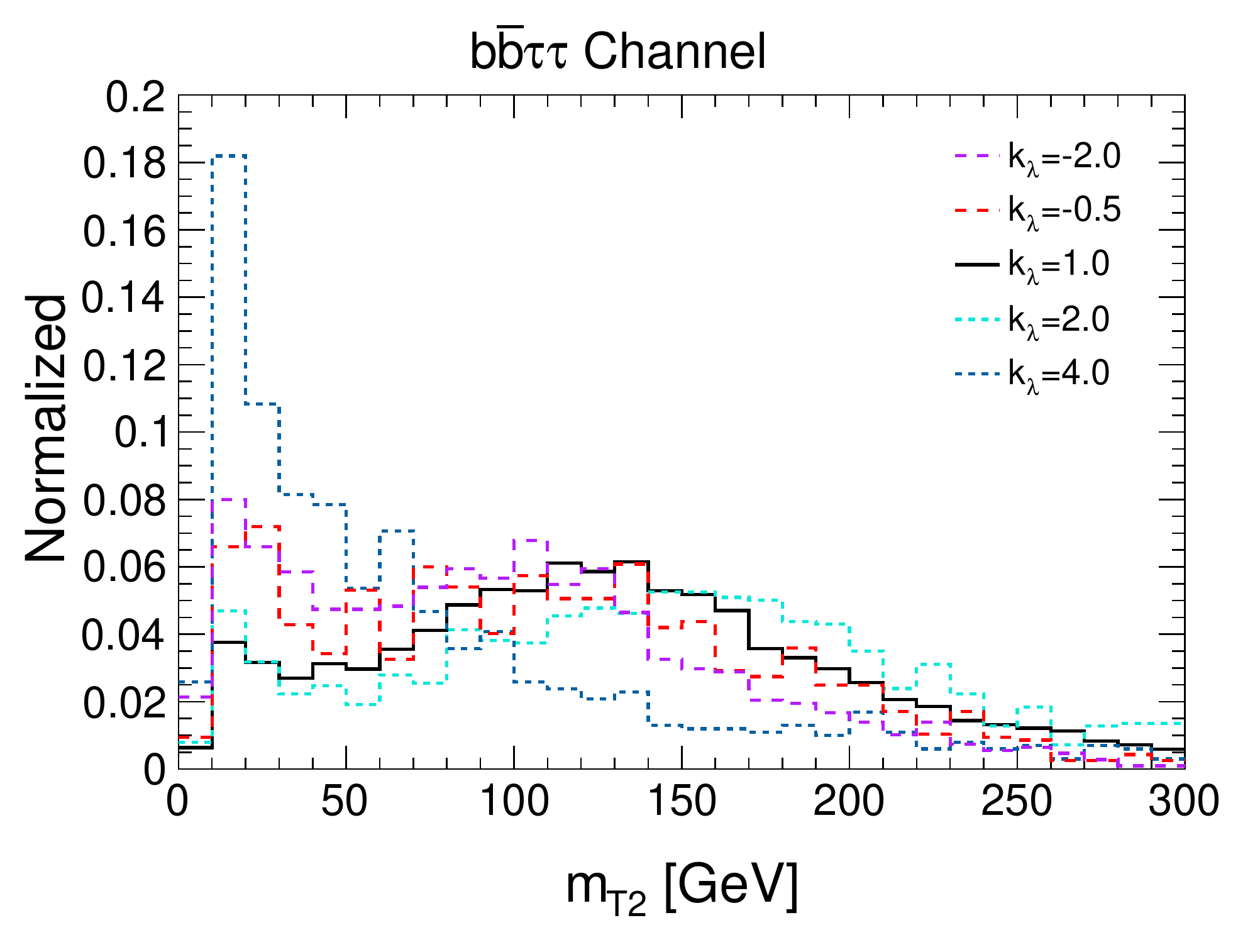}\\
\includegraphics[scale=0.37]{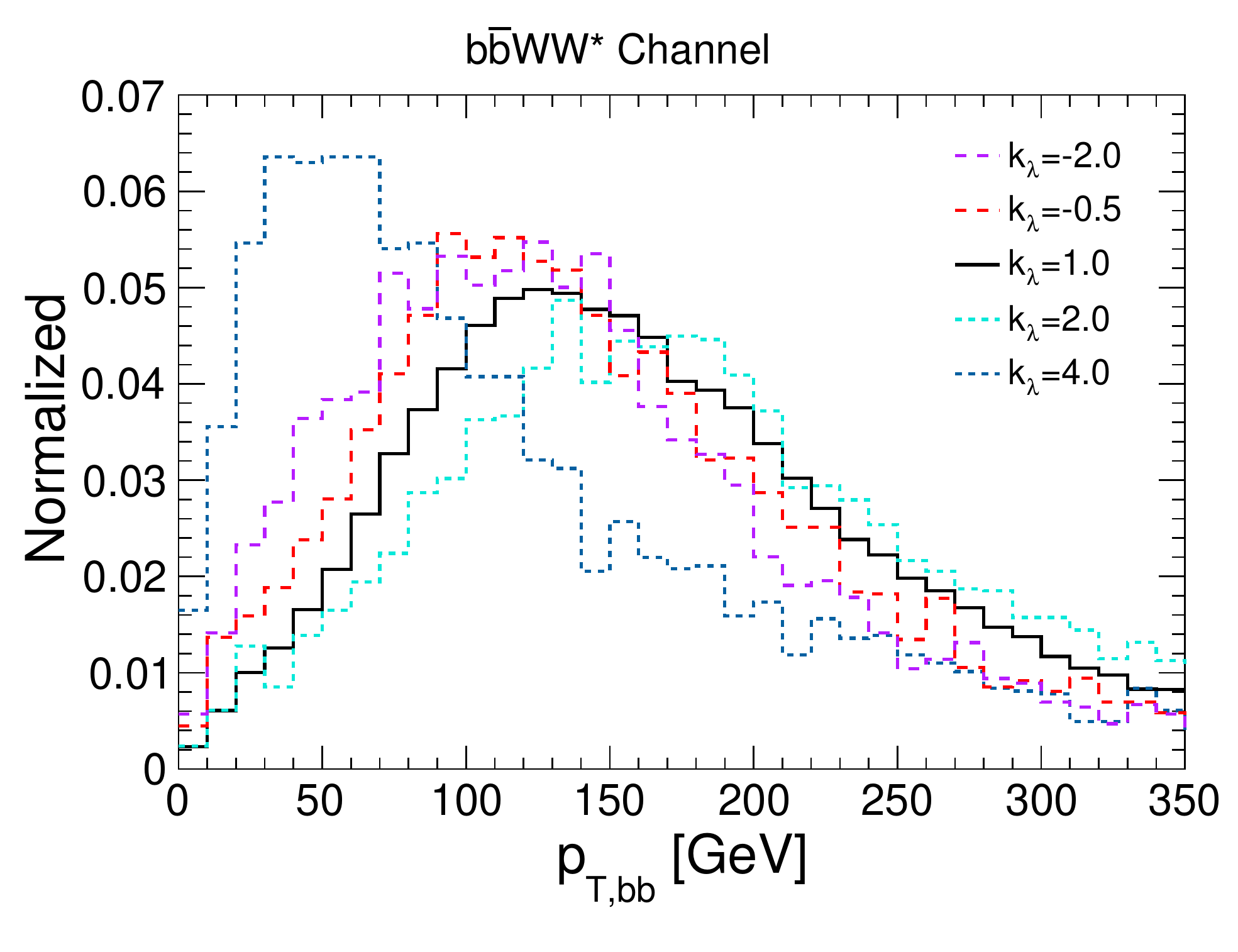}
\includegraphics[scale=0.37]{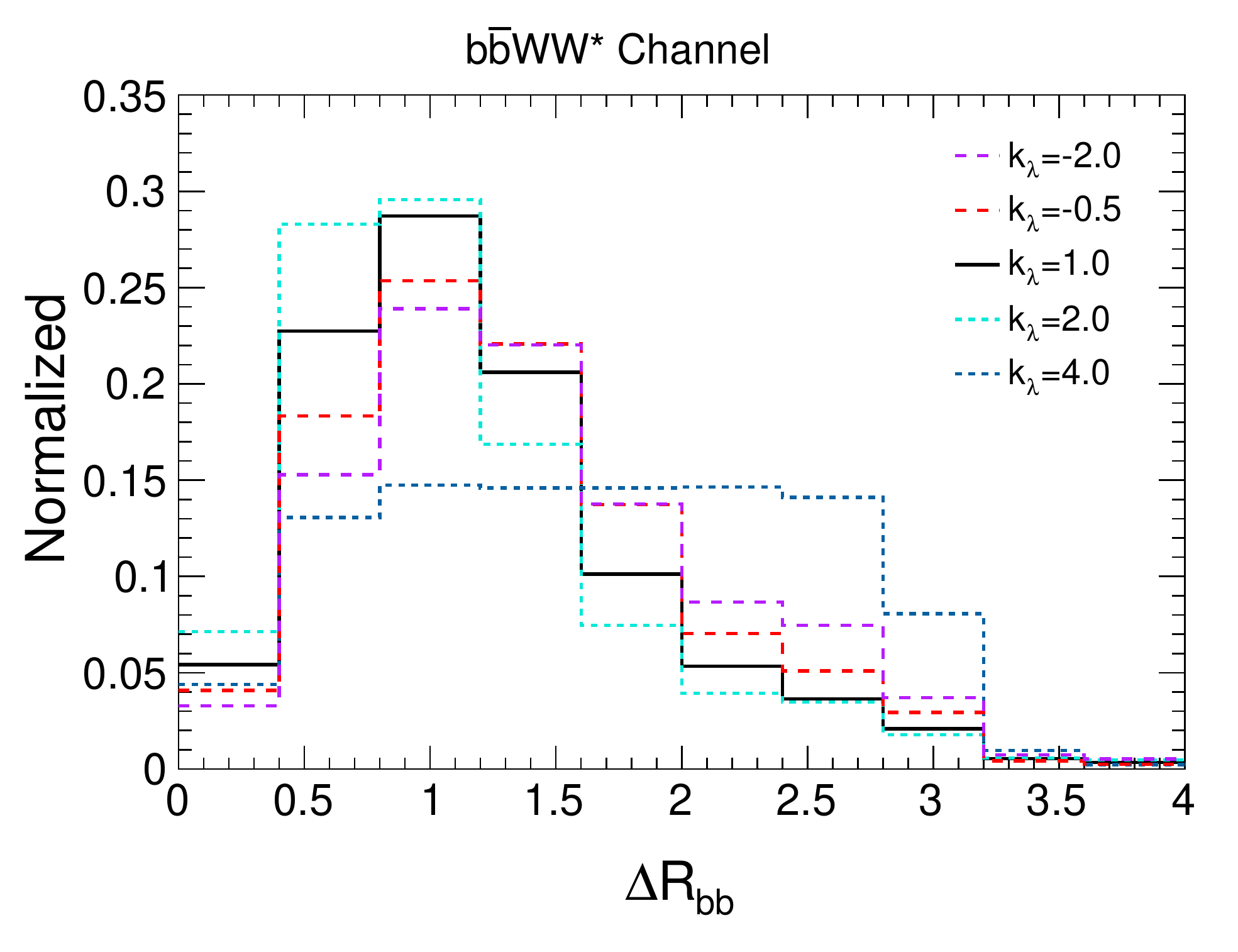}
\caption{\it Normalized distribution of (a) $p_{T,\gamma\gamma}$ and $m_{hh}$ in the $b\bar{b}\gamma\gamma$ channel (top), (b) $p_{T,\tau_h\tau_h}$ and $m_{T2}$ in the $b\bar{b}\tau\tau$ channel (middle), (c) $p_{T,bb}$ and $\Delta R_{bb}$ in the $b\bar{b}WW^{*}$ channel (bottom), for different values of $k_\lambda=\lambda/\lambda_{SM}$.}
\label{selfC:fig1}
\end{figure}

As discussed earlier, varying $k_{\lambda}$ also leads to modified kinematic distributions. In Fig.~\ref{selfC:fig1}, we illustrate the difference in kinematic distributions arising from different values of the Higgs self-coupling. In the top panel of Fig.~\ref{selfC:fig1}, we show the normalized distribution of $p_{T,\gamma\gamma}$~(left) and $m_{hh}$~(right) in the $b\bar{b}\gamma\gamma$ channel for $k_{\lambda} = -2,~-0.5,~1,~2$ and 4. In the central (bottom) panels of Fig.~\ref{selfC:fig1}, we illustrate the normalized distribution of $p_{T,\tau_{h}\tau_{h}}$~(left-side) and $m_{T2}$~(right-side) ($p_{T,bb}$~(left-side) and $\Delta R_{bb}$~(right-side)) for the $b\bar{b}\tau^{+}\tau^{-}$ ($b\bar{b}WW^{*}$) channel. Among the $p_{T}$ distributions illustrated in the left panel of Fig.~\ref{selfC:fig1}, we see that the distribution for $k_{\lambda} =4$ stands out in all three channels with a peak at relatively smaller values compared to their other counterparts. The reason behind this observation is the increased contribution from the triangle diagram. We see a roughly similar behavior in the $m_{T2}$ and $m_{hh}$ distributions. In the lower right panel of Fig.~\ref{selfC:fig1}, we observe that the $\Delta R_{bb}$ distribution for $k_{\lambda}=4$ signal becomes flatter compared to the other signals due to relatively smaller $p_{T}$ of the parent Higgs boson.

The results from this section strongly indicate that the SM di-Higgs search analyses can be very sensitive to new physics effects arising from a different value of the yet unmeasured Higgs self-coupling. It is therefore of utmost importance to have an understanding of the differences in the kinematic properties emerging from varying $\lambda_{h}$ while searching for new physics as well as the double Higgs production at the futuristic HE-LHC. We would also like to emphasize that the HE-LHC will be sensitive to the entire range of $k_{\lambda}$ considered in this section through non-resonant di-Higgs searches. The signal significances listed in Table~\ref{tab:lambda} show that the combination of search results in the $b\bar{b}\gamma\gamma$, $b\bar{b}\tau^{+}\tau^{-}$ and $b\bar{b}W^{+}W^{-}$ channels would result in a combined signal significance of $4.5\sigma$ at the HE-LHC for the $k_{\lambda} = 3$ scenario. Combination of search results in the same three channels for the $k_{\lambda}=2$ and $4$ scenarios result in a projected signal significance of $>5\sigma$ at the HE-LHC. The smaller values of $k_{\lambda}$, $viz.$ $k_{\lambda}=-0.5,~-1$ and $-2$, result in an impressive projected signal significance of $> 30\sigma$, while the SM hypothesis also exhibits a projected signal significance of $> 10\sigma$. We conclude this section by highlighting upon the need to improve the discrimination capability between the modifications arising from a variation in $\lambda_{h}$ to constrain the Higgs self-coupling more precisely. 

\section{Summary}
\label{sec:summary}

In the first part of this work, we studied the future potential of the HE-LHC in probing the non-resonant double Higgs production in the SM in several final states by performing multivariate analysis using the BDTD algorithm, the XGBoost toolkit and DNN. In this regard, we analyzed the $b\bar{b}\gamma\gamma$, fully hadronic $b\bar{b}\tau\tau$, fully leptonic $b\bar{b}WW^{*}$, fully leptonic $WW^{*}\gamma\gamma$, the $2b4l^{\prime}$ and $2b2e2\mu$ final states of $b\bar{b}ZZ^{*}$, and, the $b\bar{b}\mu\mu$ channels. The highest signal significance and $S/B$ value was observed in the $b\bar{b}\gamma\gamma$ channel, where we obtained a signal significance of $9.42,~12.46$ and $10.03$, from the BDTD, XGBoost and DNN classifiers, respectively, assuming zero systematic uncertainty. Upon assuming a systematic uncertainty of $2\%$~($5\%$), the signal significance from the BDTD, XGBoost and DNN analysis, drops down to around 5.93~(2.91), 9.31~(5.11) and 7.26~(3.9), respectively. The relatively smaller reduction in the value of signal significance from the XGBoost and the DNN analysis compared to that from the BDTD analysis, upon assuming non-zero systematic uncertainty, is a reflection of relatively larger $S/B$ value in the former two cases. The $b\bar{b}\gamma\gamma$ channel, thus, displays a discovery potential at the HE-LHC. The second most promising search channel was observed to be the $b\bar{b}\tau^{+}\tau^{-}$ final state with a signal significance of $2.77$~(BDTD), $4.78$~(XGBoost) and $4.25$~(DNN) at zero systematic uncertainty. The signal significance attains a value of [0.35,~0.14]~(BDTD), [1.13,~0.46]~(XGBoost), [1.15,~0.47]~(DNN) upon considering a systematic uncertainty of [$2\%,~5\%$]. The fully leptonic $b\bar{b}WW^{*}$ channel was also analyzed using these three classifiers and a signal significance of $1.42,~2.75$ and $1.43$ was observed from the analysis using BDTD, XGBoost and DNN, respectively, at zero systematic uncertainty. Here, the signal significance value registers a sharper drop upon assuming non-zero systematic uncertainties compared to that in the case of the $b\bar{b}\gamma\gamma$ and the $b\bar{b}\tau^{+}\tau^{-}$ channels. The signal significance value falls down to 0.09~(0.03), 0.23~(0.09) and 0.14~(0.06), from the BDTD, XGBoost and DNN analysis, respectively, upon assuming a systematic uncertainty of $2\%~(5\%)$. This drastic reduction in the signal significance can be attributed to the relatively smaller $S/B$ value. In the aforesaid channels, the performance of the XGBoost classifier was better than the BDTD and DNN. The XGBoost toolkit resulted in a signal significance which was higher than its BDTD counterpart by $\sim 32\%$, $\sim 73\%$ and $\sim 94\%$ in the $b\bar{b}\gamma\gamma$, $b\bar{b}\tau^{+}\tau^{-}$ and $b\bar{b}WW^{*}$ channels, respectively. We also analyzed the $WW^{*}\gamma\gamma$, $b\bar{b}ZZ^{*}$ and the $b\bar{b}\mu^{+}\mu^{-}$ search channels using the BDTD and the XGBoost classifiers. We obtained a signal significance of $1.64$ (BDTD) and $2.05$ (XGBoost) in the $WW^{*}\gamma\gamma$ channel and the second highest $S/B$ value after the $b\bar{b}\gamma\gamma$ channel. The $2b4l^{\prime}$ and $2b2e2\mu$ final states arising from $b\bar{b}ZZ^{*}$ were also subjected to the BDTD and XGBoost optimizations, and both of these final states resulted in a signal significance of $< 1\sigma$. However, combining the XGBoost search results for these two final states result in a signal significance of $1.3$. The $b\bar{b}ZZ^{*}$ mode also exhibited an impressive $S/B$ value. The future potential of the $b\bar{b}\mu^{+}\mu^{-}$ channel at the HE-LHC was also analyzed, and we obtained a signal significance of $0.2$ and $0.42$ using the BDTD and the XGBoost classifiers, respectively. Combination of results from all the search channels considered in this work yield a combined projected signal significance of $\sim 10$ and $\sim 14$ from the BDTD and the XGBoost classifiers, respectively. These projections indicate towards the possibility to observe the non-resonant di-Higgs signal at the HE-LHC at discovery potential.

In the second part, the implications of varying $k_{\lambda}$ on the potential reach of the $b\bar{b}\gamma\gamma$, $b\bar{b}\tau^{+}\tau^{-}$ and $b\bar{b}WW^{*}$ search analyses optimized for the SM hypothesis are studied along with the variations in the kinematics of the di-Higgs final states emerging from varying $k_{\lambda}$. In this respect, we consider the following values of $k_{\lambda}$: -2,~-1,~-0.5,~1,~2,~3 and 4. The signal events corresponding to these different values of $k_{\lambda}$ were passed through the BDTD and XGBoost classifiers trained for the case of $k_{\lambda}=1$. Both, the BDTD and the XGBoost classifiers, exhibit a signal significance of $>5\sigma$ for $k_{\lambda}=$~-2~,-1~,-0.5,~1, and 2, and a signal significance of $>2\sigma$ for $k_{\lambda}=$~3 and 4, at zero systematic uncertainty. Combination of searches in the $b\bar{b}\gamma\gamma$, $b\bar{b}\tau^{+}\tau^{-}$ and $b\bar{b}WW^{*}$ channels result in a signal significance of $>5\sigma$ even for the $k_{\lambda}=4$ scenario. Upon assuming a systematic uncertainty of $2\%$, the BDTD~(XGBoost) classifier exhibits a signal significance of $>5\sigma$ for $k_{\lambda}=$~-2,~-1,~-0.5, and 1~($k_{\lambda}=$~-2,~-1,~-0.5,~1 and 2), and a signal significance of $> 2\sigma$ for $k_{\lambda}=$~2,~3, and 4~($k_{\lambda}=$~3 and 4), in the $b\bar{b}\gamma\gamma$ channel. If the systematic uncertainty is increased to $5\%$, then both, BDTD and XGBoost classifiers, exhibit a signal significance of $> 5\sigma$ only for $k_{\lambda} =$~-2,~-1, and -0.5, and $k_{\lambda} =$~-2,~-1,~-0.5 and 1.0, respectively. For $k_{\lambda} = 1$ and $k_{\lambda} = 2$, the signal significance from the BDTD and the XGBoost classifier, respectively, is $>2\sigma$. Both the classifiers fall short of the exclusion reach for $k_{\lambda} = 3,4$. Our projections thus indicate that the HE-LHC would be sensitive to the entire range of $k_{\lambda}$ considered in this work through direct searches in the non-resonant di-Higgs search channels.

\acknowledgments
The authors thank Shankha Banerjee for the helpful discussions on this work. The work of BB is supported by the Department of Science and Technology, Government of India, under the Grant Agreement number IFA13-PH-75 (INSPIRE Faculty Award).

\bibliographystyle{JHEP}
\bibliography{refs}

\newpage 
\appendix

\section{Outlining the generation cuts and production cross sections for the signal and backgrounds}
\label{sec:appendixA}

\begin{table}[htb!]
\centering
\begin{bigcenter}
\scalebox{0.7}{%
\begin{tabular}{|c|c|c|c|c|}

\hline
Process & Signal and Backgrounds & \makecell{Generation-level cuts ($\ell=e^\pm,\mu^\pm$)\\ (NA : Not Applied)} & Cross section (fb)  \\ \hline

\multirow{13}{*}{$b \bar{b} \gamma\gamma$} & \multicolumn{1}{l|}{Signal ($hh\to b \bar{b} \gamma \gamma$)} & \multicolumn{1}{l|}{NA} & \multicolumn{1}{l|}{$0.37$}   \\\cline{2-4}

                                           & \multicolumn{1}{l|}{$b\bar{b}\gamma\gamma+$ jet} 
                                           & \multicolumn{1}{l|}{\makecell{$p_{T,j/b/\gamma}>20~\text{GeV}$, $|\eta_{j/b/\gamma}|<5.0$, $\Delta R_{j,b,\gamma}$\footnote{$\Delta R_{b,j,\gamma}$ means $\Delta R$ between all possible combination of $b/j/\gamma$.}$>0.2$, \\ $m_{bb}>50$ GeV, $110<m_{\gamma\gamma}>140$ GeV}} 
                                           & \multicolumn{1}{l|}{$134.84$}\\\cline{2-4}  

                                           & \multicolumn{1}{l|}{$c\bar{c}\gamma\gamma$} 
                                           & \multicolumn{1}{l|}{\makecell{$p_{T,j/\gamma}>20~\text{GeV}$, $|\eta_{j/\gamma}|<5.0$, $\Delta R_{j,\gamma}>0.2$, \\ $m_{jj}>50$ GeV, $110<m_{\gamma\gamma}>140$ GeV}} 
                                           & \multicolumn{1}{l|}{$705.02$\footnote{$c\to b$ average fake rate is $3.5\%$\cite{Sirunyan:2017ezt}.\label{cfakeb}}}\\\cline{2-4}                                                                                                                             
                                           
                                           & \multicolumn{1}{l|}{$jj\gamma\gamma$} 
                                           & \multicolumn{1}{l|}{same as $c\bar{c}\gamma\gamma$} 
                                           & \multicolumn{1}{l|}{$12772.97$\footnote{$j\to b$ average fake rate is $0.135\%$\cite{Sirunyan:2017ezt}.\label{jfakeb}}}\\\cline{2-4}                                                                                                                                                                      

                                           & \multicolumn{1}{l|}{$t\bar{t}h$, $h\to\gamma\gamma$} & \multicolumn{1}{l|}{$p_{T,b/\gamma}>20~\text{GeV}$, $|\eta_{b/\gamma}|<5.0$, $\Delta R_{bb/\gamma\gamma}>0.2$}  & \multicolumn{1}{l|}{$5.19$}  \\\cline{2-4}

                                           & \multicolumn{1}{l|}{$b\bar{b}h$, $h\to\gamma\gamma$} & \multicolumn{1}{l|}{\makecell{$p_{T,b/\gamma}>20~\text{GeV}$, $|\eta_{b/\gamma}|<5.0$, $\Delta R_{b,\gamma}>0.2$, \\ $m_{bb}>50$ GeV}}  & \multicolumn{1}{l|}{$0.32$}   \\\cline{2-4}                                      
                                                                                     
                                           & \multicolumn{1}{l|}{$Zh$, $Z\to b\bar{b}$, $h\to\gamma\gamma$} & \multicolumn{1}{l|}{same as $b\bar{b}h$}  & \multicolumn{1}{l|}{ $0.46$}   \\\cline{2-4}
                                           
                                           & \multicolumn{1}{l|}{$b\bar{b}jj$} 
                                           & \multicolumn{1}{l|}{\makecell{$p_{T,j/b}>20~\text{GeV}$, $|\eta_{j/b}|<5.0$, $\Delta R_{j,b}>0.2$, \\ $m_{bb}>50$ GeV, $m_{jj}>50$ GeV}} & \multicolumn{1}{l|}{$650847671.98$\footnote{$j\to\gamma$ fake rate is $0.05\%$\cite{ATL-PHYS-PUB-2017-001}.\label{jfakea}}} \\\cline{2-4}                                                                                 
                                                                                     
                                           & \multicolumn{1}{l|}{$b\bar{b}j\gamma$} 
                                           & \multicolumn{1}{l|}{\makecell{$p_{T,j/b/\gamma}>20~\text{GeV}$, $|\eta_{j/b/\gamma}|<5.0$, $\Delta R_{j,b,\gamma}>0.2$, \\ $m_{bb}>50$ GeV}}                                 & \multicolumn{1}{l|}{$1078323.46$\footref{jfakea}}                                 \\\cline{2-4}                                                                      
                                           
                                           & \multicolumn{1}{l|}{$c\bar{c}j\gamma$} 
                                           & \multicolumn{1}{l|}{$p_{T,b/\gamma}>20~\text{GeV}$, $|\eta_{b/\gamma}|<5.0$, $\Delta R_{bb/\gamma\gamma}>0.2$}                                  
                                           & \multicolumn{1}{l|}{$4399502.87$\footref{cfakeb}\footref{jfakea}} \\\cline{2-4}                                                     
                                                       
                                           & \multicolumn{1}{l|}{$Z\gamma\gamma+$ jet, $Z\to b\bar{b}$} & {same as $b\bar{b}\gamma\gamma+$ jet} & \multicolumn{1}{l|}{$3.56$}   \\\cline{2-4}                                                                                                                                                                  
                                           
                                           & \multicolumn{1}{l|}{$hc\bar{c}$, $h\to\gamma\gamma$} & {\makecell{$p_{T,j/\gamma}>20~\text{GeV}$, $|\eta_{j/\gamma}|<5.0$, $\Delta R_{j,\gamma}>0.2$, \\ $m_{jj}>50$ GeV}} & \multicolumn{1}{l|}{$0.14$\footref{cfakeb}}   \\\cline{2-4}                                                                                                                                                                  
    
                                           & \multicolumn{1}{l|}{$hjj$, $h\to\gamma\gamma$} & \multicolumn{1}{l|}{same as $hc\bar{c}$} & \multicolumn{1}{l|}{$20.19$\footref{jfakeb}}   \\\hline


\multirow{13}{*}{$b \bar{b} \tau \tau$} 
                                           & \multicolumn{1}{l|}{Signal ($hh\to b \bar{b} \tau \tau$)} & \multicolumn{1}{l|}{NA} & \multicolumn{1}{l|}{$10.22$}   \\\cline{2-4}

                                           & \multicolumn{1}{l|}{$t\bar{t}$ had} & \multicolumn{1}{l|}{\makecell{$p_{T,j/b}>20~\text{GeV}$, $|\eta_{j/b}|<5.0$, $\Delta R_{j,b}>0.2$, \\ $m_{bb}>50$ GeV }}  & \multicolumn{1}{l|}{$657474.63$\footnote{$j\to\tau$ fake rate is $0.35\%$~\cite{CMS-PAS-TAU-16-002}.\label{jfaketau}}}  \\\cline{2-4}
                                                                                      
                                           & \multicolumn{1}{l|}{$t\bar{t}$ semi-lep} & \multicolumn{1}{l|}{\makecell{$p_{T,j/b}>20~\text{GeV}$, $p_{T,\ell}>15~\text{GeV}$, $|\eta_{j/b/\ell}|<5.0$, \\ $\Delta R_{j,b,\ell}>0.2$, $m_{bb}>50$ GeV }}  & \multicolumn{1}{l|}{$803269.61$}   \\\cline{2-4}

                                           & \multicolumn{1}{l|}{$t\bar{t}$ lep} & \multicolumn{1}{l|}{same as $t\bar{t}$ semi-lep}  & \multicolumn{1}{l|}{$245010.26$}   \\\cline{2-4}
                                                                                                                                 
                                           & \multicolumn{1}{l|}{$b\bar{b}\tau\tau$} & \multicolumn{1}{l|}{\makecell{$p_{T,b}>20~\text{GeV}$, $p_{T,\ell}>15~\text{GeV}$, $|\eta_{b/\ell}|<5.0$, \\ $\Delta R_{b,\ell}>0.2$, $m_{bb}>50$ GeV, $m_{\ell\ell}>30$ GeV}}  & \multicolumn{1}{l|}{$8280.06$}   \\\cline{2-4}
                                                                                      
                                            & \multicolumn{1}{l|}{$b\bar{b}h$, $h\to\tau\tau$} & \multicolumn{1}{l|}{\makecell{$p_{T,b}>20~\text{GeV}$, $p_{T,\ell}>15~\text{GeV}$, $|\eta_{b/\ell}|<5.0$, \\ $\Delta R_{b,\ell}>0.2$, $m_{bb}>50$ GeV}}  & \multicolumn{1}{l|}{$6.14$}  \\\cline{2-4}
                                                                                                                                                                             
                                           & \multicolumn{1}{l|}{Zh} & \multicolumn{1}{l|}{NA}  & \multicolumn{1}{l|}{$56.39$}   \\\cline{2-4}
                                           & \multicolumn{1}{l|}{$t\bar{t}h$} & \multicolumn{1}{l|}{NA}  & \multicolumn{1}{l|}{$2860.00$}  \\\cline{2-4}
                                           & \multicolumn{1}{l|}{$t\bar{t}Z$} & \multicolumn{1}{l|}{NA}  & \multicolumn{1}{l|}{$3477.02$}  \\\cline{2-4}
                                           & \multicolumn{1}{l|}{$t\bar{t}W$} & \multicolumn{1}{l|}{NA}  & \multicolumn{1}{l|}{$986.58$}   \\\cline{2-4}
                                           & \multicolumn{1}{l|}{$bbjj$} & \multicolumn{1}{l|}{\makecell{$p_{T,j/b}>20~\text{GeV}$, $|\eta_{j/b}|<5.0$, $\Delta R_{j,b}>0.2$, \\ $m_{bb}>50$ GeV, $m_{jj}>50$ GeV}}  & \multicolumn{1}{l|}{$650847671.98$\footref{jfaketau}}   \\\hline    
\end{tabular}}
\end{bigcenter}
\caption{Generation level cuts and cross-sections for the various backgrounds used in the analyses.}
\label{app1:1}
\end{table}

\begin{table}[htb!]
\centering
\begin{bigcenter}
\scalebox{0.6}{%
\begin{tabular}{|c|c|c|c|c|}

\hline
Process & Backgrounds & \makecell{Generation-level cuts ($\ell=e^\pm,\mu^\pm$)\\ (NA : Not Applied)} & Cross section (fb)  \\ \hline


\multirow{7}{*}{$b \bar{b} WW^*$}          
                                           & \multicolumn{1}{l|}{Signal ($hh\to b\bar{b}WW^*$, $W\to \ell\nu$)} & \multicolumn{1}{l|}{NA}  & \multicolumn{1}{l|}{$34.82$}  \\\cline{2-4}
                                           
                                           & \multicolumn{1}{l|}{$t\bar{t}$ lep} & \multicolumn{1}{l|}{\makecell{$p_{T,j/b}>20~\text{GeV}$, $p_{T,\ell}>15~\text{GeV}$, $|\eta_{j/b/\ell}|<5.0$, \\ $\Delta R_{j,b,\ell}>0.2$, $m_{bb}>50$ GeV}}  & \multicolumn{1}{l|}{$245010.26$}  \\\cline{2-4}
                                                                                      
                                           & \multicolumn{1}{l|}{$\ell\ell b\bar{b}$} & \multicolumn{1}{l|}{\makecell{$p_{T,b}>20~\text{GeV}$, $p_{T,\ell}>15~\text{GeV}$, $|\eta_{b/\ell}|<5.0$, \\ $\Delta R_{b,\ell}>0.2$, $m_{bb}>50$ GeV}}  & \multicolumn{1}{l|}{$25794.58$}   \\\cline{2-4}                                                                            
                                                                                     
                                           & \multicolumn{1}{l|}{$t\bar{t}h$} & \multicolumn{1}{l|}{NA}  & \multicolumn{1}{l|}{$2860.00$}  \\\cline{2-4}
                                           & \multicolumn{1}{l|}{$t\bar{t}Z$} & \multicolumn{1}{l|}{NA}  & \multicolumn{1}{l|}{$3477.02$}  \\\cline{2-4}
                                           & \multicolumn{1}{l|}{$t\bar{t}W$} & \multicolumn{1}{l|}{NA}  & \multicolumn{1}{l|}{$986.58$}   \\\cline{2-4} 
                                           & \multicolumn{1}{l|}{$tW$} & \multicolumn{1}{l|}{\makecell{$p_{T,j/b}>20~\text{GeV}$, $p_{T,\ell}>10~\text{GeV}$, $|\eta_{j/b/\ell}|<5.0$, \\ $\Delta R_{j,b,\ell}>0.2$, $m_{bb}>50$ GeV}}  & \multicolumn{1}{l|}{$6.59$}   \\\hline

\multirow{6}{*}{$WW^*\gamma\gamma$ }       & \multicolumn{1}{l|}{Signal ($hh\to WW^*\gamma\gamma$, $W\to \ell\nu$)} & \multicolumn{1}{l|}{NA} & \multicolumn{1}{l|}{$0.14$}  \\\cline{2-4}                                                                   
                                                                                                                                 
                                           & \multicolumn{1}{l|}{$t\bar{t}h$, $h\to \gamma\gamma$, $t\to b~W\to b~\ell\nu$} & \multicolumn{1}{l|}{\makecell{$p_{T,\gamma}>20~\text{GeV}$, $p_{T,\ell}>10~\text{GeV}$, $|\eta_{\gamma/\ell}|<5.0$, \\ $\Delta R_{\gamma\gamma/\ell\ell}>0.2$}}  & \multicolumn{1}{l|}{$1.74$}   \\\cline{2-4}
                                           
                                           & \multicolumn{1}{l|}{$Zh$ + jets, $Z\to \ell\ell$, $h\to\gamma\gamma$} & \multicolumn{1}{l|}{\makecell{$p_{T,j/\gamma}>20~\text{GeV}$, $p_{T,\ell}>10~\text{GeV}$, $|\eta_{j/\gamma/\ell}|<5.0$, \\ $\Delta R_{\gamma\gamma/\ell\ell}>0.2$}}  & \multicolumn{1}{l|}{$0.29$}  \\\cline{2-4}
                                                                                            
                                           & \multicolumn{1}{l|}{$\ell\ell\gamma\gamma$ + jet} & \multicolumn{1}{l|}{\makecell{$p_{T,j/\gamma}>20~\text{GeV}$, $p_{T,\ell}>10~\text{GeV}$, $|\eta_{j/\gamma/\ell}|<5.0$, $m_{\ell\ell}>20$ GeV, \\ $\Delta R_{\gamma\gamma/\ell\ell/\gamma\ell/\gamma j}>0.2$, $120<m_{\gamma\gamma}<130$ GeV}}  & \multicolumn{1}{l|}{$3.77$}  \\\hline  


\multirow{9}{*}{$b\bar{b}ZZ^*$}          & \multicolumn{1}{l|}{Signal ($hh\to b\bar{b}ZZ^*$, $Z\to \ell\ell$)} & \multicolumn{1}{l|}{NA} & \multicolumn{1}{l|}{$0.0218$}  \\\cline{2-4}                                                                                                        
                                                                                                                                 
                                           & \multicolumn{1}{l|}{$t\bar{t}h$ (all categories combined)}                       & \multirow{6}{*}{$p_{T,b}>20~\text{GeV}$, $|\eta_{b}|<5.0$, $\Delta R_{bb}>0.2$, $m_{bb}>50$ GeV}  & \multicolumn{1}{l|}{$171.47$}   \\\cline{2-2}\cline{4-4}
                                           
                                           & \multicolumn{1}{l|}{$t\bar{t}Z$, ($t\to bW, W\to \ell\nu$), ($Z\to \ell\ell$)}                       &   & \multicolumn{1}{l|}{$36.92$}   \\\cline{2-2}\cline{4-4}
                                           
                                           & \multicolumn{1}{l|}{$Zh$, $Z\to b\bar{b}$, $h\to ZZ^*\to 4\ell$}                             &   & \multicolumn{1}{l|}{$0.06$}   \\\cline{2-2}\cline{4-4}
                                           
                                           & \multicolumn{1}{l|}{$Wh$, $W\to cs$, $h\to ZZ^*\to 4\ell$}    &   & \multicolumn{1}{l|}{$0.21$\footref{cfakeb}\footref{jfakeb}}  \\\cline{2-2}\cline{4-4}
                                                       
                                           & \multicolumn{1}{l|}{$Whc$, $W\to cs$, $h\to ZZ^*\to 4\ell$}    &   & \multicolumn{1}{l|}{$0.003$\footref{cfakeb}}  \\\cline{2-2}\cline{4-4}                                                  
                                                       
                                           & \multicolumn{1}{l|}{ggF-$hb\bar{b}$, $h\to ZZ^*\to 4\ell$}             &   & \multicolumn{1}{l|}{$0.04$}   \\\cline{2-2}\cline{4-4}                                                 
                                                                                            
                                           & \multicolumn{1}{l|}{ggF-$hc\bar{c}$, $h\to ZZ^*\to 4\ell$}             &   & \multicolumn{1}{l|}{$0.02$\footref{cfakeb}}   \\\cline{2-2}\cline{4-4}                                                 

                                           & \multicolumn{1}{l|}{VBF-$hjj$, $h\to ZZ^*\to 4\ell$}             &   & \multicolumn{1}{l|}{\footref{jfakeb}$0.44\times 10^{-5}$}   \\
                                           
                                           \hline

     
\multirow{8}{*}{$b\bar{b}\mu\mu$}          & \multicolumn{1}{l|}{Signal ($hh\to b\bar{b}\mu\mu$)} & \multicolumn{1}{l|}{NA} & \multicolumn{1}{l|}{$0.0355$}  \\\cline{2-4}                                                                   
                                                                                                              
                                           & \multicolumn{1}{l|}{$t\bar{t}$, $t\to bW,~W\to\mu\nu_\mu$}  & \multicolumn{1}{l|}{\makecell{$p_{T,b/\ell}>20~\text{GeV}$, $|\eta_{b/\ell}|<5.0$, $\Delta R_{b,\ell}>0.2$ \\ $m_{bb}>50$ GeV, $m_{\ell\ell}>100$ GeV}}  & \multicolumn{1}{l|}{$12952.01$}   \\\cline{2-4}                                       
                                                                                                                                 
                                           & \multicolumn{1}{l|}{$b\bar{b}\mu\mu$}                       & \multirow{6}{*}{same as $t\bar{t}$}  & \multicolumn{1}{l|}{$351.28$}   \\\cline{2-2}\cline{4-4}
                                           
                                           & \multicolumn{1}{l|}{$c\bar{c}\mu\mu$}                       &   & \multicolumn{1}{l|}{$313.79$\footref{cfakeb}}   \\\cline{2-2}\cline{4-4}
                                           
                                           & \multicolumn{1}{l|}{$jj\mu\mu$}                             &   & \multicolumn{1}{l|}{$13930.93$\footref{jfakeb}}   \\\cline{2-2}\cline{4-4}
                                           
                                           & \multicolumn{1}{l|}{$Zh$, $Z\to b\bar{b}$, $h\to\mu\mu$}    &   & \multicolumn{1}{l|}{$0.04$}  \\\cline{2-2}\cline{4-4}
                                                       
                                           & \multicolumn{1}{l|}{$b\bar{b}h$, $h\to \mu\mu$}             &   & \multicolumn{1}{l|}{$0.02$}   \\\cline{2-2}\cline{4-4}                                                 
                                                                                            
                                           & \multicolumn{1}{l|}{$t\bar{t}h$, $h\to \mu\mu$}             &   & \multicolumn{1}{l|}{$0.46$}   \\\hline

\end{tabular}}
\end{bigcenter}
\caption{Generation level cuts and cross-sections for the various backgrounds used in the analyses.}
\label{app1:2}
\end{table}

\end{document}